\documentclass[reprint,superscriptaddress,amsmath,amssymb,aps,prb,citeautoscript]{revtex4-1}
\usepackage{subfig}
\usepackage{rotating}
\usepackage{color}
\usepackage{multirow}
\usepackage{xcolor}
\usepackage{appendix}
\usepackage{dsfont}

\DeclareMathOperator{\Tr}{Tr}

\DeclareGraphicsExtensions{.png,.pdf,.eps}
\graphicspath{{figures/}}
\usepackage{epstopdf}

\begin{document}
\title{Elastic dipole tensors and relaxation volumes of point defects in concentrated random magnetic Fe-Cr alloys}
\author{Jan S. Wr\'obel}
\email[Corresponding author: ]{jan.wrobel@pw.edu.pl}
\affiliation{Faculty of Materials Science and Engineering, Warsaw University of Technology, ul. Wo\l{}oska 141, 02-507 Warsaw, Poland}
\author{Marcin R. Zem\l{}a}
\affiliation{Faculty of Materials Science and Engineering, Warsaw University of Technology, ul. Wo\l{}oska 141, 02-507 Warsaw, Poland}
\author{Duc Nguyen-Manh}
\affiliation{UK Atomic Energy Authority, Culham Centre for Fusion Energy, Abingdon, Oxfordshire OX14~3DB, United Kingdom}
\author{P\"ar Olsson}
\affiliation{KTH Royal Institute of Technology, Department of Physics, 106 91 Stockholm,   Sweden}
\author{Luca Messina}
\affiliation{KTH Royal Institute of Technology, Department of Physics, 106 91 Stockholm, Sweden}
\author{Christophe Domain}
\affiliation{EdF-R$\&$D, D\'epartement Mat\'eriaux et M\'ecanique des Composants, Les Renardi\`eres, F-77250 Moret sur Loing, France}
\author{Tomasz Wejrzanowski}
\affiliation{Faculty of Materials Science and Engineering, Warsaw University of Technology, ul. Wo\l{}oska 141, 02-507 Warsaw, Poland}
\author{Sergei L. Dudarev}
\affiliation{UK Atomic Energy Authority, Culham Centre for Fusion Energy, Abingdon, Oxfordshire OX14~3DB, United Kingdom}
\date{\today}

\begin{abstract}
Point defects in body-centred cubic Fe, Cr and concentrated random magnetic Fe-Cr  are investigated using density functional theory and theory of elasticity. The volume of a substitutional Cr atom in ferromagnetic bcc Fe is approximately 18\% larger than the volume of a host Fe atom, whereas the volume of a substitutional Fe atom in antiferromagnetic bcc Cr is 5\% smaller than the volume of a host Cr atom. Elastic dipole $\boldsymbol{P}$ and relaxation volume $\boldsymbol{\Omega}$ tensors of vacancies and self-interstitial atom (SIA) defects exhibit large fluctuations, with vacancies having negative and SIA large positive relaxation volumes. Dipole tensors of vacancies are nearly isotropic across the entire alloy composition range, with diagonal elements $P_{ii}$ decreasing as a function of Cr content. Fe-Fe and Fe-Cr SIA dumbbells are more anisotropic than Cr-Cr dumbbells. Fluctuations of elastic dipole tensors of SIA defects are primarily associated with the variable crystallographic orientations of the dumbbells. Statistical properties of tensors $\boldsymbol{P}$ and $\boldsymbol{\Omega}$ are analysed using their principal invariants, suggesting that point defects differ significantly in alloys containing below and above 10\% at.  Cr. The relaxation volume of a vacancy depends sensitively on whether it occupies a Fe or a Cr lattice site. A correlation between elastic relaxation volumes and magnetic moments of defects found in this study suggests that magnetism is a significant factor influencing elastic fields of defects in Fe-Cr alloys.

\end{abstract}


\maketitle

\section{Introduction}
Defects are the stable strong local distortions of regular atomic order that form in crystalline metals and alloys under irradiation or during mechanical deformation \cite{CaiNix}.
Defects not only have an effect on how a material responds to the applied stress and deformation, but they also affect electronic properties, including thermal and electrical conductivity, and magnetism.  

Microstructural evolution of an alloy occurring as a result of accumulation of defects is driven by short- and long-range interactions of alloying elements with dislocations, surfaces, grain boundaries, and point defects. Short-range interactions involving variation of chemical compositions in the vicinity of defects can be investigated using Density Functional Theory (DFT) \cite{Ruban2008a, Wrobel2015, Lavrentiev2007, Wrobel2017, Fernandez-Caballero2017,Fedorov2020}. Long-range interaction between the defects is elastic, and it is mediated by the distortions that defects generate in the crystal lattice \cite{Leibfried,Bacon1979, Freysoldt2014, Clouet2008, Varvenne2013, Varvenne2017, Dudarev2018, Ma2019, Ma2019a, Ma2019b}. 

The fundamental quantities, describing elastic fields and long-range elastic interaction between defects, are the elastic dipole and relaxation volume tensors \cite{Bacon1979, Freysoldt2014, Clouet2008, Varvenne2013, Varvenne2017, Dudarev2018, Ma2019, Ma2019a, Ma2019b, Domain2001}. These quantities can be computed using DFT calculations or other atomic level simulations, and can then be used in the context of larger scale models, for example where defects and ensembles of defects are treated as objects of continuum elasticity \cite{Dudarev2010, Dudarev2018a}.
So far, elastic dipole and relaxation volume tensors of point defects have been investigated primarily for pure metals \cite{Bacon1979, Freysoldt2014, Clouet2008, Varvenne2013, Varvenne2017, Dudarev2018, Clouet2018,Ma2019, Ma2019a, Ma2019b, Domain2001}. 

In Refs. \cite{Dudarev2018,Ma2019} it was shown that the elastic field of an isotropic point defect, for example a vacancy, is fully defined by a single parameter, the elastic relaxation volume of the defect. On the other hand, a self-interstitial atom (SIA) defect often adopts an anisotropic dumbbell configuration, and the treatment of its elastic field requires using several independent parameters defining the relaxation volume of the defect and its orientation \cite{Dudarev2018,Ma2019}.

A vacancy, because of the isotropic nature of its dipole tensor, does not interact with a shear strain field even in an elastically anisotropic cubic material, whereas the anisotropic structure of an SIA defect enables elastic interaction with shear strain, applied externally or generated by other defects or dislocations \cite{Clouet2008}.
The investigation of elastic dipole tensors and relaxation volumes, as well as other properties of point defect in concentrated alloys, is a challenging task since these quantities depend on the alloy composition, atomic short-range order as well as the local environment of a defect \cite{Wrobel2017a,Samin2019}. In a magnetic alloy the structure of a defect is also affected by the non-linear magneto-volume effects.

Here, we focus on the investigation of point defects in concentrated Fe-Cr alloys, which are the base alloy system underpinning the composition of many industrial steels. The phase stability and properties of magnetic Fe-Cr alloys were extensively explored both theoretically \cite{Ruban2008a,Klaver2006,Klaver2007,Lavrentiev2007,Lavrentiev2009,Lavrentiev2011,Nguyen-Manh2007,Nguyen-Manh2008,Nguyen-Manh2012,Olsson2003,Olsson2006,Olsson2007,Wrobel2015,Senninger2016}
and experimentally \cite{Mirebeau2010, Hardie2013, Porollo1998, Kuksenko2011}. The analysis performed in Ref. \cite{Lavrentiev2018} showed that vacancies attract Cr atoms and hence may form vacancy-Cr clusters in dilute bcc Fe-Cr alloys. Investigation of point defects in dilute Fe-Cr alloys \cite{Olsson2007, Klaver2007, Nguyen-Manh2007, Nguyen-Manh2012, Klaver2016,Becquart2018} shows that the formation energy of self-interstitial atom (SIA) dumbbells depends on the local configuration of Cr atoms surrounding a defect.
However, to the best of authors' knowledge, elastic dipole and relaxation volume tensors of point defects, and the long-range elastic fields of such defects in concentrated Fe-Cr alloys have never been systematically explored. 

In this paper, we study point defects in concentrated random Fe-Cr alloys, with Cr concentration up to 35\%. Since estimating the relaxation volume of a defect using the stress method, which is described below, requires information about elastic constants of the material, which vary with alloy composition, elastic properties of random Fe-Cr alloys are investigated as a function of Cr content. To find the most stable point defect configurations, formation energies of defects were determined using concentration-dependent chemical potentials of Fe and Cr. Relaxation volumes of dumbbells are also correlated with the magnetic moments of atoms forming these defects. We also assess the difference between relaxation volumes of point defects computed using the stress method and full cell relaxation method \cite{Hofmann2015,Mason2019}.

\section{Methodology}
\subsection{Elastic dipole tensors and relaxation volumes}
A point defect induces a long range elastic field in the surrounding lattice. The energy of interaction between a localised defect and external homogeneous strain filed $\epsilon_{ij}^{\mathrm{ext}}$, arising from the quadratic cross-terms in the volume integral of the density of elastic energy of the defect {\it and} external field,   is \cite{Leibfried}
\begin{equation}
E= -P_{ij}\epsilon_{ij}^{\mathrm{ext}},
\label{eq:en_interaction}
\end{equation}
where repeated indeces imply summation, and $P_{ij}$ is the $ij$-th element of the elastic dipole tensor, $\boldsymbol{P}$, of the defect. This second-rank tensor is a fundamental quantity relating the elastic field of a defect and its atomic structure. Tensor $\boldsymbol{P}$ fully characterizes all the elastic properties of a localised defect. 

Elements of the dipole tensor can be computed using the equation \cite{Clouet2008,Clouet2018, Dudarev2018,Dudarev2018a,Ma2019,CALANIE}  
\begin{equation}
P_{ij}=-V_{\mathrm{cell}}\bar{\sigma}_{ij},
\label{eq:dipole_tensor_fixed_vol}
\end{equation}
where $V_{\mathrm{cell}}$ is the volume of the simulation cell and $\bar{\sigma}_{ij}$ is the average, macroscopic, stress in the cell containing the defect. If the cell contains no defect then $\bar{\sigma}_{ij}=0$. 

In practice, the elements of an elastic dipole tensor are determined using either the above stress method, Eq. (\ref{eq:dipole_tensor_fixed_vol}), where the average {\it strain} in the simulation cell is zero \cite{Dudarev2018}, and hence the cell volume and its shape remain fixed and only the positions of ions are relaxed. Alternatively, $P_{ij}$ can be computed using the full cell relaxation method, where the cell volume and its shape are relaxed to the zero macroscopic stress condition \cite{Clouet2018,Mason2019}. The main difference between the two methods is that the latter one takes into account not only the elastic relaxation effects but also non-elastic non-linear relaxation occurring in the core of the defect as well as everywhere in the simulation cell \cite{Mura1987}. The stress and cell relaxation methods are reviewed together with other possible methods for computing elastic dipole tensors in Refs. \cite{Clouet2018,Varvenne2017,CALANIE,Mason2019}. 

In the full cell relaxation method,
the dipole tensor is computed from the elements of macroscopic strain associated with the relaxation of the cell to the zero stress condition
\begin{equation}
P_{ij}=V_{\mathrm{cell}}C_{ijkl}\epsilon_{kl}^{\mathrm{app}},
\label{eq:dipole_tensor_full_relax}
\end{equation}
where $C_{ijkl}$ is the fourth-rank tensor of elastic stiffness and $\epsilon_{kl}^{\mathrm{app}}$ is the macroscopic strain developing as a result of full relaxation of atomic positions and the shape of the simulation cell.

The dipole tensor is related to another fundamental tensor entity, also characterising the defect, \textit{via} the following relation
\begin{equation}
P_{ij}=C_{ijkl}\Omega_{kl},
\label{eq:dipole_tensor}
\end{equation}
where $\Omega_{kl}$ is the $kl$-th element of the so-called relaxation volume tensor, $\boldsymbol{\Omega}$. It is related to the elastic dipole tensor through the tensor of elastic compliance, $S_{ijkl}$:
\begin{equation}
\Omega_{ij}=S_{ijkl}P_{kl}.
\label{eq:relax_vol_tensor} 
\end{equation}
Tensors $S_{ijkl}$ and $C_{ijkl}$ are related as \cite{CaiNix}
$$
C_{ijmn}S_{nmkl}={1\over 2}\left(\delta_{ik}\delta_{jl}+\delta_{il}\delta_{jk} \right).
$$
The energy of interaction between a defect and external elastic field can be expressed in terms of either the elastic dipole or relaxation volume tensor as \cite{Ma2019a}
\begin{equation}
E= -P_{ij}\epsilon_{ij}^{\mathrm{ext}}=-\Omega_{ij}\sigma_{ij}^{\mathrm{ext}},
\label{eq:en_interaction1}
\end{equation}
where $\sigma_{ij}^{\mathrm{ext}}=C_{ijkl}\epsilon ^{\mathrm{ext}}_{kl}$ is the stress tensor of external elastic field.

The elastic relaxation volume of a defect $\Omega_{rel}$ can be computed by taking the trace of the relaxation volume tensor
\begin{equation}
\Omega_{rel}=\Tr\boldsymbol{\Omega}=\Omega_{11}+\Omega_{22}+\Omega_{33}.
\label{eq:relax_vol}
\end{equation}

$\Omega_{rel}$ is a convenient parameter characterizing the degree of macroscopic expansion or contraction of the material due to the presence of defects in it \cite{Dudarev2018a}. Also, it describes the ``size'' interaction between the defects, whereas the deviatoric component of the relaxation volume tensor, \textit{i.e}. its off-diagonal terms and differences between diagonal components, gives rise to the so-called ``shape'' interaction. In the limit where elastic relaxation around the defect is isotropic and the relaxation volume tensor of a defect is diagonal \cite{Dudarev2018a} $\Omega_{ij}={1\over3}\Omega_{rel}\delta_{ij}$, where $\delta _{ij}$ is the Kronecker delta-symbol, equation (\ref{eq:en_interaction1}) can be further simplified as \cite{HealdSpeight1975}
\begin{equation}
E=-\Omega_{ij}\sigma_{ij}^{\mathrm{ext}}=
-{1\over3}\sigma _{ii}^{\mathrm{ext}}\Omega_{rel}=p\Omega_{rel},
\label{eq:en_interaction2}
\end{equation}
where $p$ is the hydrostatic pressure, $p=-{1\over3}\sigma _{ii}^{\mathrm{ext}}$. To derive the above formula, we noted that since repeated indeces imply summation, $\delta _{ii}=3$. 

To analyse elastic dipole and relaxation volume tensors of point defects, it is convenient to use the notion of principal invariants, which are the quantities independent of the orientation of Cartesian coordinate axes. The formula relating a second-rank tensor ($\boldsymbol{A}$) and its principal invariants is
\begin{equation}
    \boldsymbol{A}^{3}-I_1\boldsymbol{A}^2+I_2\boldsymbol{A}-I_3\mathds{1}=\bar{\bar{0}},
\label{eq:PI}
\end{equation}
where $\mathds{1}$ is the identity tensor, $\bar{\bar{0}}$ is zero matrix, and $I_1$, $I_2$, $I_3$ are the principal invariants that can be expressed as
\begin{equation}
I_1=\Tr{\boldsymbol{A}},
\label{eq:I1}
\end{equation}
\begin{equation}
I_2=\frac{1}{2}\left[ (\Tr{\boldsymbol{A}})^2-\Tr({\boldsymbol{A}^2)} \right],
\label{eq:I2}
\end{equation}
\begin{equation}
I_3=\det{\boldsymbol{A}}.
\label{eq:I3}
\end{equation}

The above relations apply to both elastic dipole and relaxation volume tensors ($\boldsymbol{A}=\boldsymbol{P}$ or $\boldsymbol{\Omega}$). In what follows, the invariants of an elastic dipole tensor will be denoted by $I_1^{P}$, $I_2^{P}$ and $I_3^{P}$, whereas those of the relaxation volume tensor by $I_1^{\Omega}$, $I_2^{\Omega}$ and $I_3^{\Omega}$.
It is worth noting that $I_1^{\Omega}$ is nothing but the relaxation volume of a defect, whereas the invariants of the elastic dipole tensor are directly related to the von Mises condition for the general state of stress $\sigma_{vM}$, which is used for predicting the yield point of a material under multi-axial loading conditions. This relationship, describing a critical stress state of a material, containing homogeneously distributed identical defects, can be defined as follows 
\begin{equation}
\sigma_{vM}={1\over V_{\mathrm{cell}}}\sqrt{\left(I_1^P\right)^2-3I_2^P}.
\label{eq:vm}
\end{equation}


\subsection{Elastic properties of alloys}

Bulk elastic constants are required for finding the elements of elastic dipole tensor using full cell relaxation, see Eq. (\ref{eq:dipole_tensor_full_relax}). Analysis performed in Ref. \cite{Mason2019} shows that relaxation volumes of clusters of point defects (voids and interstitial loops) may vary significantly, depending on the interatomic potential. Hence, having a correct starting estimate for the elastic stiffness parameters of Fe-Cr alloys is important for the investigation of elastic dipole and relaxation volume tensors of defects in these alloys.

For pure elemental cubic crystals, the tensor of elastic constants $C_{ijkl}$ can be parameterized using only three independent parameters, $C_{11}$, $C_{12}$ and $C_{44}$, see \cite{Nye1985}. Elastic properties of alloys are more complicated and generally there can be up to twenty one non-zero independent elastic constants. Elastic constants of disordered alloys adopting crystal lattice with cubic symmetry can be approximated as
\begin{equation}
\bar{C}_{11}=\frac{C_{11}+C_{22}+C_{33}}{3},
\label{eq:average_c11}
\end{equation}
\begin{equation}
\bar{C}_{12}=\frac{C_{12}+C_{13}+C_{23}}{3},
\label{eq:average_c12}
\end{equation}
\begin{equation}
\bar{C}_{44}=\frac{C_{44}+C_{55}+C_{66}}{3},
\label{eq:average_c44}
\end{equation}
\begin{eqnarray}
C_{14}=C_{15}=C_{16}=C_{24}=C_{25}=C_{26}= \nonumber \\
C_{34}=C_{35}=C_{36}=C_{45}=C_{46}=C_{56}=0.
\label{eq:c14_non-diagonal}
\end{eqnarray}

In this study, the second-order elastic constants were computed by deforming an unstrained equilibrium structure and analysing the corresponding variation of the total energy $E_{tot}$ as a function of components of strain. Applied deformation changes the total energy as follows \cite{Nye1985}
\begin{equation}
U=\frac{E_{tot}-E_0}{V_0}=\frac{1}{2}\sum_{i=1}^6\sum_{j=1}^6 C_{ij}\epsilon_{i}\epsilon_{j},
\label{eq:energy_vs_strain}
\end{equation}
where $E_0$ is the total energy of the unstrained lattice, $V_0$ is the volume of an undistorted cell and $C_{ij}$ are the elements of the elastic constant matrix in the Voigt notation. Indices $i$ and $j$ vary from 1 to 6 following the sequence {$xx, yy, zz, yz, xz, xy$} \cite{Nye1985}.

For each deformation, eight values of strain ($\pm0.5\%, \pm1.0\%, \pm1.5\%, \pm2.0\%$) were considered and the corresponding energies computed. Each curve showing how the total energy varies as a function of deformation was then fitted to a quadratic form and the respective elastic constants obtained. 

The anisotropy of elastic properties of Fe-Cr alloys was studied in order to gain understanding of stress concentration at grain boundaries, which may cause cracking in brittle materials \cite{Yamamoto2008,Yang2008}, as well as to enable the evaluation of strain fields of point defects, clusters of point defects or dislocations \cite{Clouet2018}. We also note that in an elastically isotropic material, any isotropic defects, such as vacancies, do not interact. The strength of their interaction in an elastically anisotropic material depends on the degree of elastic anisotropy \cite{Eshelby1955,Lie1968,Hudson2005}.

The variation of Young's modulus as a function of the orientation with respect to the crystal lattice was investigated using a method developed in Ref. \cite{Wrobel2012}. Following Refs. \cite{Zhang2007,Zhang2007a}, the Young modulus $E(hkl)$ projected onto a direction normal to a lattice plane system described by Miller indices ($hkl$) is defined as  
\begin{equation}
E(hkl)=\frac{1}{S'_{3333}},
\label{eq:E_hkl}
\end{equation}
where $S'_{3333}$ is a component of the elastic compliance tensor in the new reference system, \textit{i.e}. the new lattice plane system rotated from the original one and described by Miller indices ($hkl$) written using the fourth-order tensor notation and transformed from the compliances of the crystal, $S_{mnop}$, in the original cubic crystal axes. Hence
\begin{equation}
S'_{ijkl}=a_{im}a_{jn}a_{ko}a_{lp}S_{mnop},
\label{eq:s_ijkl}
\end{equation}
where $a_{im}, a_{jn}, a_{ko}$, and $a_{lp}$ are the components of a matrix $(a_{rt})$ describing a transformation from the original cubic crystal axes to the new lattice plane system described by the Miller indices ($hkl$) that can be expressed explicitly in terms of the Miller indices of this plane as \cite{Wrobel2012}
\begin{eqnarray}
 && (a_{rt}) = \\
 && \left[ \begin{array}{ccc} 
 \frac{hl}{\sqrt{h^2+k^2}\sqrt{h^2+k^2+l^2}} & \frac{kl}{\sqrt{h^2+k^2}\sqrt{h^2+k^2+l^2}} & -\frac{\sqrt{h^2+k^2}}{\sqrt{h^2+k^2+l^2}} \\
 -\frac{k}{\sqrt{h^2+k^2}} & \frac{h}{\sqrt{h^2+k^2}} & 0 \\
 \frac{h}{\sqrt{h^2+k^2+l^2}} & \frac{k}{\sqrt{h^2+k^2+l^2}} & \frac{l}{\sqrt{h^2+k^2+l^2}} 
 \end{array} \right],\nonumber
\label{eq:a_rt}
\end{eqnarray}
where $h, k, l$ are the direction cosines.

In the method described in Ref. \cite{Wrobel2012}, the transformation matrix $(a_{rt})$ is evaluated for each orientation chosen from a uniformly distributed set of directions and $E(hkl)$ is obtained from Eq. (\ref{eq:E_hkl}).

\subsection{Formation energies of point defects}

The formation energy of a vacancy or a self-interstitial atom (SIA) in an alloy is defined as
\begin{equation}
E_f^{vac,A}=E_{vac}-(E_{ref}-\mu_{A})+E_{el}^{corr},
\label{eq:form_en_vacancy}
\end{equation}
\begin{equation}
E_f^{SIA, A}=E_{SIA}-(E_{ref}+\mu_{A})+E_{el}^{corr},
\label{eq:form_en_dumbbell}
\end{equation}
where $E_{vac}$ and $E_{SIA}$ are the total energies of structures containing a vacancy and a self-interstitial atom, respectively, and  $E_{ref}$ is the total energy of the corresponding reference structure containing no defect. $\mu_A$ is the chemical potential of atom A (here, a Cr or Fe atom), which was removed or inserted into the original structure in order to form a vacancy or a self-interstitial atom defect, respectively. $E_{el}^{corr}$ is a correction term resulting from the conditions of vanishing average macroscopic strain (in the stress method) {\it and} periodicity \cite{Dudarev2018,Varvenne2013,Varvenne2017,CALANIE}. 
Methods  for evaluating $E_{el}^{corr}$ are described in Refs. \cite{Varvenne2013,Varvenne2017,Clouet2018,Dudarev2018,Ma2019a,CALANIE}. It should be noted that the origin of $E_{el}^{corr}$ is purely elastic \cite{CALANIE}, and it does not include non-elastic effects \cite{Mura1987}. Therefore, the formation energies of defects computed using full cell relaxation are usually lower than those computed using the stress method where the boundaries of the simulation cell are assumed fixed \cite{Varvenne2013,CALANIE}, even if the $E_{el}^{corr}$ term is taken into account \cite{Mason2019}. 

Chemical potentials of Fe and Cr atoms are estimated from the total energy of the system, where at $T=0$ and $p=0$ in the thermodynamic limit \cite{LandauStatPhys} $E=\mu _{Fe}N_{Fe}+\mu _{Cr}N_{Cr}$, where $N_{Fe}$ and $N_{Cr}$ are the numbers of Fe and Cr atoms in the corresponding structure, respectively. Using this expression, we find the difference between the minimum substitutional energies $\Delta E^{Fe\rightarrow Cr}$ and $\Delta E^{Cr\rightarrow Fe}$ \cite{Piochaud2014} as
\begin{equation}
\mu_{Cr}-\mu_{Fe}=\frac{1}{2}(\Delta E^{Fe\rightarrow Cr} - \Delta E^{Cr\rightarrow Fe}).
\label{eq:chem_pot_Fe_Cr}
\end{equation}
For each composition of the alloy, the minimum substitution energies are evaluated from the total energy difference between the reference structure and three structures, for each element, where a randomly chosen Fe (or Cr) atom has been replaced by a Cr (or Fe) atom. 

\subsection{Computational details}
All the total-energy calculations were performed using density functional theory in the plane-wave basis, and pseudopotentials derived within the projector augmented wave (PAW) method \cite{PAW,BlochPAW} implemented in the Vienna Ab-inito Package (VASP) code \cite{Kresse1,Kresse2}. The PAW pseudopotentials used here did not include the semicore electrons. Exchange and correlation effects were treated in the generalized gradient approximation with the Perdew-Burke-Ernzerhof \cite{PBE} parametrization. Collinear spin-polarized calculations, with a Vosko-Wilk-Nusair spin interpolation of the correlation potential, were carried out assuming that the initial magnetic moments of Fe and Cr atoms were 3 and -1 Bohr magnetons ($\mu_B$), respectively. The magnetic moments of Cr atoms were treated as being initially antiferromagnetically aligned with respect to the ferromagnetically ordered magnetic moments of Fe atoms. The structures contained 250 ({$\pm 1$} Fe/Cr) atoms in the form of 5$\times$5$\times$5 supercells with conventional body-centred cubic structure. Non-collinear magnetic effects \cite{Nguyen-Manh2015} were not treated in this study. The total energies were found using the Monkhorst-Pack \cite{Monkhorst} scheme to sample the Brillouin zone. A 3$\times$3$\times$3 \textit{k}-point grid was used when performing atomic relaxations. Structures of point defects in concentrated random Fe-Cr alloys, with concentrations up to 35\% at. Cr, were taken from Ref. \cite{Castin2017} where a DFT database of point-defect relaxation energies and migration barriers was used for training neural-network models. Fixed volume DFT simulations of structures without defects as well as those containing point defects in bcc Fe and bcc Cr were performed using the same parameters as in Ref. \cite{Castin2017}, namely the plane-wave energy cut-off of 300 eV and convergence criteria of $10^{-3}$ eV and $10^{-4}$ eV set for the total relaxation energies of ions and electrons, respectively. The energies of structures containing defects and the residual stresses given in Ref. \cite{Castin2017} were directly comparable with results of calculations performed in this study, and they were used for determining the formation energies as well as elastic dipole and relaxation volume tensors of point defects. Since Cr atoms are distributed randomly in the alloy structures included in the database, this study describes properties of point defects in concentrated {\it random} Fe-Cr alloys. Short-range order effects are not considered in this work. 

In DFT calculations involving full cell relaxation, which are required for the derivation of chemical potentials and elastic properties, as well as for comparison with the fixed volume results, the energy cut-off was set at 400 eV. We remind the reader that at $T=0$, accurate evaluation of chemical potentials requires using the condition $p=0$, implying full relaxation of the simulation cell. The total energy convergence criterion was set to $10^{-6}$ eV/cell, and atomic force components in the final relaxed structures were below $5\cdot 10^{-3}$ eV/\AA. 

\section{Results}
\subsection{Chemical potentials and formation energies of defects}

\begin{figure}
\includegraphics[width=0.95\columnwidth]{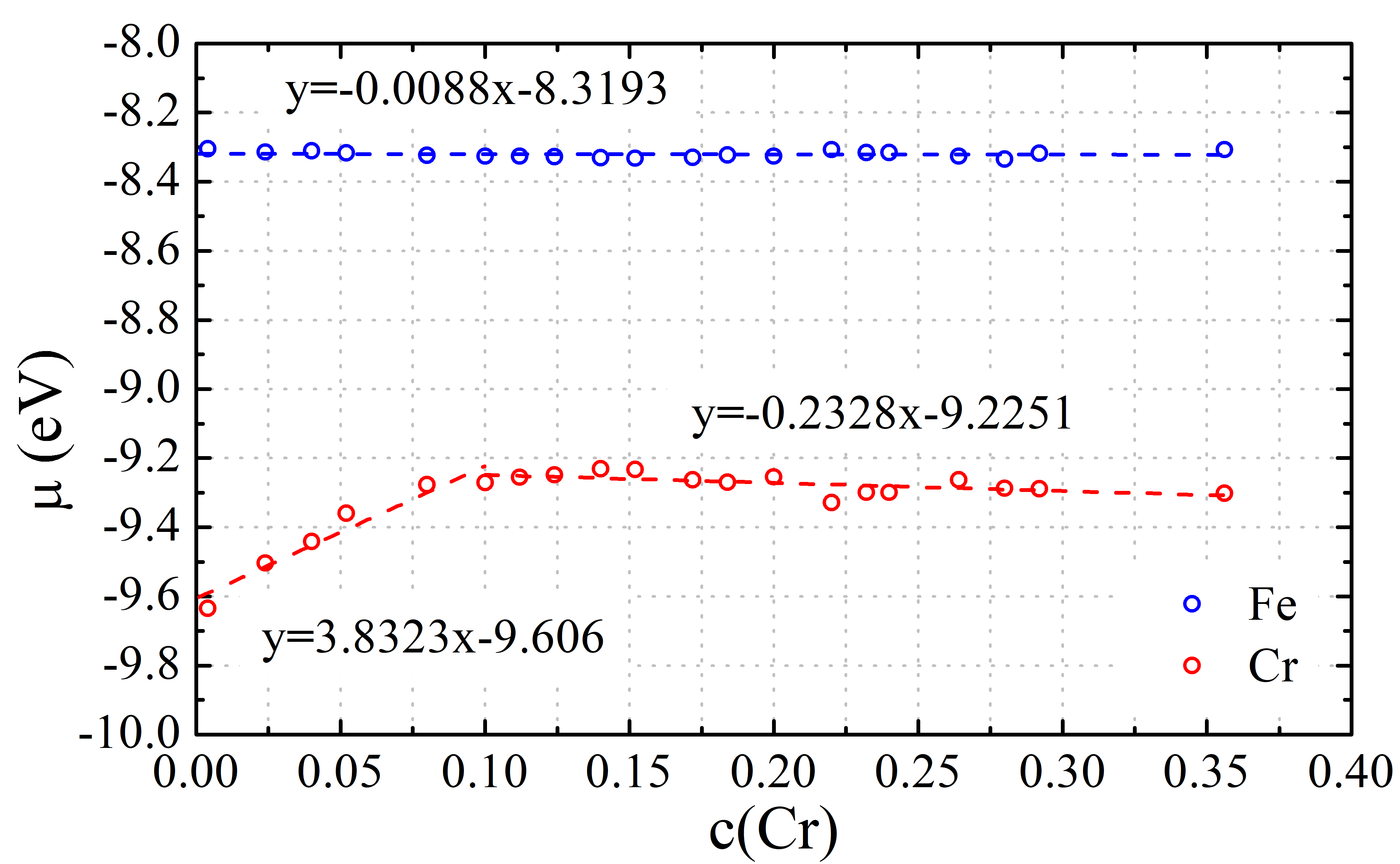}
\caption{Chemical potentials of Fe and Cr derived from fixed-volume DFT simulations. Dashed blue and red lines show the interpolated values of $\mu_{Fe}$ and $\mu _{Cr}$ as functions of Cr content. Similar trends are found in calculations involving full cell relaxation.
        \label{fig:chempot}}
\end{figure}

Chemical potentials of Fe ($\mu_{Fe}$) and Cr ($\mu_{Cr}$) atoms in random Fe-Cr alloys were estimated from DFT simulations assuming either a fixed volume of the simulation cell, or full atomic and volume relaxation. Simulations were performed for twenty alloy structures with concentrations chosen approximately evenly across the range of Cr concentrations. Fig. \ref{fig:chempot} shows that the chemical potential of Fe in Fe-Cr alloys remains almost constant over the entire range of compositions explored in this study, and its value is close to the chemical potential of pure bcc Fe, which is -8.31 eV. 
The chemical potential of Cr atoms behaves differently below and above approximately 10\% at. Cr, which corresponds to the Cr solubility limit in Fe-Cr alloys. Below the solubility limit, $\mu_{Cr}$ increases as a function of Cr content, whereas above the solubility limit it slowly decreases as a function of Cr concentration. 

We note that the results shown in Fig. \ref{fig:chempot} are insensitive to the energy cut-off and the internal degrees of freedom, for example  the chemical potentials derived from fixed-volume DFT simulations are virtually identical to those derived from simulations involving full cell relaxation –- the difference is smaller than 0.1\%.  Bearing this in mind, still only the values obtained with full relaxation of simulation cells, corresponding to vanishing pressure $p=0$, are shown in Fig. 1. Interpolated values of chemical potentials of Fe and Cr shown by dashed lines in Fig. \ref{fig:chempot} were used as a reference when evaluating the formation energies of point defects in Fe-Cr alloys. Values of $\mu_{Fe}$ for pure bcc Fe and $\mu_{Cr}$ for pure anti-ferromagnetic bcc Cr were derived from the total energies of bcc Fe and Cr, respectively. Values of $\mu_{Cr}$ in bcc Fe and $\mu_{Fe}$ in anti-ferromagnetic bcc Cr were computed using the method described in Section II.D for the structures containing one Cr atom in bcc Fe and one Fe atom in bcc Cr, respectively.
The computed formation energies of defects in bcc Fe matrix and bcc anti-ferromagnetic Cr matrix are given in Table \ref{tab:elastic_dipole_tensors_pure}.

Since most of the results for Fe-Cr alloys were obtained using fixed volume simulations cell, defined by the lattice parameter of pure Fe $a = 2.831$ \AA, all the results for Cr given in this study were also computed assuming this lattice parameter. The computed formation energies include the correction term resulting from periodic boundary conditions and the requirement of vanishing average strain \cite{CALANIE,Dudarev2018}.

In agreement with earlier studies \cite{Domain2001,Dudarev2013}, the computed formation energies of vacancies are significantly smaller than those of self-interstitial atom defects (SIAs). The formation energies of defects in bcc Cr are notably larger than in bcc Fe. In accord with Refs. \cite{Domain2001,Willaime2005}, the most stable configuration of a SIA defect in pure Fe is a $\left\langle 110\right\rangle$ dumbbell, with the energy of formation of $E_{form} = 4.019$ eV found in our calculations. This formation energy is more than 0.7 eV lower than the formation energy of a self-interstitial atom defect with a $\left\langle 111\right\rangle$ orientation.

In agreement with Refs. \cite{Olsson2009,Ma2019}, we find that the most stable configuration of a Cr-Cr dumbbell in pure anti-ferromagnetic bcc Cr is a symmetry-broken $\left\langle 11\xi\right\rangle$ dumbbell, where $\xi$ is an irrational number varying from 0 to approximately 2.2. The difference between $E_{form}$ of  $\left\langle 11\xi\right\rangle$ SIA and $E_{form}$ of $\left\langle 110\right\rangle$ and $\left\langle 111\right\rangle$ SIAs in pure Cr is 0.14 eV and 0.23 eV, respectively. This shows that the difference between energies of various SIA dumbbell configurations in bcc Cr is smaller than those in bcc Fe.

A symmetry-broken $\left\langle 11\xi\right\rangle$ configuration is also the most stable one for a Cr-Cr dumbbell in Fe matrix. This agrees with results from Ref. \cite{Klaver2007} showing that a Cr-Cr $\left\langle 110\right\rangle$ dumbbell configuration in the presence of additional Cr atom in the neighbourhood (lowering the symmetry of a structure) may transform into a lower energy configuration, for example a $\left\langle 22\overline 1\right\rangle$ Cr-Cr dumbbell.

A mixed Fe-Cr $\left\langle 110\right\rangle$ dumbbell is the most stable mixed SIA defect configuration in bcc Fe matrix. Figs. \ref{fig:Fe-Cr_dumb_vac_scheme}a and  \ref{fig:Fe-Cr_dumb_vac_scheme}b show that it can be formed either by adding a Cr atom to a Fe site or by adding a Fe atom to a Cr site. The formation energies of a Fe-Cr $\left\langle 110\right\rangle$ dumbbell in the former and latter cases are 3.964 eV and 3.975 eV, respectively. In both cases, formation energies of Fe-Cr dumbbells were more than 0.04 eV lower than that of a $\left\langle 110\right\rangle$ Fe-Fe, in agreement with Ref. \cite{Messina2020}, and they were 0.48 eV lower than the formation energy of a $\left\langle 11\xi\right\rangle$ Cr-Cr dumbbell.

In bcc Cr matrix, the difference between the energies of the most and least stable dumbbell configurations is significantly larger than in bcc Fe. The formation energy of a $\left\langle 110\right\rangle$ Fe-Fe SIA in bcc Cr equals 4.057 eV, and it is more than 1 eV and 2 eV smaller than that of the most stable Fe-Cr and Cr-Cr dumbbells, respectively.

\begin{table*}[]
\centering
\caption{Formation energies of defects and elements of elastic dipole tensors $P_{ij}$ (in eV) of defects, relaxation volume tensors $\Omega_{ij}$ (in \AA$^3$), relaxation volumes of defects and substitutional atoms $\Omega_{rel}$ (in \AA$^3$) and relaxation volumes $\Omega_{rel}^{at}$ expressed in the units of atomic volume $\Omega_0=a^3/2$. The reference atomic volume $\Omega_0=11.345$ \AA$^3$ corresponds to the bcc lattice parameter of $a=$2.831 \AA. }
\begin{tabular}{ccccccccccccccccc}
\hline
\hline
          & $E_{form}$    & $P_{11}$    & $P_{22}$    & $P_{33}$    & $P_{12}$   & $P_{23}$   & $P_{31}$ & $\frac{P_{11}}{P_{22}}$   & $\Omega_{11}$    & $\Omega_{22}$    & $\Omega_{33}$    & $\Omega_{12}$  & $\Omega_{23}$  & $\Omega_{31}$  & $\Omega_{rel}$ & $\Omega_{rel}^{at}$ \\
\hline
\multicolumn{16}{c}{Fe}                                                                                                           \\
(Vac)$_{Fe}$   & 2.183 & -3.682 & -3.682 & -3.682 & 0.000 & 0.000 & 0.000 & 1.00 & -1.015 & -1.015 & -1.015 & 0.000 & 0.000 & 0.000 & -3.045 & -0.268 \\
Ref. \cite{Ma2019}  &  2.190 & -3.081 & -3.081 & -3.081 & 0.000 & 0.000 & 0.000 & 1.00 & -0.831 & -0.831 & -0.831 & 0.000 & 0.000 & 0.000 &        & -0.220 \\
(Cr)$_{Fe}$   &  & 2.531 & 2.531 & 2.531 & 0.000 & 0.000 & 0.000 & 1.00 & 0.698 & 0.698 & 0.698 & 0.000 & 0.000 & 0.000 & 2.093 & 0.184 \\
(Fe-Fe)$^{\left\langle 110\right\rangle}_{Fe}$ & 4.019 & 24.853 & 20.534 & 20.534 & 0.000 & 4.620 & 0.000 & 1.21 & 6.851  & 5.660  & 5.660  & 0.000 & 1.274 & 0.000 & 18.171 & 1.602  \\
Ref. \cite{Ma2019}  &  4.321 & 25.832 & 21.143 & 21.143 & 0.000 & 5.122 & 0.000 & 1.22 & 9.777  & 4.294  & 4.302  & 0.000 & 3.819 & 0.000 &        & 1.620  \\
(Fe-Fe)$^{\left\langle 111\right\rangle}_{Fe}$ & 4.762 & 21.596 & 21.596 & 21.596 & 5.204 & 5.204 & 5.204 & 1.00 & 5.953  & 5.953  & 5.953  & 1.435 & 1.435 & 1.435 & 17.859 & 1.574  \\
(Fe-Cr)$^{\left\langle 110\right\rangle}_{Fe}$ & 3.964 & 23.756 & 21.826 & 21.826 & 0.000 & 4.691 & 0.000 & 1.09 & 6.548  & 6.016  & 6.016  & 0.000 & 1.293 & 0.000 & 18.581 & 1.638  \\
(Fe-Cr)$^{\left\langle 110\right\rangle}_{Cr}$ & 3.975 & 21.065 & 19.136 & 19.136 & 0.000 & 4.691 & 0.000 & 1.10 & 5.807  & 5.275  & 5.275  & 0.000 & 1.293 & 0.000 & 16.356 & 1.442  \\
(Cr-Cr)$^{\left\langle 110\right\rangle}_{Cr}$ & 4.501 & 19.472 & 22.269 & 22.269 & 0.000 & 6.160 & 0.000 & 0.87 & 5.367  & 6.138  & 6.138  & 0.000 & 1.698 & 0.000 & 17.644 & 1.555  \\
(Cr-Cr)$^{\left\langle 11\xi\right\rangle}_{Cr}$ & 4.465 & 20.693 & 21.048 & 21.048 & 1.576 & 5.045 & 1.576 & 0.98 & 5.704  & 5.802  & 5.802  & 0.434 & 1.391 & 0.434 & 17.307 & 1.526  \\
(Cr-Cr)$^{\left\langle 111\right\rangle}_{Cr}$ & 4.554 & 20.092 & 20.092 & 20.092 & 4.585 & 4.585 & 4.585 & 1.00 & 5.538  & 5.538  & 5.538  & 1.264 & 1.264 & 1.264 & 16.614 & 1.462  \\
(Cr-Cr)$^{\left\langle 110\right\rangle}_{Fe}$ & 4.481 & 22.145 & 24.960 & 24.960 & 0.000 & 6.160 & 0.000 & 0.89 & 6.104  & 6.880  & 6.880  & 0.000 & 1.698 & 0.000 & 19.864 & 1.751  \\
(Cr-Cr)$^{\left\langle 11\xi\right\rangle}_{Fe}$ & 4.446 & 23.384 & 23.738 & 23.738 & 1.576 & 5.045 & 1.576 & 0.98 & 6.446  & 6.543  & 6.543  & 0.434 & 1.391 & 0.434 & 19.532 & 1.722  \\
(Cr-Cr)$^{\left\langle 111\right\rangle}_{Fe}$ & 4.535 & 22.782 & 22.782 & 22.782 & 4.585 & 4.585 & 4.585 & 1.00 & 6.280  & 6.280  & 6.280  & 1.264 & 1.264 & 1.264 & 18.839 & 1.661  \\
\hline
\multicolumn{16}{c}{Cr}                                                                                                           \\
(Vac)$_{Cr}$     & 2.717 & -7.753 & -7.753 & -7.753 & 0.000 & 0.000 & 0.000 & 1.00 & -2.225 & -2.225 & -2.225 & 0.000 & 0.000 & 0.000 & -6.675 & -0.588 \\
Ref. \cite{Ma2019}    & 3.004  & -5.777 & -5.777 & -5.777 & 0.000 & 0.000 & 0.000 & 1.00 & -1.618 & -1.618 & -1.618 & 0.000 & 0.000 & 0.000 &        & -0.414 \\
(Fe)$_{Cr}$       &       & -0.726 & -0.726 & -0.726 & 0.000 & 0.000 & 0.000 & 1.00 & -0.208 & -0.208 & -0.208 & 0.000 & 0.000 & 0.000 & -0.625 & -0.055 \\
(Cr-Cr)$^{\left\langle 110\right\rangle}_{Cr}$ & 6.262 & 16.410 & 21.083 & 21.083 & 0.000 & 4.886 & 0.000 & 0.78 & 4.709  & 6.050  & 6.050  & 0.000 & 1.402 & 0.000 & 16.809 & 1.482  \\
Ref. \cite{Ma2019}      &  6.515  & 18.955 & 20.530 & 20.530 & 0.000 & 4.790 & 0.000 & 0.92 & 5.166  & 5.820  & 5.820  & 0.000 & 3.757 & 0.000 &        & 1.434  \\
(Cr-Cr)$^{\left\langle 11\xi\right\rangle}_{Cr}$ & 6.116 & 19.755 & 18.445 & 18.445 & 1.098 & 3.629 & 1.098 & 1.07 & 5.669  & 5.293  & 5.293  & 0.315 & 1.041 & 0.315 & 16.256 & 1.433  \\
Ref. \cite{Ma2019}      &  6.361  & 21.882 & 18.389 & 18.389 & 2.058 & 4.040 & 2.058 & 1.19 & 6.436  & 4.987  & 4.987  & 1.614 & 3.168 & 1.614 &        & 1.400  \\
(Cr-Cr)$^{\left\langle 111\right\rangle}_{Cr}$ & 6.354 & 18.056 & 18.056 & 18.056 & 3.682 & 3.682 & 3.682 & 1.00 & 5.182  & 5.182  & 5.182  & 1.057 & 1.057 & 1.057 & 15.545 & 1.370  \\
Ref. \cite{Ma2019}    & 6.617  & 18.728 & 18.728 & 18.728 & 4.617 & 4.617 & 4.617 & 1.00 & 5.244  & 5.244  & 5.244  & 3.622 & 3.622 & 3.622 &        & 1.343  \\
(Fe-Cr)$^{\left\langle 110\right\rangle}_{Cr}$ & 5.085 & 22.180 & 16.622 & 16.622 & 0.000 & 3.753 & 0.000 & 1.33 & 6.365  & 4.770  & 4.770  & 0.000 & 1.077 & 0.000 & 15.905 & 1.402  \\
(Fe-Cr)$^{\left\langle 110\right\rangle}_{Fe}$ & 5.108 & 23.048 & 17.489 & 17.489 & 0.000 & 3.753 & 0.000 & 1.32 & 6.614  & 5.019  & 5.019  & 0.000 & 1.077 & 0.000 & 16.652 & 1.468  \\
(Fe-Fe)$^{\left\langle 110\right\rangle}_{Fe}$ & 4.057 & 25.438 & 15.277 & 15.277 & 0.000 & 4.337 & 0.000 & 1.67 & 7.300  & 4.384  & 4.384  & 0.000 & 1.245 & 0.000 & 16.068 & 1.416 \\ 
\hline
\hline
\end{tabular}
\label{tab:elastic_dipole_tensors_pure}
\end{table*}

\begin{figure*}
\centering
    \begin{minipage}{.33\textwidth}
	   \centering
	  a)\includegraphics[width=.95\linewidth]{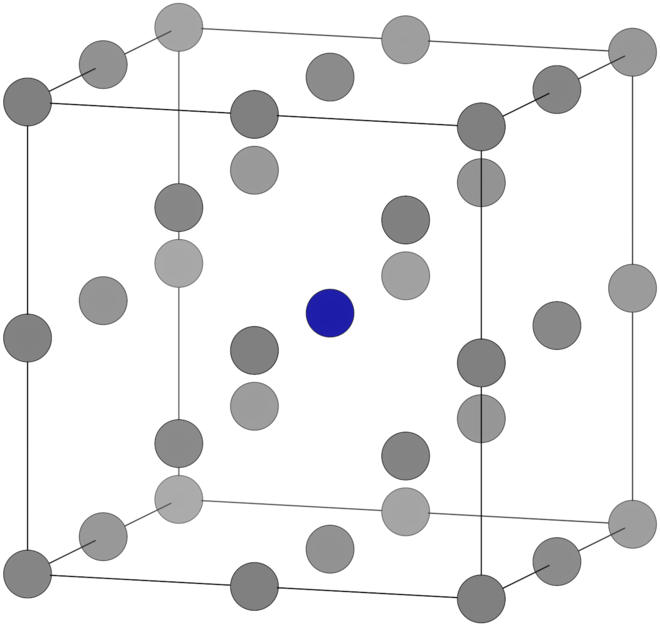}
	\end{minipage}%
	\begin{minipage}{.33\textwidth}
	  	\centering
	  	b)\includegraphics[width=.95\linewidth]{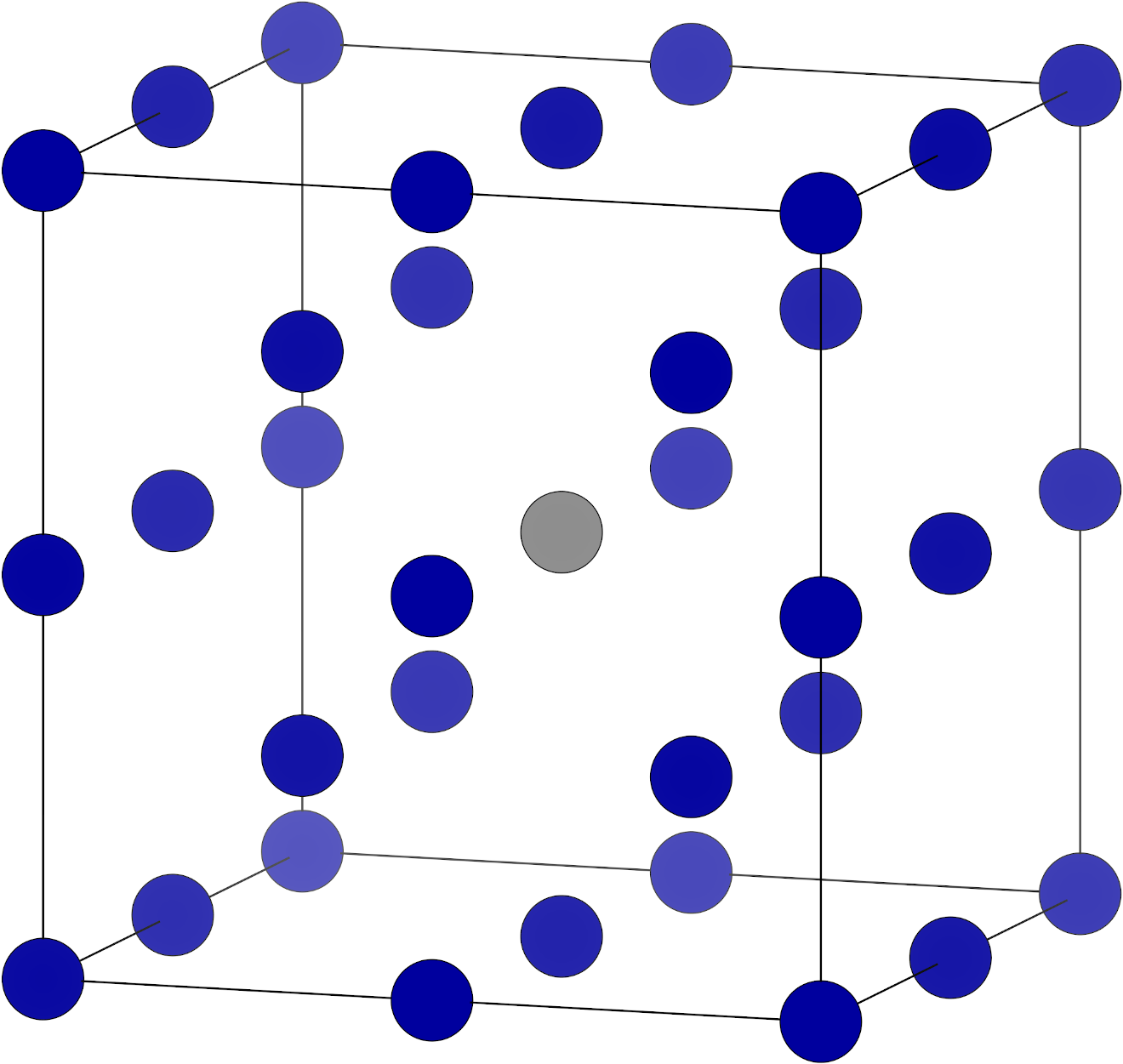}
	\end{minipage}
	\begin{minipage}{.33\textwidth}
	  	\centering
	  	c)\includegraphics[width=.95\linewidth]{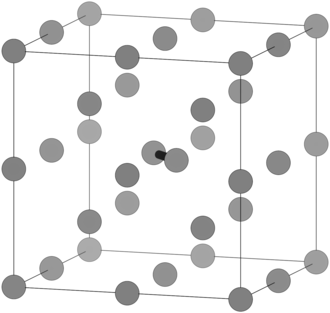}	  	
	\end{minipage}%
	\newline
	\begin{minipage}{.33\textwidth}
	  	\centering
	  	d)\includegraphics[width=.95\linewidth]{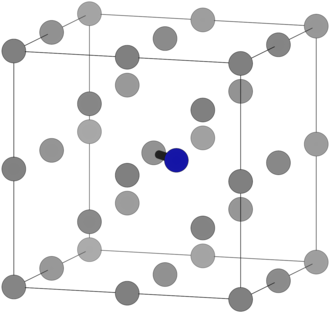}
	\end{minipage}%
	\begin{minipage}{.33\textwidth}
	  	\centering
	  	e)\includegraphics[width=.95\linewidth]{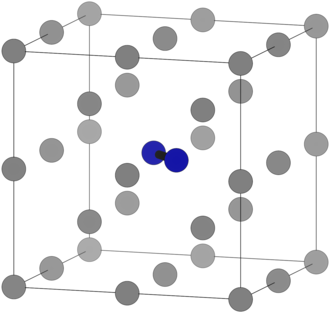}	  	
	\end{minipage}%
	\begin{minipage}{.33\textwidth}
	  	\centering
	  	f)\includegraphics[width=.95\linewidth]{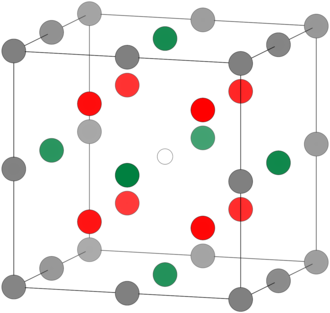}
	\end{minipage}%
	\newline
\caption{Schematic representation of structures: (a) a Cr atom in bcc Fe, (b) a Fe atom in bcc Cr, (c) Fe-Fe, (d) Fe-Cr and (e) Cr-Cr dumbbells in bcc Fe. Fe and Cr atoms are shown as gray and blue spheres, respectively. (f) Schematic representation of atoms in the neighbourhood of a defect (white sphere). Atoms in the first and second nearest neighbour shells are shown as red and green spheres, respectively.} 
        \label{fig:structures}
\end{figure*}

\begin{figure}
a) \includegraphics[width=0.95\columnwidth]{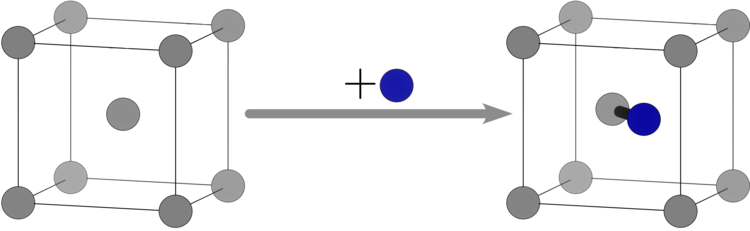}
b) \includegraphics[width=0.95\columnwidth]{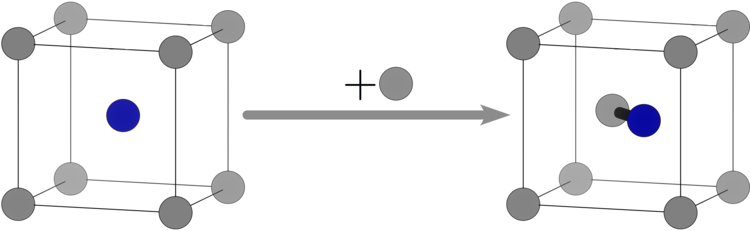}
c) \includegraphics[width=0.95\columnwidth]{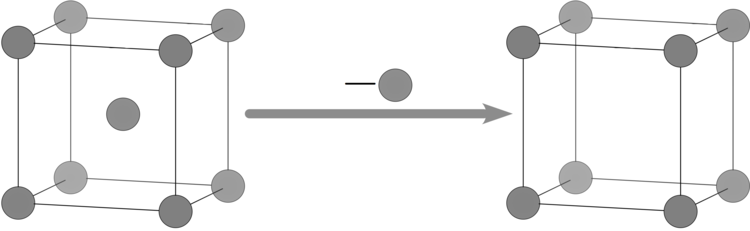}
d) \includegraphics[width=0.95\columnwidth]{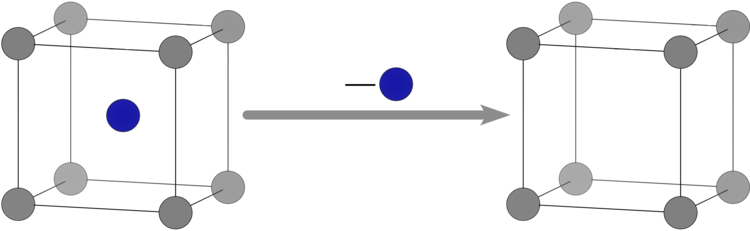}
\caption{Schematic representation of the process of formation of a mixed Fe-Cr dumbbell (a) by adding a Cr atom to a Fe site and (b) by adding a Fe atom to a Cr site. Schematic representation of formation of a vacancy (c) on a Fe site and (d) on a Cr site. Fe and Cr atoms are shown as gray and blue spheres, respectively. }
        \label{fig:Fe-Cr_dumb_vac_scheme}
\end{figure}

\begin{figure*}
\centering
    \begin{minipage}{.50\textwidth}
	   \centering
	  a)\includegraphics[width=.95\linewidth]{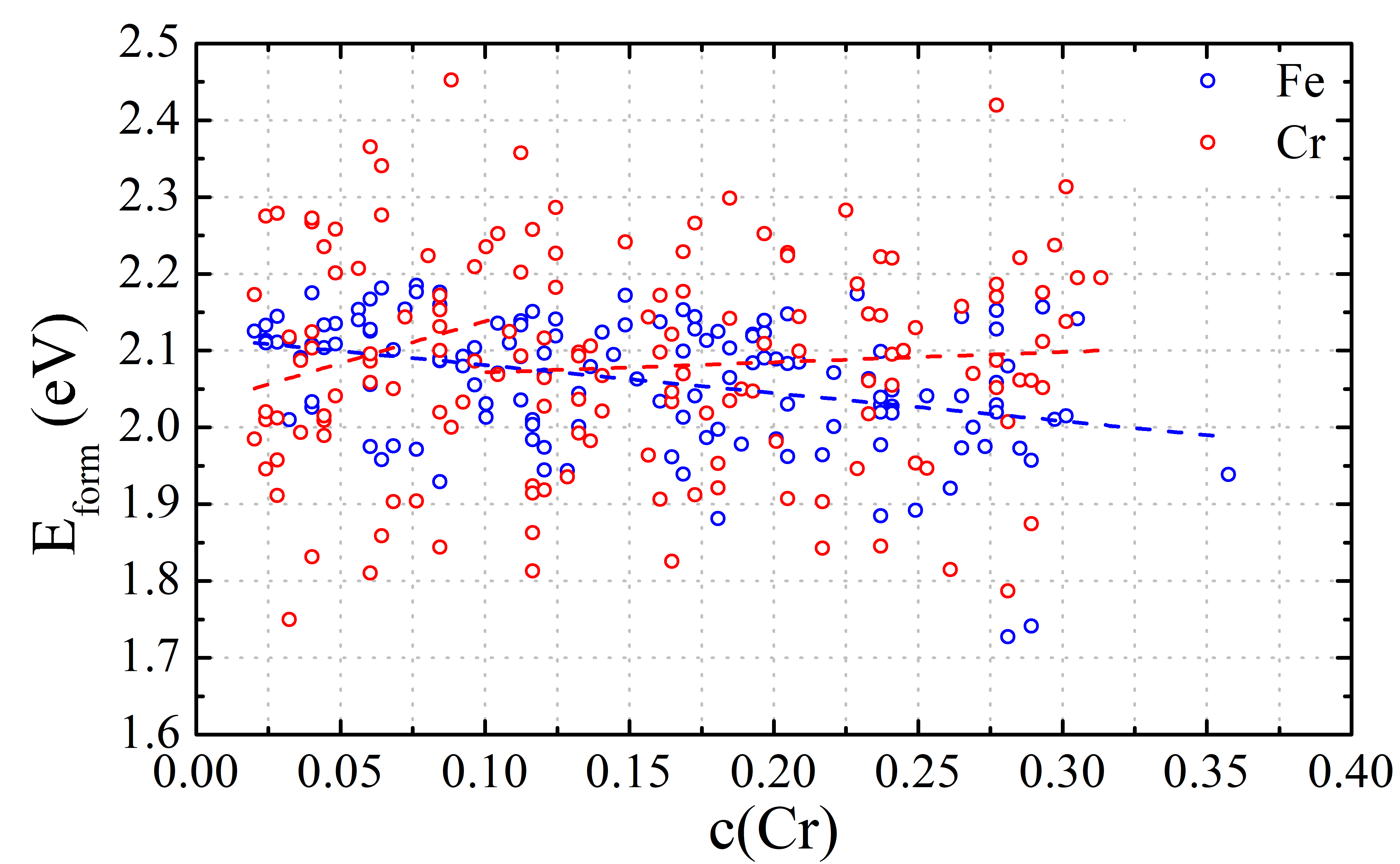}
	\end{minipage}%
	\begin{minipage}{.50\textwidth}
	  	\centering
	  	b)\includegraphics[width=.95\linewidth]{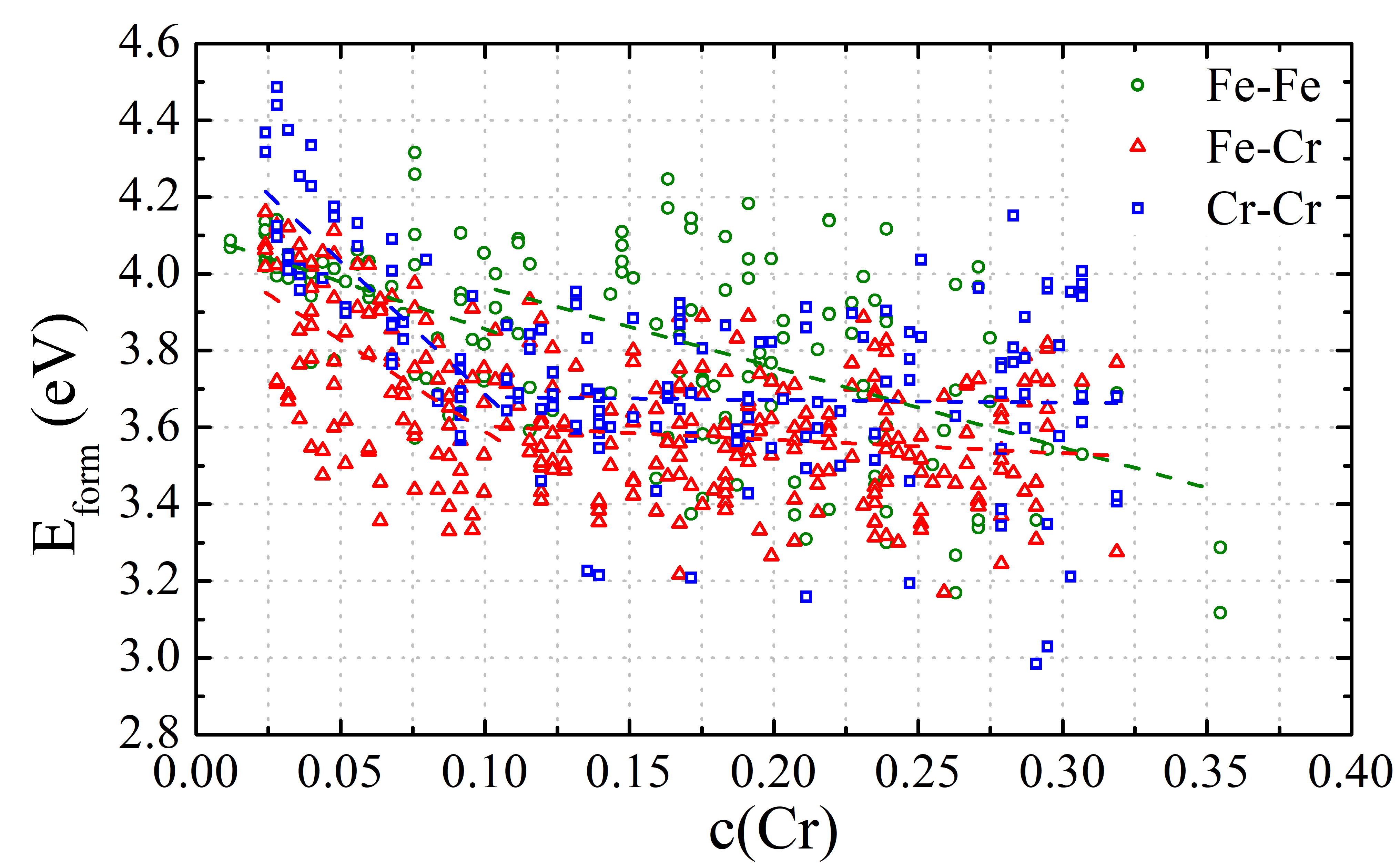}
	\end{minipage}
	\newline
	\begin{minipage}{.50\textwidth}
	  	\centering
	  	c)\includegraphics[width=.95\linewidth]{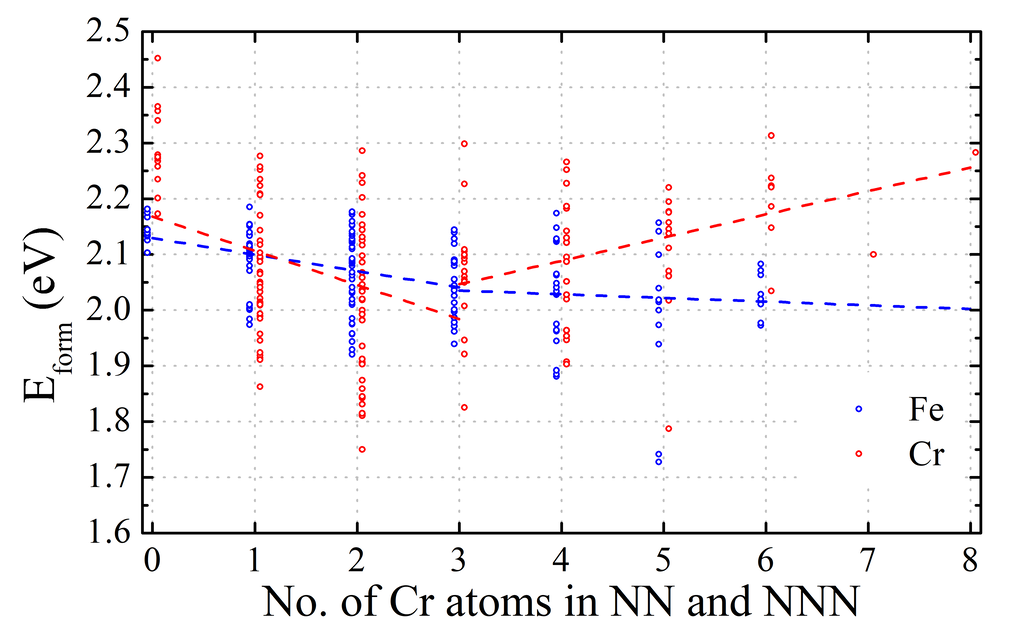}	  	
	\end{minipage}%
	\begin{minipage}{.50\textwidth}
	  	\centering
	  	d)\includegraphics[width=.95\linewidth]{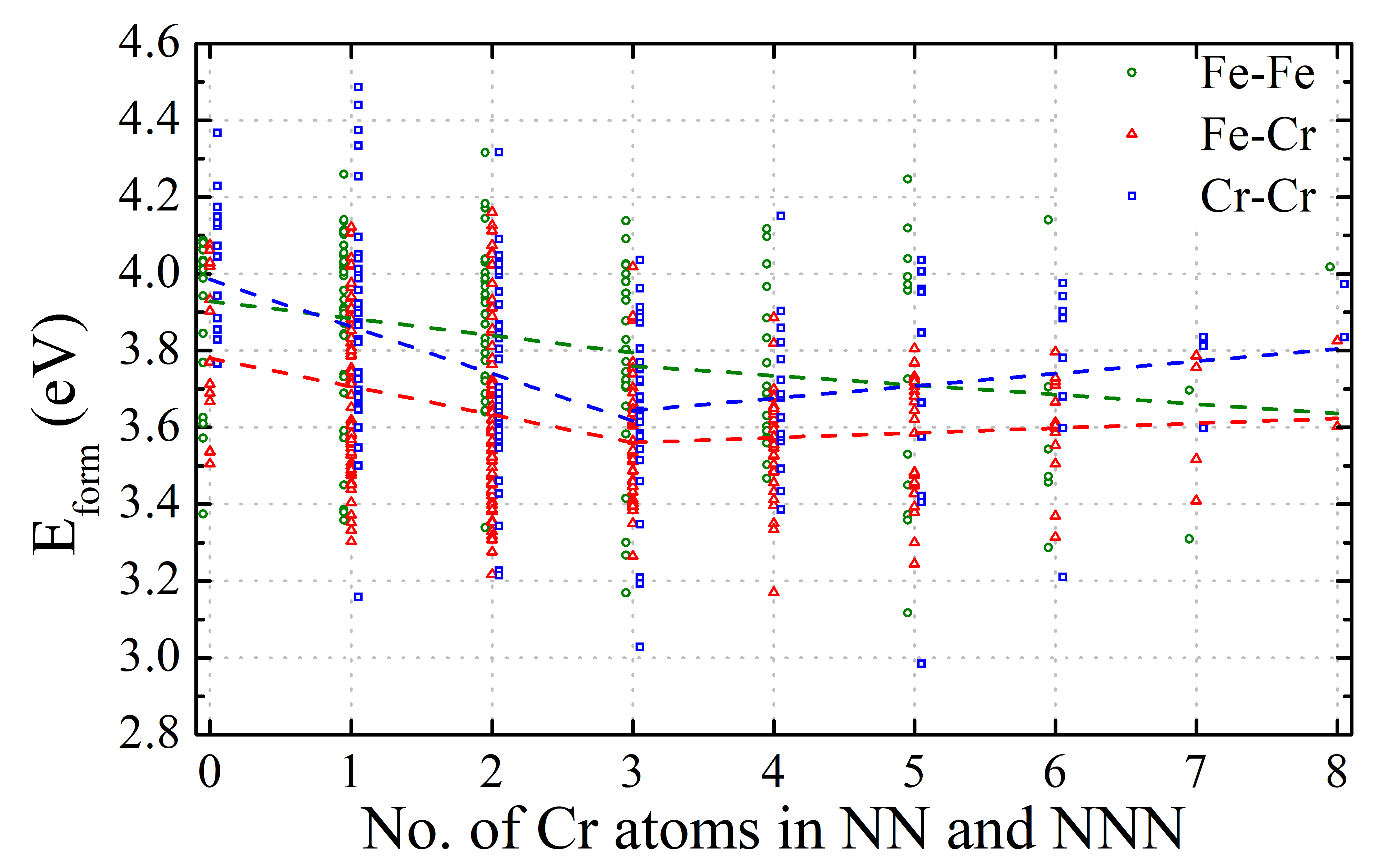}
	\end{minipage}%
	\newline
	\begin{minipage}{.50\textwidth}
	  	\centering
	  	e)\includegraphics[width=.95\linewidth]{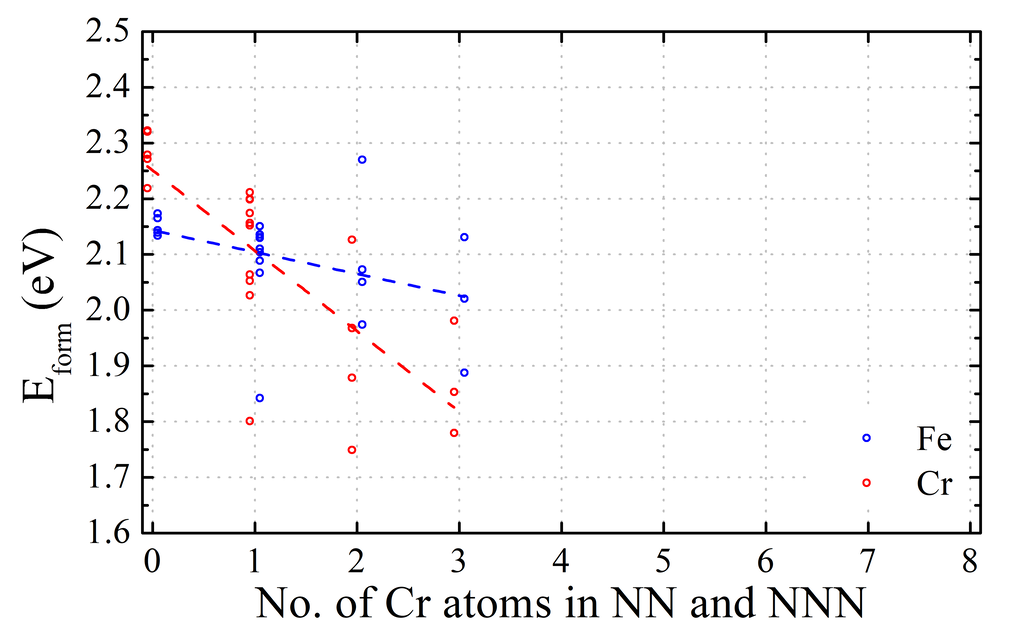}	  	
	\end{minipage}%
	\begin{minipage}{.50\textwidth}
	  	\centering
	  	f)\includegraphics[width=.95\linewidth]{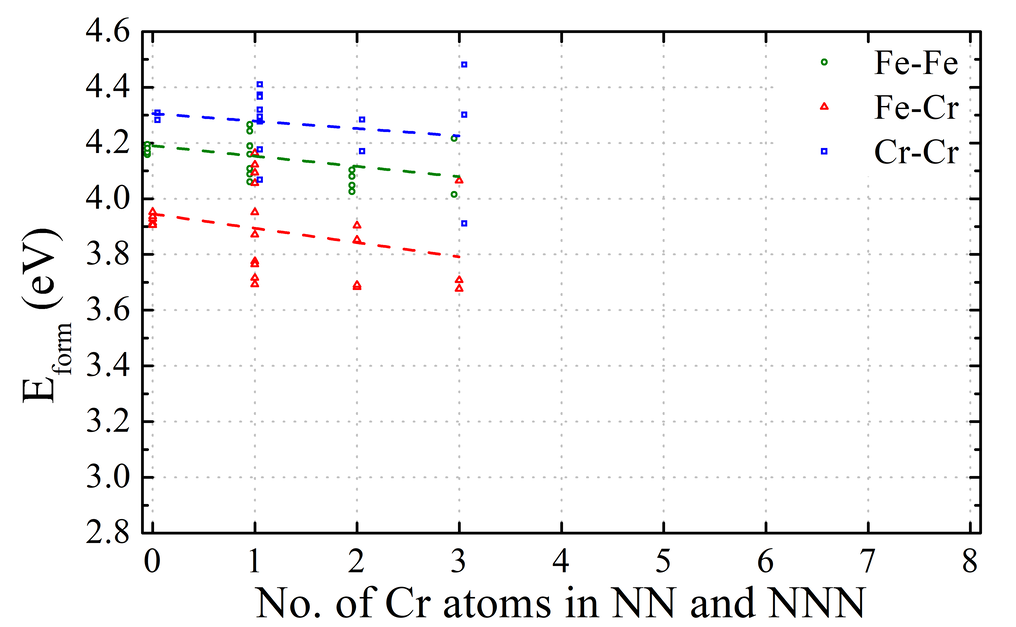}
	\end{minipage}%
	\newline
\caption{Formation energy of vacancies (a,c,e) and (b,d,f) SIA dumbbells in random Fe-Cr alloys shown over the entire range of concentrations (a-d) and for an alloy with 5\% at. Cr (e,f) plotted as a function of Cr concentration (a,b) and the total number of Cr atoms in the NN and NNN coordination shells of a defect (c-f). Linear trends are indicated by dashed lines. 
        \label{fig:form_en}}
\end{figure*}

The formation energies of approximately 300 vacancy and 400 dumbbell configurations derived from fixed cell volume DFT simulations are shown, as functions of Cr content and the number of Cr atoms in the local environment of a defect, in Figs. \ref{fig:form_en}a-d. The figures show that the formation energies of vacancies and SIA dumbbells fluctuate significantly, depending on the alloy composition and the local chemical environment of a defect. 

To separate the role of the two effects and investigate properties of defects only as functions of the number of Cr atoms in their local environment, further 120 DFT calculations were performed for the defect-free structures of Fe-Cr alloys containing 5\% at. Cr, and the same structures containing defects. Even for one alloy composition and the same number of Cr atoms in the 1st and 2nd nearest-neighbour (NN and NNN, see Fig. \ref{fig:structures}f) coordination shells around a defect, formation energies fluctuate by as much as 1 eV. This shows that the defect formation energies depend not only on parameters like the average alloy composition or the number of Cr atoms in the NN and NNN coordination shells, but also on the configuration of Cr atoms around a defect. 

Vacancies in Fe-Cr alloys can be formed by removing either a Fe atom or a Cr atom from a lattice site, see Figs. \ref{fig:Fe-Cr_dumb_vac_scheme}c and \ref{fig:Fe-Cr_dumb_vac_scheme}d. Figs. \ref{fig:form_en}a, \ref{fig:form_en}c and \ref{fig:form_en}e, show that there is a notable difference between the formation energies of vacancies on Fe and Cr sites. The average formation energy of a vacancy on a Fe site decreases slightly as a function of Cr content from approx. 2.1 eV at low Cr concentration to approx. 2.0 eV at 30\% at. Cr. On the other hand, the average value of $E_{form}$ for a vacancy on a Cr site increases with Cr content. The increase is more rapid in the range of Cr concentration below 10\% at. Cr. Formation energies of vacancies on Cr sites are also more scattered than those associated with Fe sites, an effect that  is probably related to the magnetic frustration of Cr atoms in bcc Fe matrix. 

Figs. \ref{fig:form_en}c shows the formation energy of a vacancy as a function of the number of Cr atoms $N_{Cr}^{def}$ in the NN and NNN shells around a defect. The data span the entire range of alloy compositions considered here, with a separate Figure \ref{fig:form_en}e showing the data for Fe-5\%Cr alloys. Since the variation of formation energies differs for configurations involving  small and large values of $N_{Cr}^{def}$, and also since defects in Fe-5\%Cr alloys are surrounded by up to three Cr atoms in the NN and NNN shells, the results are divided into two intervals where $N_{Cr}^{def}$ is smaller and larger than 3. The variation of the average formation energy of vacancies in Fe-5\%Cr alloy is similar to the variation found for other Cr concentrations. For the smaller number of Cr atoms, Figs. \ref{fig:form_en}c and \ref{fig:form_en}e show that the formation energy $E_{form}$ of a vacancy on either Fe and Cr sites decreases with increasing $N_{Cr}^{def}$. The rate of variation is more rapid for vacancies on Cr sites. For $N_{Cr}^{def}$ larger than 3, the formation energy of a vacancy on a Fe site slightly decreases whereas that on a Cr site increases.

The variation formation energies of dumbbells as a function of Cr content is significantly different below and above approximately 10\% at. Cr, see Fig. \ref{fig:form_en}b for more detail. Above 10\% at. Cr concentration, the average values of $E_{form}$ of Fe-Cr and Cr-Cr SIAs remain almost constant, whereas below that concentration there is a rapid decrease of $E_{form}$ as a function of Cr content. Only the slope of the trend line for $E_{form}$ computed for Fe-Fe SIAs remains similar over the whole considered range of Cr concentrations. Similarly to bcc Fe matrix, Fe-Cr dumbbells are generally the most stable interstitial defects in Fe-Cr alloys in the range of alloys compositions explored in this study. They exhibit the lowest mean values of $E_{form}$ over the concentration range up to approximately 32\% at. Cr. For each composition up to approximately 10-12\% at. Cr, the most stable Fe-Cr SIA exhibits the lowest $E_{form}$ among all the computed dumbbell configurations. For larger Cr concentrations, the Cr-Cr and Fe-Fe dumbbells may be more stable than the Fe-Cr SIAs. 

Similarly to the variation of the formation energy of dumbbells as a function of Cr content, values of $E_{form}$ shown in Fig. \ref{fig:form_en}d vary differently for smaller and larger values of $N_{Cr}^{def}$. For Fe-Cr and Cr-Cr SIAs, the average value of $E_{form}$ decreases and then slightly increases as a function of $N_{Cr}^{def}$ when $N_{Cr}^{def}$ is smaller and larger than 3, respectively. For Fe-Fe SIAs, the mean value of $E_{form}$ decreases as a function of $N_{Cr}^{def}$ over the range of $N_{Cr}^{def}$. For every value of $N_{Cr}^{def}$, Fe-Cr SIAs have the lowest mean $E_{form}$. However, for the majority of $N_{Cr}^{def}$, the most stable Cr-Cr dumbbells have smaller $E_{form}$ than the most stable Fe-Cr and Fe-Fe SIAs. 

Similarly to the case of vacancies, the trend lines of mean $E_{form}$ for Fe-Fe and Fe-Cr dumbbells as a function of $N_{Cr}^{def}$ in Fe-5\%Cr alloy  are generally similar to those found for other Cr concentrations, however the values are usually larger, as seen from the comparison of Figs. \ref{fig:form_en}d and \ref{fig:form_en}f. The largest difference is found for Cr-Cr dumbbells, for which the mean $E_{form}$ in the Fe-5\%Cr alloy does not decrease as a function of $N_{Cr}^{def}$ as rapidly as for other Cr concentrations. As a result, the mean value of $E_{form}$ for a structure with three Cr atoms in the local environment of a defect in the Fe-5\%Cr alloy is approximately 0.5 eV larger than the one averaged over structures with the same $N_{Cr}^{def}$ value in all the other Fe-Cr alloys. This may stem from the fact that the magnitudes of magnetic moments of Cr atoms vary significantly as a function of Cr composition in Fe-Cr alloys \cite{Olsson2006,Klaver2006,Wrobel2015}, and this may affect the value of $E_{form}$ for Cr-Cr dumbbells. The strong dependence on Cr concentration of the formation energies of Cr-Cr interstitial defect may also explain the larger spread of their values in comparison with Fe-Fe and Fe-Cr dumbbell defects, see Figs. \ref{fig:form_en}b and \ref{fig:form_en}d.

Equations interpolating the variation of formation energies of vacancies and dumbbells as a function of Cr concentration and a number of Cr atoms in NN and NNN are given in Table III in Appendix.

\subsection{Elastic properties}

\begin{figure*}
\centering
    \begin{minipage}{.50\textwidth}
	    \centering
	    a)\includegraphics[width=.95\linewidth]{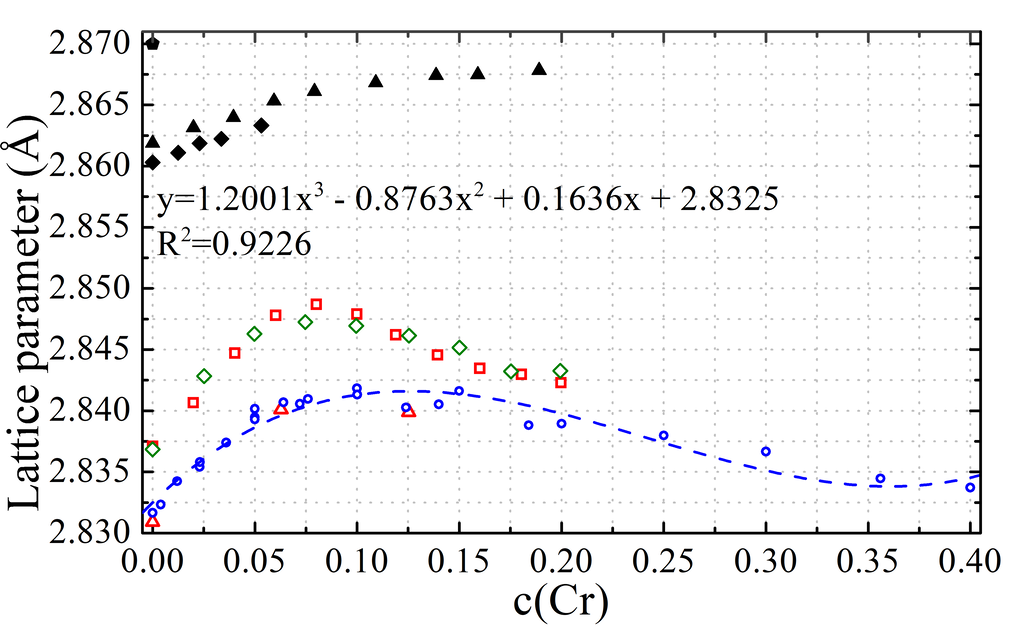}			  	
	\end{minipage}%
	\begin{minipage}{.50\textwidth}
	  	\centering
	  	b)\includegraphics[width=.95\linewidth]{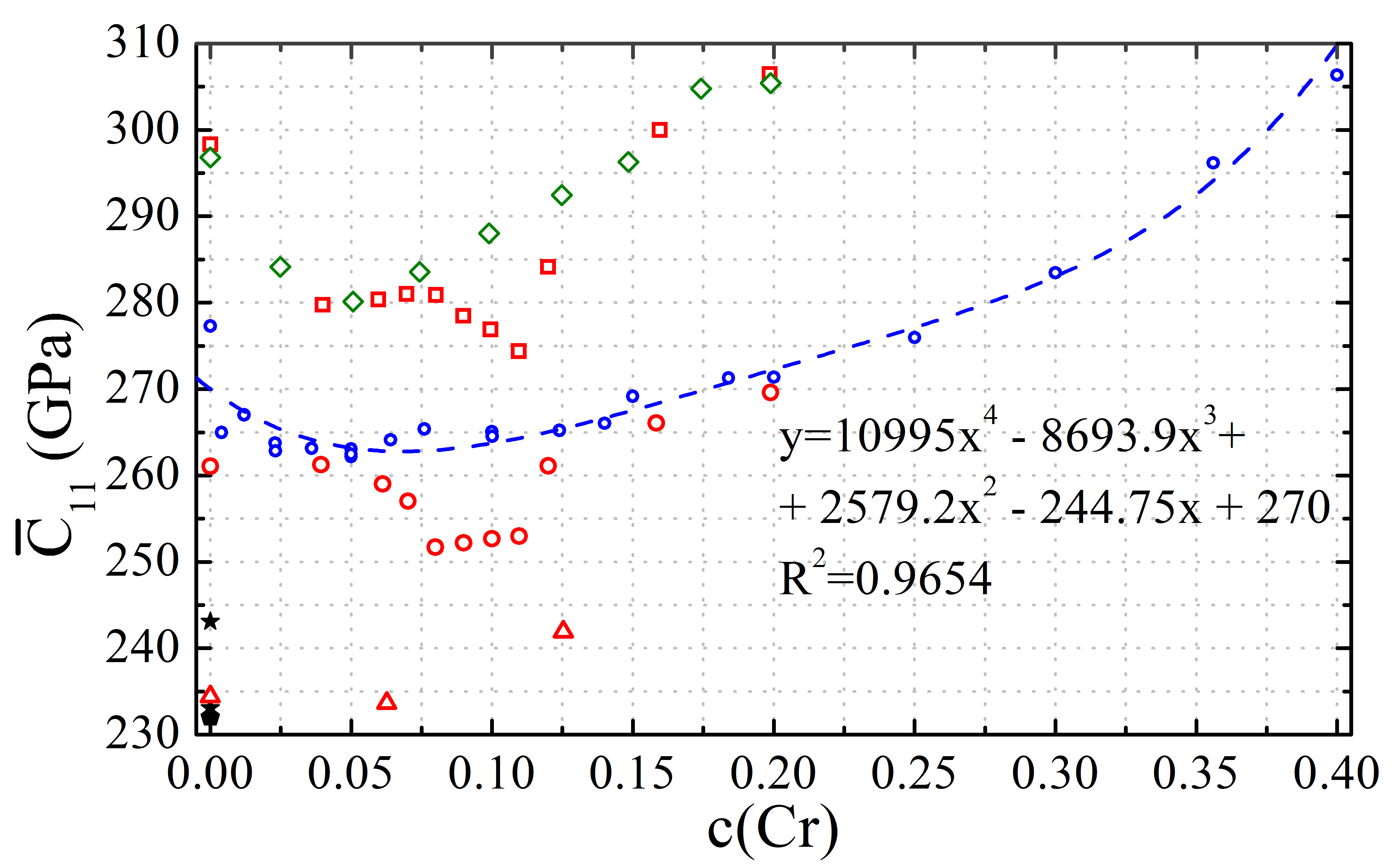}
	\end{minipage}
	\newline
	\begin{minipage}{.50\textwidth}
	  	\centering
	  	c)\includegraphics[width=.95\linewidth]{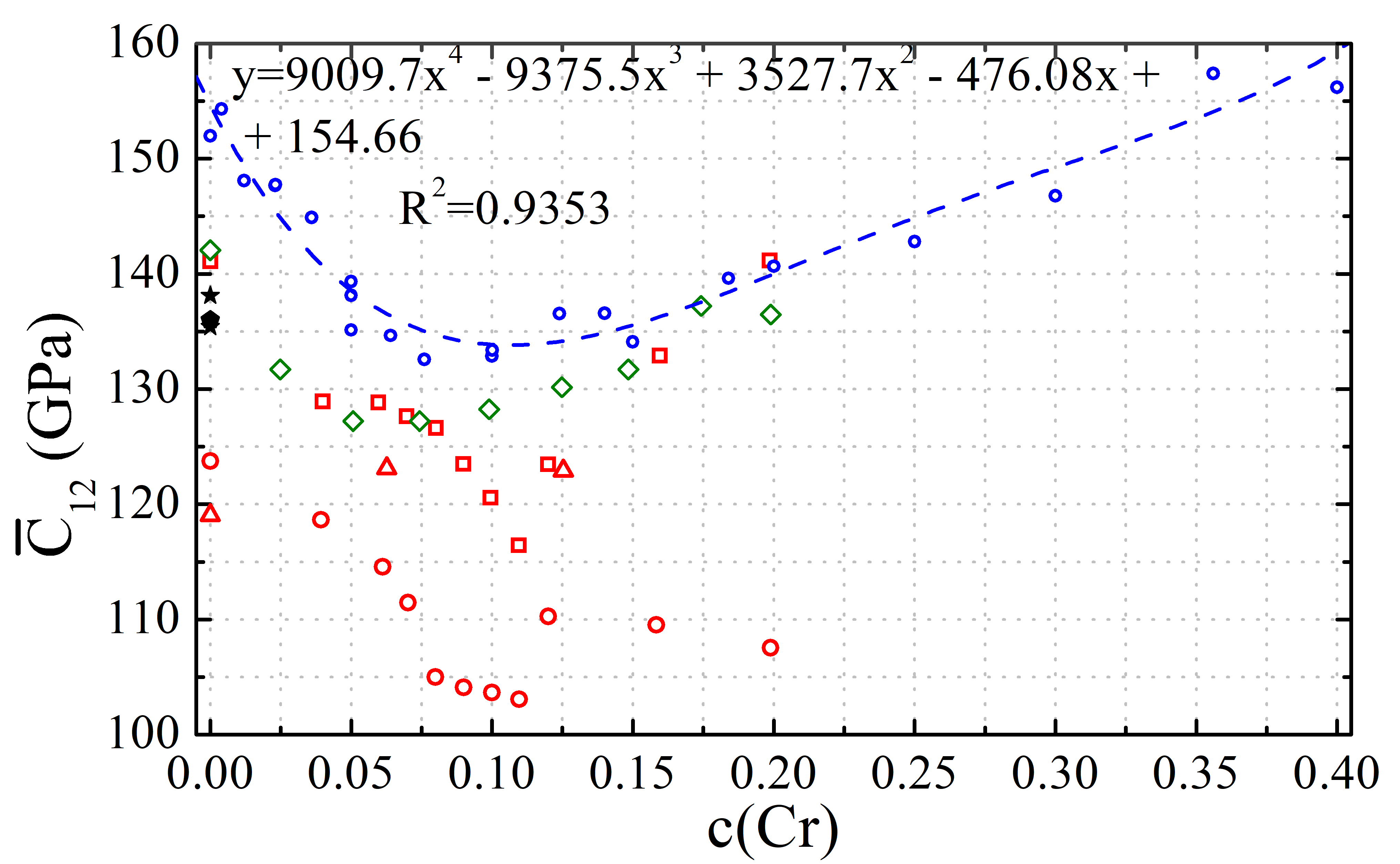}			  	
	\end{minipage}%
	\begin{minipage}{.50\textwidth}
	  	\centering
	  	d)\includegraphics[width=.95\linewidth]{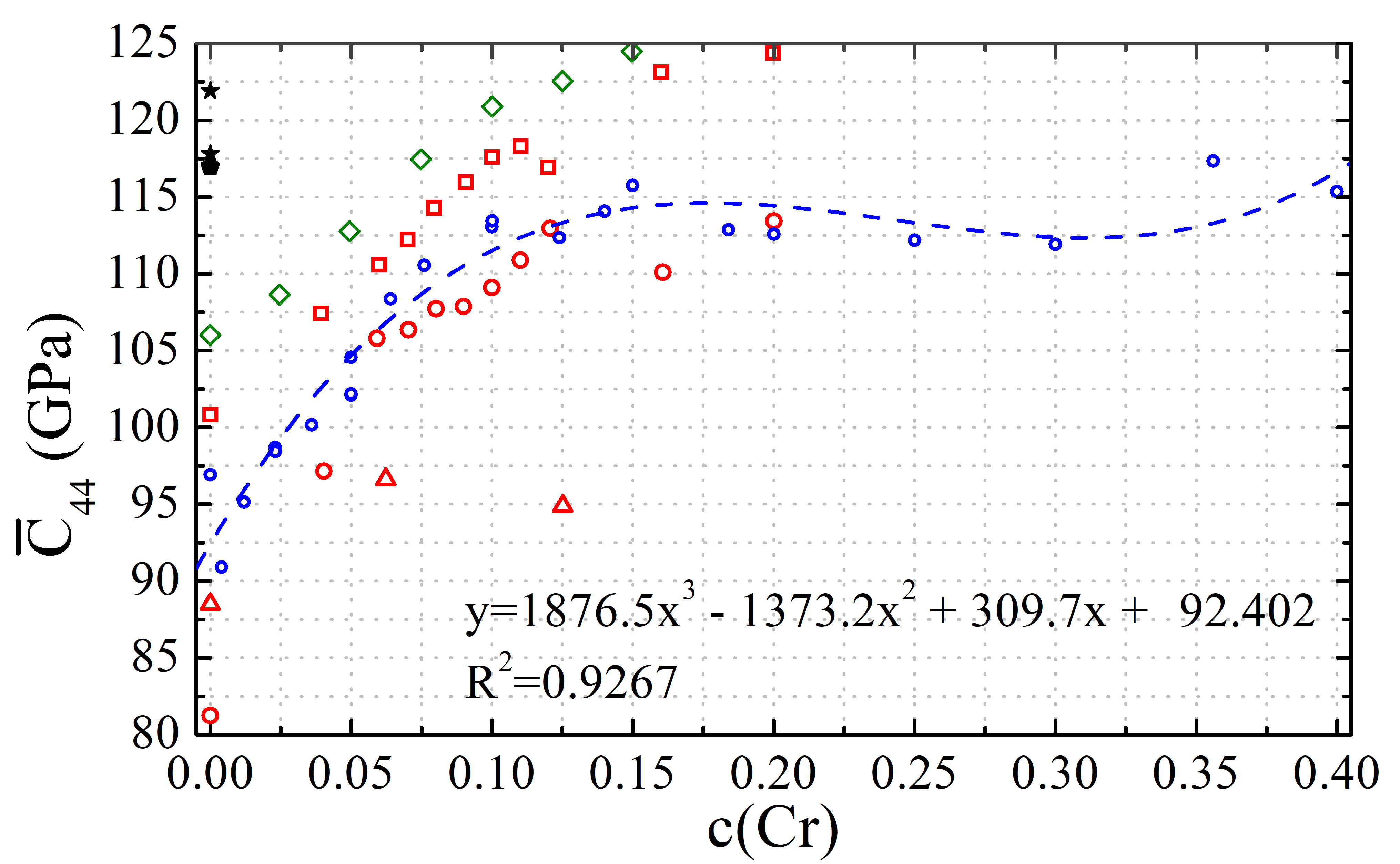}
	\end{minipage}%
	\newline
	\begin{minipage}{1.0\textwidth}
	\centering
	\includegraphics[width=0.8\linewidth]{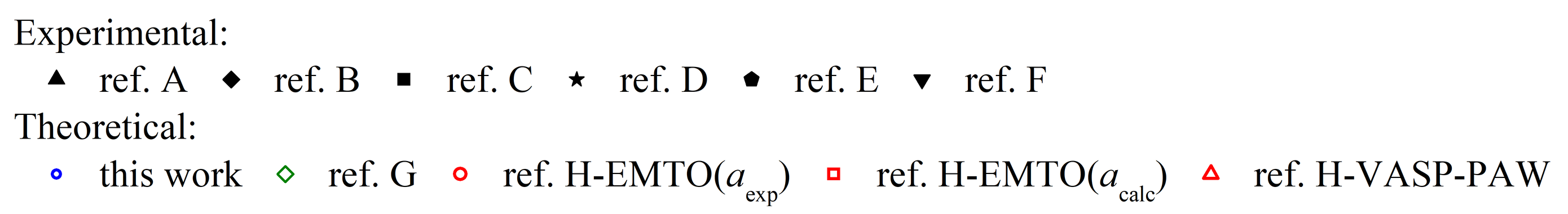}			  	
	\end{minipage}%
\caption{(a) Average lattice parameter of fully relaxed Fe-Cr structures and average elastic moduli (a) $\bar{C}_{11}$, (b) $\bar{C}_{12}$ and (c) $\bar{C}_{44}$ plotted as functions of Cr content. Experimental: ref. A \cite{Preston1932}, ref. B \cite{Sutton1955}, ref. C \cite{Speich1972}, ref. D \cite{Rayne1961}, ref. E \cite{Dever1972}, ref. F \cite{Ghosh2002}; Theoretical: ref. G \cite{Zhang2009}, ref. H \cite{Razumovskiy2011}.
        \label{fig:elast_const}}
\end{figure*}

To investigate elastic properties of disordered Fe-Cr alloys, and their variation as a function of Cr content, 21 random structures with Cr content up to 40\% at. were fully relaxed by simultaneously minimizing atomic forces and components of the global stress tensor. Average lattice parameters of the structures are shown in Fig. \ref{fig:elast_const}a. The values found in our calculations are in agreement with earlier DFT results obtained using special quasi-random structures \cite{Razumovskiy2011} and are smaller than the values obtained using the coherent potential approximation (CPA) \cite{Zhang2009,Razumovskiy2011}. We note that the experimental lattice parameters \cite{Preston1932,Sutton1955} are significantly higher than all the predicted values. This is likely associated with the approximations involved in the exchange-correlation functionals \cite{Zhang2009,Razumovskiy2011}. Also, calculations predict a visible maximum of the lattice parameter for alloys with Cr content between 7 and 12\% at. Cr, which is less well pronounced in the experimental data.  

\begin{figure*}
\centering
    \begin{minipage}{.50\textwidth}
	    \centering
	    a)\includegraphics[width=.95\linewidth]{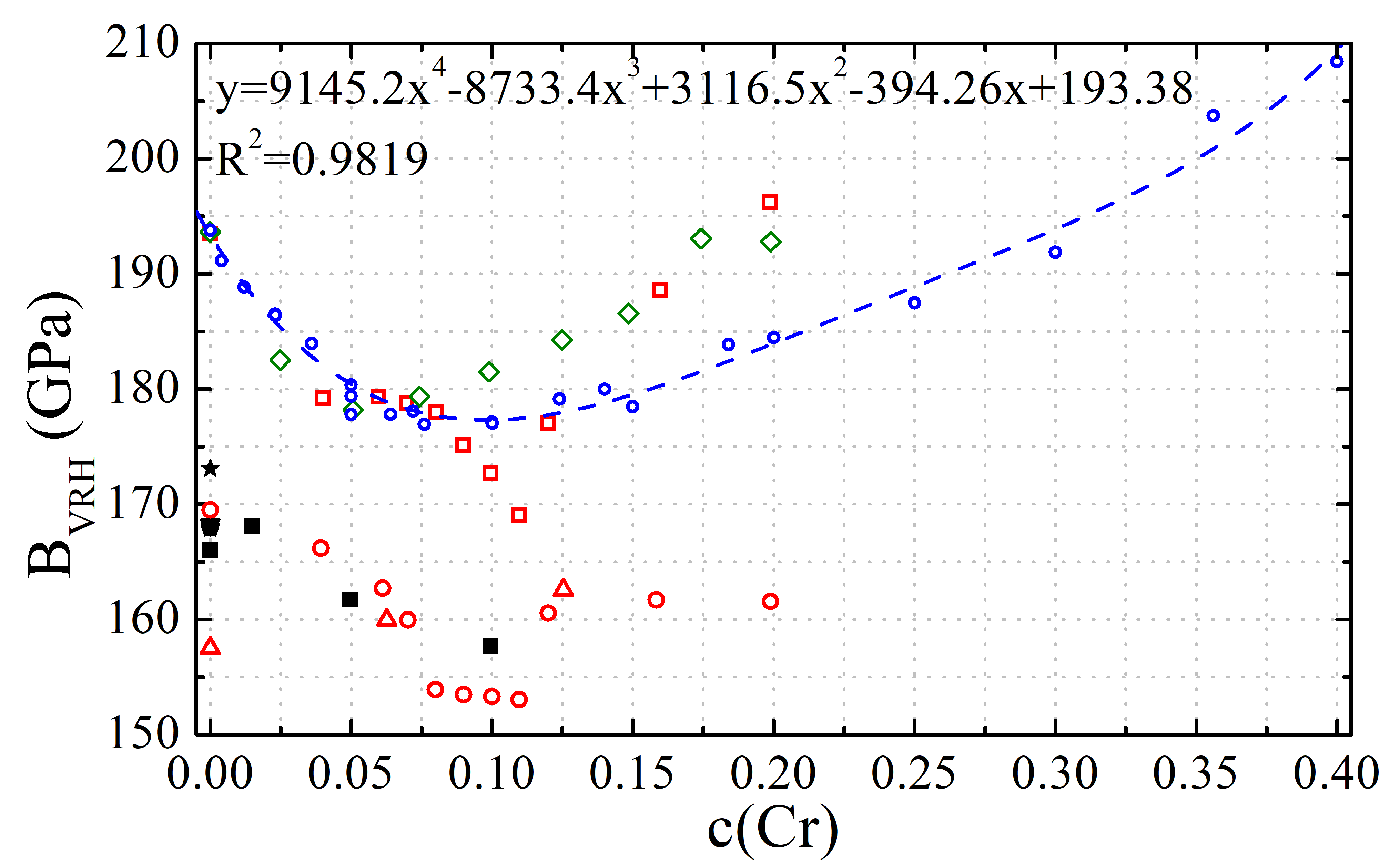}			  	
	\end{minipage}%
	\begin{minipage}{.50\textwidth}
	  	\centering
	  	b)\includegraphics[width=.95\linewidth]{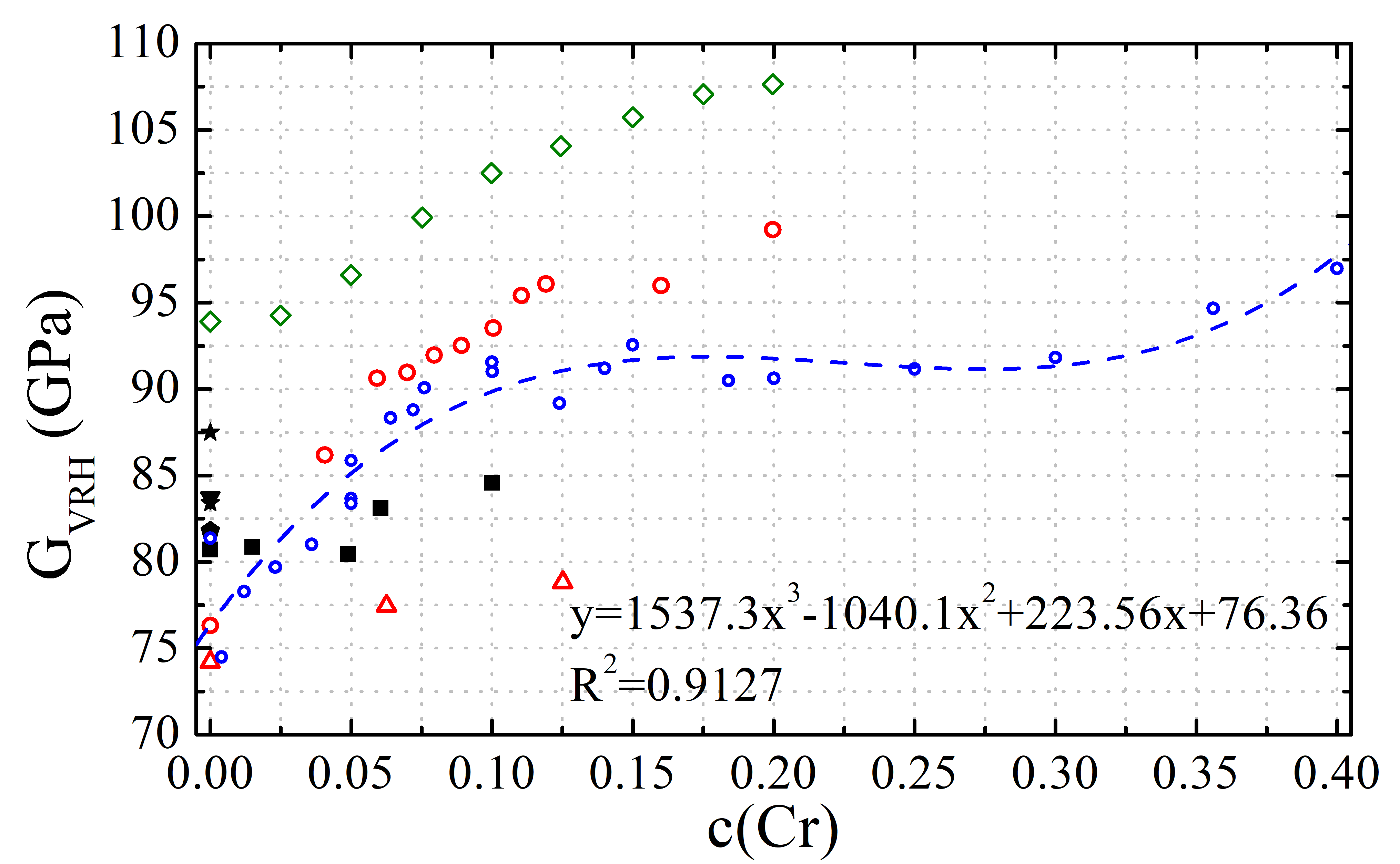}
	\end{minipage}
	\newline
	\begin{minipage}{.50\textwidth}
	  	\centering
	  	c)\includegraphics[width=.95\linewidth]{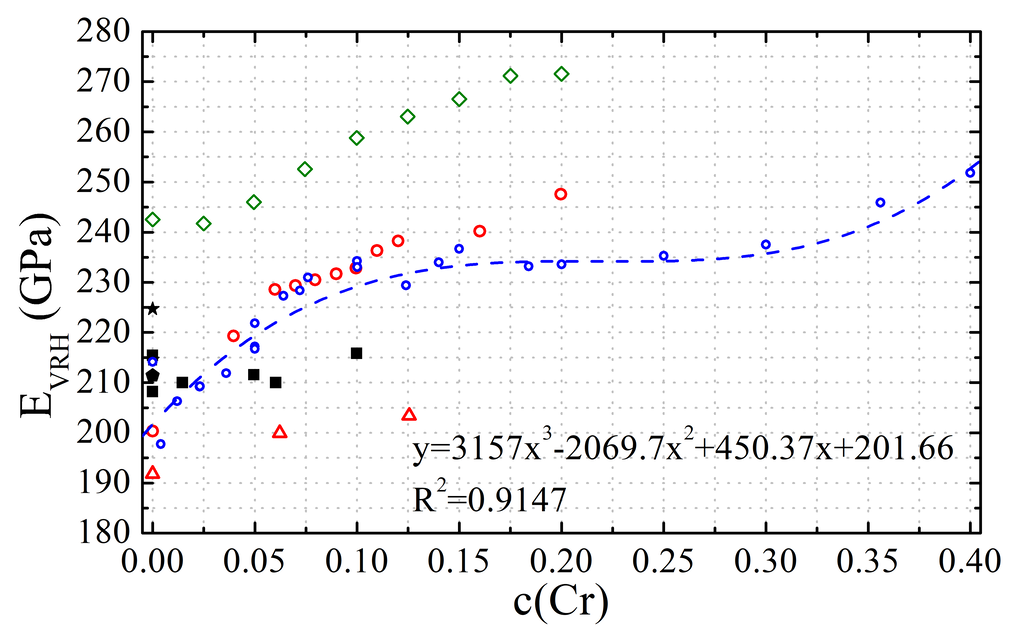}			  	
	\end{minipage}%
	\begin{minipage}{.50\textwidth}
	  	\centering
	  	d)\includegraphics[width=.95\linewidth]{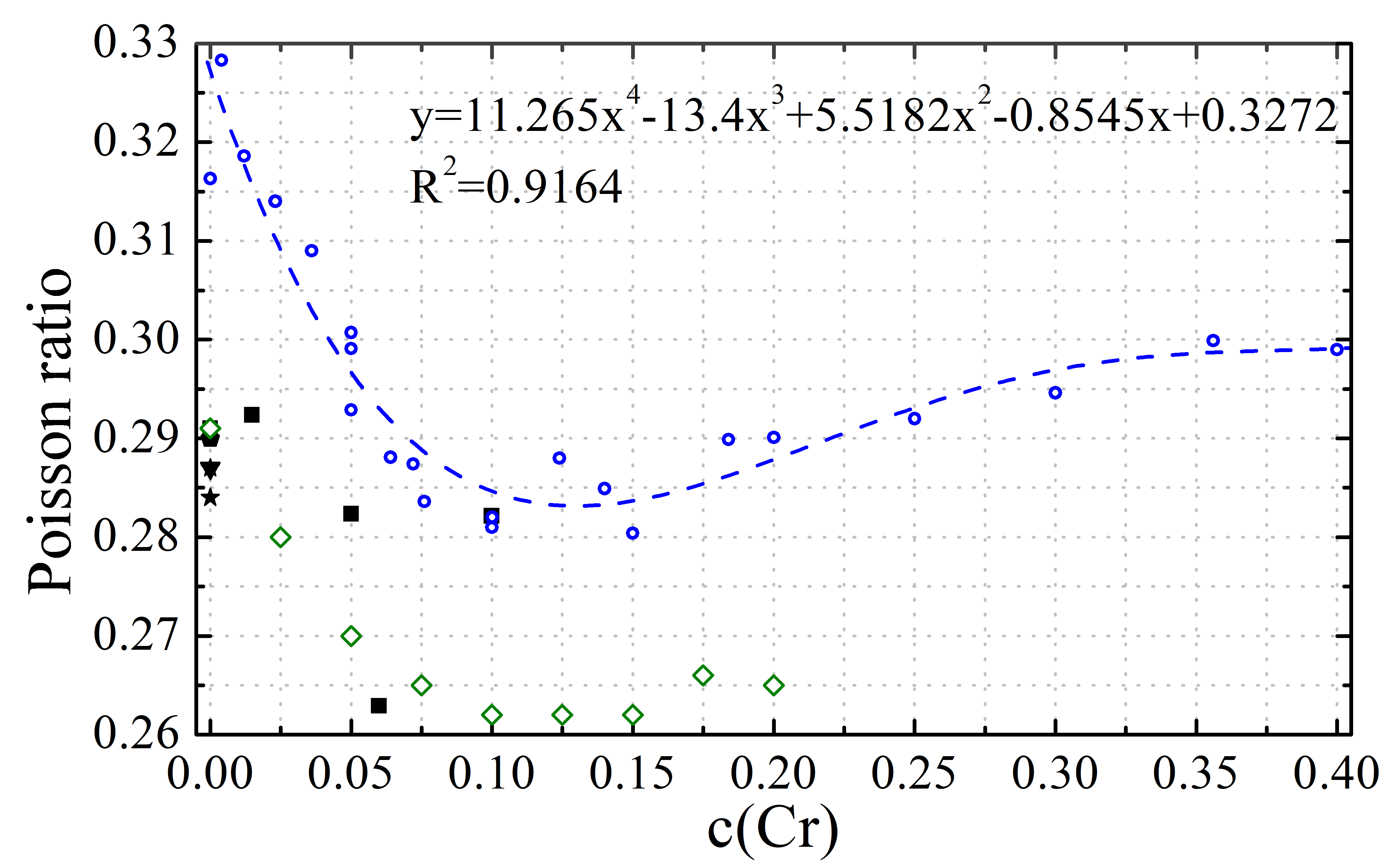}
	\end{minipage}%
		\newline
	\begin{minipage}{1.0\textwidth}
	\centering
	\includegraphics[width=0.8\linewidth]{Legenda.png}			  	
	\end{minipage}%
\caption{(a) Bulk modulus, (b) shear modulus, (c) Young's modulus and (d) Poisson's ratio calculated using the Voigt-Reuss-Hill method and average elastic moduli for random Fe-Cr structures as a function of Cr concentration. Experimental: ref. A \cite{Preston1932}, ref. B \cite{Sutton1955}, ref. C \cite{Speich1972}, ref. D \cite{Rayne1961}, ref. E \cite{Dever1972}, ref. F \cite{Ghosh2002}; Theoretical: ref. G \cite{Zhang2009}, ref. H \cite{Razumovskiy2011}.
        \label{fig:elast_prop}}
\end{figure*}

Elastic properties of a disordered Fe-10\%Cr alloy structure evaluated using various approximations are summarised in Table \ref{tab:elastic_prop_Fe10Cr}. The difference between elastic properties calculated using different approaches does not exceed 1\%.  Therefore, it is appropriate to use the elastic constants of disordered Fe-Cr alloys derived from Eqs. (\ref{eq:average_c11},\ref{eq:average_c12},\ref{eq:average_c44},\ref{eq:c14_non-diagonal}). To verify how the elastic properties vary depending on the specific atomic configurations of random Fe-Cr structures, calculations were performed for three additional structures of Fe-5\%Cr alloy. As Figs. \ref{fig:elast_const} and \ref{fig:elast_prop} show, the difference between the maximum and minimum values of each elastic constant and elastic bulk property does not exceed 3\%. Since the differences between elastic properties of Fe-Cr alloys with different compositions can be an order of magnitude larger, the effect of atomic arrangement in random Fe-Cr structures can be safely neglected in the context of this study.

\begin{table}[]
\centering
\caption{Elastic properties of disordered structure of Fe-10\%Cr alloy calculated using various approximations. $E_{min}$, $E_{max}$, $E_{av}$ are the minimum, maximum and average values of Young's moduli, $E_{SD}$ is the standard deviation and $\frac{E_{SD}}{E_{av}}$ is the coefficient of variation. $B_{\mathrm{VRH}}$, $G_{\mathrm{VRH}}$, $E_{\mathrm{VRH}}$ are the average bulk, shear, and Young's moduli obtained using the Voigt-Reuss-Hill method \cite{Hill1952}. $\nu_{av}$ is the average Poisson's ratio. All the moduli and standard deviations are given in GPa units, the coefficient of variation is in percent [\%].  }
\begin{tabular}{cccc}
 & Full & Approx. using & Approx. using \\
 & matrix & Eq. (\ref{eq:c14_non-diagonal}) & Eqs. (\ref{eq:average_c11},\ref{eq:average_c12},\ref{eq:average_c44},\ref{eq:c14_non-diagonal}) \\
     \hline
$E_{min}$  & 175.81      & 175.84                & 176.09                   \\
$E_{max}$  & 283.25      & 281.60                 & 281.62                   \\
$E_{av}$  & 232.61      & 232.68                & 232.71                   \\
$\frac{E_{max}}{E_{min}} $  & 1.611       & 1.601                 & 1.599                    \\
$E_{SD}$  & 28.15       & 28.14                 & 28.14                    \\
$\frac{E_{SD}}{E_{av}} $  & 12.10\%     & 12.09\%               & 12.09\%                  \\
$B_{\mathrm{VRH}}$  & 177.15      & 177.13                & 177.18                   \\
$G_{\mathrm{VRH}}$  & 91.53       & 91.55                 & 91.55                    \\
$E_{\mathrm{VRH}}$  & 234.21      & 234.25                & 234.26                   \\
$\nu_{av}$  & 0.2811      & 0.2810                 & 0.2811                  
\end{tabular}
\label{tab:elastic_prop_Fe10Cr}%
\end{table}

Average elastic constants $\bar{C}_{11}, \bar{C}_{12}, \bar{C}_{44}$ of random Fe-Cr structures plotted as functions of Cr concentration are shown in Fig. \ref{fig:elast_const}b-d. They were computed for 21 random structures with Cr content up to 40\%. For each fully relaxed structure, nine elastic constants were computed and average elastic constants $\bar{C}_{11}, \bar{C}_{12}, \bar{C}_{44}$ were evaluated using Eqs. (\ref{eq:average_c11},\ref{eq:average_c12},\ref{eq:average_c44},\ref{eq:c14_non-diagonal}). Results for $\bar{C}_{11}, \bar{C}_{12}, \bar{C}_{44}$ were interpolated using analytical formula in order to then use them in the calculations of elastic interactions and relaxation volumes for each alloy composition, see Fig. \ref{fig:elast_const}b-d. Analysis of earlier theoretical studies shows that the computed elastic constants of Fe-Cr alloys can vary depending on method used and the chosen value of the lattice parameter. The difference between the calculated values can be as large as 30-40 GPa (see Figs. \ref{fig:elast_const}a and \ref{fig:elast_const}b). For pure Fe, theoretical predictions often overestimate the experimental values of $\bar{C}_{11}$ and usually $\bar{C}_{12}$, and underestimate $\bar{C}_{44}$.

In calculations of relaxation volume tensors and relaxation volumes of point defects in bcc Fe and bcc Cr we used the following computed values of elastic constants: $C_{11}=277.29$ GPa, $C_{12}=151.29$ GPa and $C_{44}=96.93$ for bcc Fe and $C_{11}=459.73$ GPa, $C_{12}=49.29$ GPa and $C_{44}=93.65$ for bcc Cr.

Using the above values of elastic constants, elastic properties of random Fe-Cr structures were evaluated, see for example Figs. \ref{fig:elast_prop}a-d showing the bulk, shear and the Young moduli as well as Poisson's ratio evaluated using the Voigt-Reuss-Hill method \cite{Hill1952}, all as functions of Cr content.  The lowest bulk modulus is found for Fe-Cr random alloys with 10\% at. Cr , which corresponds to the solubility limit of Cr. The shear and the Young moduli increase rapidly as functions of Cr content up to approx. 10\% at. Cr. For larger Cr concentrations they vary slowly. The Poisson ratio decreases rapidly as a function of Cr content up to approx. 10\% at. Cr. For larger Cr concentrations it increases but only slightly. In Ref. \cite{Zhang2013} it was proposed that the anomalous behaviour of elastic properties of Fe-Cr alloys as a function of Cr content results from the interplay of magnetic and chemical effects. The effect of alloying on elastic properties is different at low and high Cr concentrations due to the rapid increase of magnitude of magnetic moments of Cr atoms in Fe-rich alloys as Cr content is lowered. Equations, interpolating elastic properties of Fe-Cr alloys over a range of Cr concentrations, are given in Table IV of the Appendix.

When compared to experimental data, the values of the bulk modulus ($B_{\mathrm{VRH}}$) and Poisson's ratio ($\nu$) of Fe-Cr alloys computed in this study appear overestimated, whereas the computed values of the shear ($G_{\mathrm{VRH}}$) and Young's ($E_{\mathrm{VRH}}$) moduli agree with experiment fairly well. The calculated concentration dependence of $B_{\mathrm{VRH}}$, $G_{\mathrm{VRH}}$, $E_{\mathrm{VRH}}$, and $\nu$ shows the same trends as those observed experimentally for Fe-Cr alloys, namely that the values of $B_{\mathrm{VRH}}$ and $\nu$ decrease whereas the values of $G_{\mathrm{VRH}}$ and $E_{\mathrm{VRH}}$ increase as a function of Cr content in the interval from 0\% to 10\% at. Cr. As was noted in Ref. \cite{Zhang2009}, the overestimation of $B_{\mathrm{VRH}}$ results mainly from the underestimation of the computed equilibrium lattice parameter.

\begin{figure*}
\centering
      \begin{minipage}{1.0\textwidth}
	    \centering
	    a)\includegraphics[width=.475\linewidth]{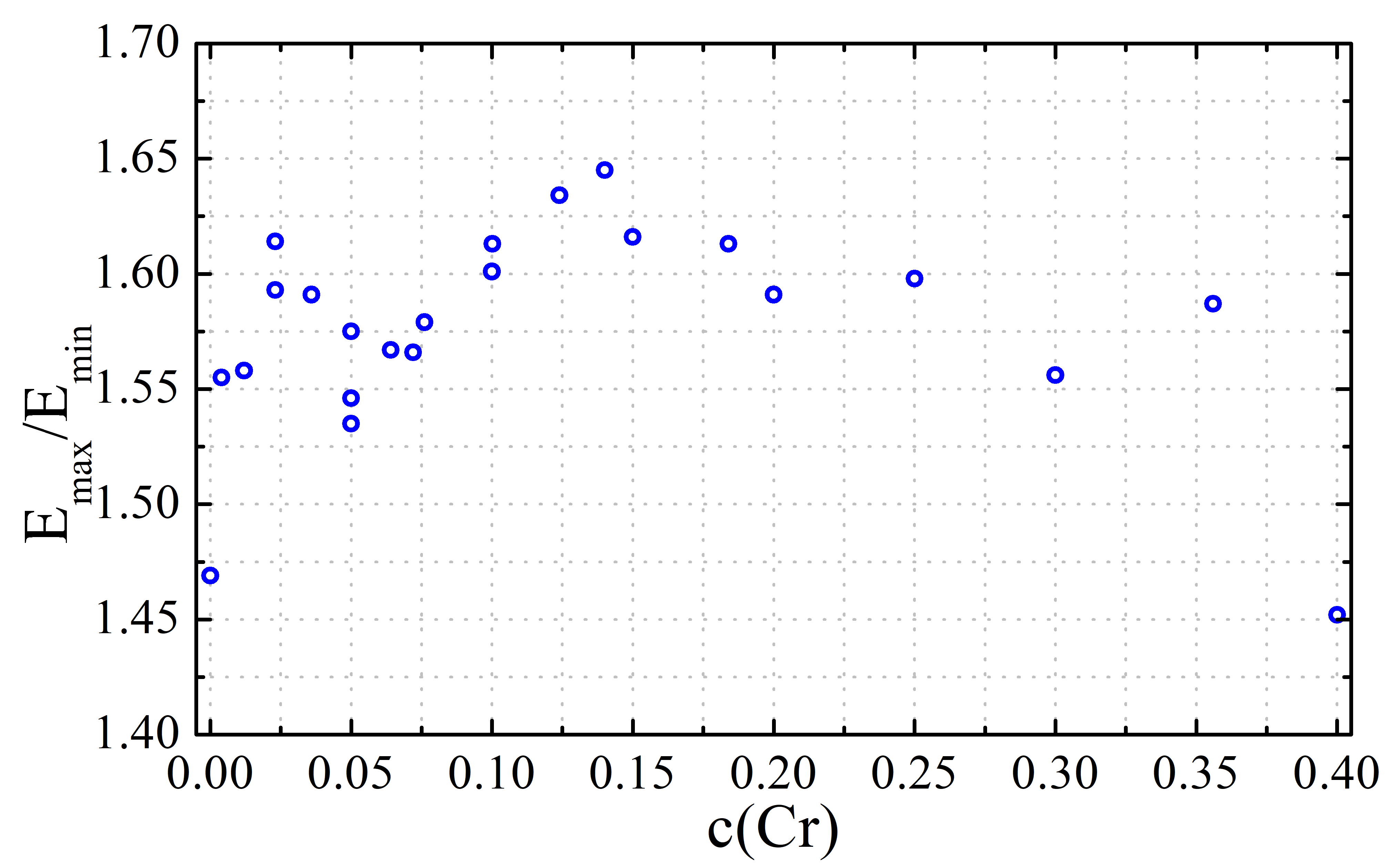}			  	
	\end{minipage}%
	\newline
		\begin{minipage}{1.0\textwidth}
	  	\centering
	  	\includegraphics[width=.4\linewidth]{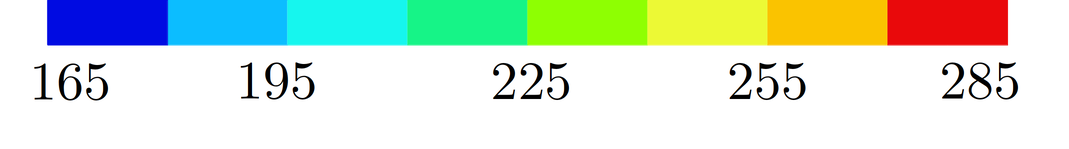}
	\end{minipage}%
	\newline
	\begin{minipage}{.33\textwidth}
	  	\centering
	  	b)\includegraphics[width=.9\linewidth]{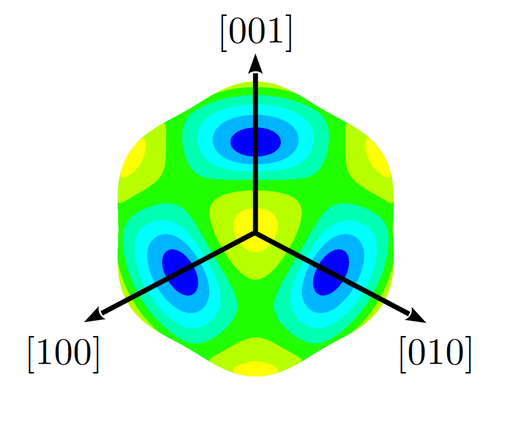}
	\end{minipage}%
	\begin{minipage}{.33\textwidth}
	  	\centering
	  	c)\includegraphics[width=.9\linewidth]{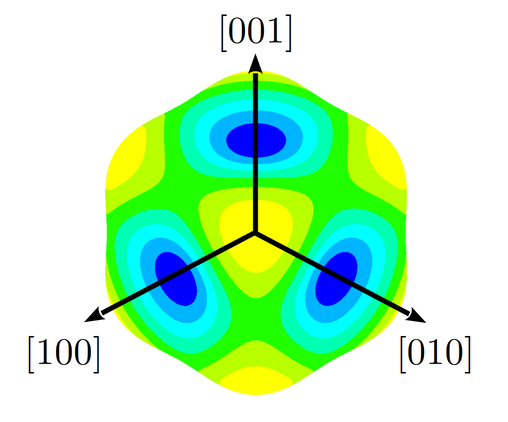}			  	
	\end{minipage}%
	\begin{minipage}{.33\textwidth}
	  	\centering
	  	d)\includegraphics[width=.9\linewidth]{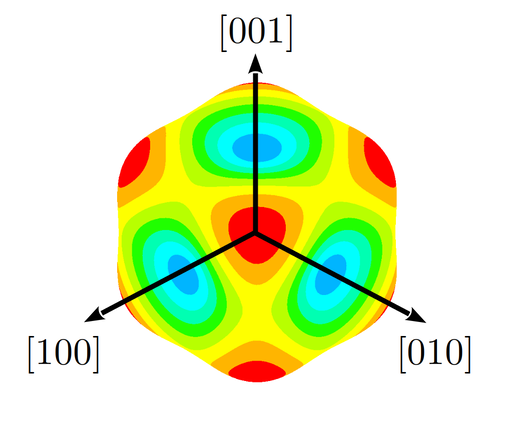}
	\end{minipage}%
	\newline
	\begin{minipage}{.33\textwidth}
	  	\centering
	  	e)\includegraphics[width=.9\linewidth]{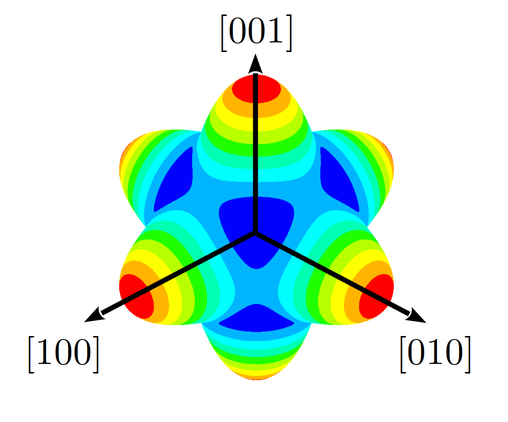}
	\end{minipage}%
	\begin{minipage}{0.25\textwidth}
	  	\centering
	  	\includegraphics[width=.75\linewidth]{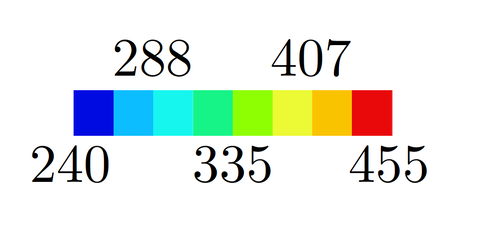}
	\end{minipage}
\caption{Anisotropy of elastic properties of random Fe-Cr structures: (a) ratio of the maximum to minimum values of Young's moduli as a function of Cr content; the Young modulus crystallographic direction dependence surfaces for (b) pure bcc Fe, (c) Fe-5\%Cr alloy, (d) Fe-30\%Cr alloy, and (e) pure bcc Cr. The scale bar above refers to (b), (c) and (d), the scale bar below refers to (e).
        \label{fig:elast_anisotropy}}
\end{figure*}

There is a significant variation of anisotropy of elastic properties as a function of Cr content (\textit{e.g}. the Young modulus can differ depending on the choice of crystallographic orientation). Fig. \ref{fig:elast_anisotropy}a shows the ratio of maximum to minimum values of Young's moduli ($E_{max}/E_{min}$). The lowest value of the $E_{max}/E_{min}$ ratio of 1.469 is observed for pure bcc Fe, whereas the highest anisotropy of Young's modulus for the random Fe-Cr structure is found in 14\% at. Cr alloy,  where $E_{max}/E_{min} = 1.645$. For Cr concentrations above 14\%, the $E_{max}/E_{min}$ ratio decreases as a function of Cr content. In alloys with low Cr content, the maximum value of the Young modulus is about 55-60\% larger than the minimum value. We note also that even for the same alloy composition, different structures exhibit slightly different elastic anisotropies. For example, in the Fe-5\%Cr alloy, values of $E_{max}/E_{min}$ vary from 1.536 to 1.575. Fig. \ref{fig:elast_anisotropy}b-d shows Young's modulus surfaces of pure bcc Fe, bcc Cr and random Fe-Cr structures containing 5\% and 30\% at. Cr, generated using the method described in Ref. \cite{Wrobel2012}. We note that the Young modulus along [111] is significantly larger than that along [100] for all the Fe-rich structures, whereas the crystallographic directions corresponding to maximum and minimum values of Young’s moduli in bcc Cr are reversed in comparison with Fe. The difference in elastic anisotropy of bcc Fe and bcc Cr is in agreement with the analysis given in Refs. \cite{Zhang2007,Zhang2007a}.

\subsection{Elastic dipole and relaxation volume tensors of point defects in bcc Fe and bcc Cr}

Elements of elastic dipole tensor $P_{ij}$, relaxation volume tensor $\Omega_{ij}$, and relaxation volumes $\Omega_{rel}$ and $\Omega_{rel}^{at}$ of a vacancy as well as Fe-Fe, Fe-Cr, and Cr-Cr $\left\langle 110\right\rangle$ dumbbells in bcc Fe and bcc Cr are summarised in Table \ref{tab:elastic_dipole_tensors_pure}. Values of relaxation volume $\Omega_{rel}$ are given in \AA$^3$ units, whereas $\Omega_{rel}^{at}$ are given in the units of atomic volume ($\Omega_0=a^3/2$), where the reference atomic volume $\Omega_{0}$ = 11.345 \AA$^3$ corresponds to the bcc lattice parameter of $a$ = 2.831 \AA. In agreement with the analysis given in Ref. \cite{Ma2019}, $P_{ij}$, $\Omega_{ij}$, $\Omega_{rel}$ and $\Omega_{rel}^{at}$ for vacancies are negative in both pure Fe and Cr, whereas for dumbbells they are positive and  their magnitudes are significantly larger than those for vacancies. The fact that SIA defects have large relaxation volumes shows that self-interstitial atom defects are primarily responsible for the swelling occurring in these metals under irradiation, as a result of formation of Frenkel vacancy -- self-interstitial pairs, and the subsequent clustering of SIA defects \cite{Derlet2020}. 

For vacancies in pure metals, all the diagonal elements of elastic dipole tensors and relaxation volume tensors are equal, and the off-diagonal elements vanish. Hence, the elastic properties of vacancies can be described by only one parameter. The values of $P_{ij}$, $\Omega_{ij}$, and $\Omega_{rel}$ are approximately twice as large for bcc Cr as for bcc Fe. For example, the relaxation volume of a vacancy in Cr is $-6.513$ \AA$^3$ and in Fe it is $-3.045$ \AA$^3$. These values are larger (\textit{i.e}. more negative) than the values found in Ref. \cite{Ma2019}. The difference is larger for the vacancy in bcc Cr. This is mainly due to the fact vacancy calculations in bcc Cr in Ref. \cite{Ma2019} were performed for the equilibrium lattice parameter of 2.862 \AA, whereas all the fixed-volume calculations in this work, including those for bcc Cr, were performed assuming the lattice parameter of bcc Fe of $a=2.831$ \AA.

We note that substitutional atoms in bcc Fe and bcc Cr, namely Cr in Fe, see Fig. \ref{fig:structures}a), and Fe in Cr (see Fig. \ref{fig:structures}b), can also be treated using the relaxation volume formalism developed for point defects. 
For example, $\Omega_{rel}^{at}$ of a substitutional Cr atom in ferromagnetic bcc Fe is equal to 0.184 atomic volume units, which means that its volume is approximately 18\% larger than the volume of a host Fe atom. Interestingly, this value is about four times larger than the value obtained from the comparison of metallic radii of Fe and Cr, which are 1.26 \AA $\,$ and 1.28 \AA, respectively \cite{Greenwood1997}. The origin of the difference is likely related to the magnetism of a Cr atom, which is different between in anti-ferromagnetic bcc Cr and in a ferromagnetic bcc Fe matrix. The magnitude of the magnetic moment of a substitutional Cr atom in bcc Fe matrix (1.80 $\mu_B$) is 70\% larger than the magnitude of magnetic moment of a Cr atom in chromium metal, where it equals 1.07 $\mu_B$, according to DFT calculations \cite{Wrobel2015}.

We note also that the absolute values of $P_{ij}$, $\Omega_{ij}$, and $\Omega_{rel}$ of a Cr atom in bcc Fe are only approximately 30\% smaller than those of a vacancy. It means that the scale of elastic distortions caused by a vacancy or a Cr substitutional atom in bcc Fe is broadly similar. The signs of $P_{ij}$, $\Omega_{ij}$, and $\Omega_{rel}$ for a vacancy and substitutional Cr are opposite, $\Omega_{rel}$ for a vacancy is negative and $\Omega_{rel}$ for a substitutional Cr is positive. The latter is important as it shows that a Cr atom in bcc Fe matrix is oversized. As a consequence, it should be expected to bind to the outside part of an interstitial dislocation loop where strain is tensile. The positive value of $\Omega_{rel}$ for a Cr atom in bcc Fe is in an agreement with experimental data obtained using atom probe tomography by Jiao and Was \cite{Jiao2011}, showing that Cr segregates to the outside of an interstitial dislocation loop. It should be noted that the agreement between our calculations and the above experimental results is not fully supported by the DFT results from Ref. \cite{Domain2018}, where it was found that binding of a Cr atom to a $\left\langle 111\right\rangle$ interstitial loop in bcc Fe is insignificant on either the compressive or tensile side of the perimeter of the loop. 

As opposed to a Cr atom in bcc Fe, a substitutional Fe atom in bcc Cr has a negative relaxation volume. It means that, similarly to a vacancy, a Fe atom in bcc Cr gives rise to lattice contraction. Still,  the absolute scale of $P_{ij}$, $\Omega_{ij}$, and $\Omega_{rel}$ characterising a Fe atom in bcc chromium matrix is almost 10 times smaller than that of a vacancy. For example, $\Omega_{rel}^{at}$ for a Fe atom and a vacancy in bcc Cr equals $-0.055$ and $-0.588$ atomic volume units, respectively. The relaxation volume of a substitutional Fe atom in the Cr matrix is similar to the value that can be derived by comparing the metallic radii of Fe and Cr \cite{Greenwood1997}. This means that, as opposed to the case of a Cr substitutional atom in bcc Fe matrix, the relaxation volume of a Fe atom in bcc Cr is not significantly affected by the magneto-volume effects.

When treating $\left\langle 110\right\rangle$ Fe-Fe, Fe-Cr and Cr-Cr dumbbells (see Figs. \ref{fig:structures}c, \ref{fig:structures}d and \ref{fig:structures}e, respectively) in bcc Fe and Cr, we find that only two diagonal elements of the elastic dipole tensor or the relaxation volume tensor are equal ($P_{22}=P_{33}$ and $\Omega_{22}=\Omega_{33}$). For a Fe-Fe dumbbell in bcc Fe, the first element $P_{11}$ is larger than either $P_{22}$ or $P_{33}$, whereas the first element is smaller than the other two elements for a Cr-Cr dumbbell in bcc Cr. The $P_{11}/P_{22}$ ratio is 1.21 and 0.78 in the former and latter cases, respectively. This effect is likely caused by the significantly different anisotropy of  elastic properties of bcc Fe and bcc Cr, illustrated in Figs. \ref{fig:elast_anisotropy}b and \ref{fig:elast_anisotropy}e.  In bcc Fe and bcc Cr, the lowest and the largest $P_{11}/P_{22}$ ratios are observed for Cr-Cr and Fe-Fe dumbbells, respectively.  
As opposed to vacancies, all the dumbbells in bcc Fe and Cr have non-vanishing  off-diagonal elements $P_{23}$ and $\Omega_{23}$ of elastic dipole and relaxation volume tensors. In bcc Fe and bcc Cr, the largest value of $P_{23}$ is found for Cr-Cr dumbbells.

In general, relaxation volumes of dumbbells in bcc Fe are larger than in bcc Cr. For example, the relaxation volume of a Fe-Fe $\left\langle 110\right\rangle$ dumbbell in bcc Fe is 18.181 \AA$^3$, which is larger than the relaxation volume of a Cr-Cr dumbbell in bcc Cr, where it is equal to 16.402 \AA$^3$. Finally, we note that the values of $P_{ij}$, $\Omega_{ij}$, and $\Omega_{rel}$ for mixed Fe-Cr dumbbells vary, depending on the type of the atom, Cr or Fe, on the defect site in the pristine structure (see Figs. \ref{fig:Fe-Cr_dumb_vac_scheme}a and \ref{fig:Fe-Cr_dumb_vac_scheme}b). For example, a Fe-Cr $\left\langle 110\right\rangle$ dumbbell on a Fe or a Cr site has the relaxation volume of 18.581 \AA$^3$ and 16.356 \AA$^3$, respectively. 

\begin{figure}
 \includegraphics[width=\linewidth]{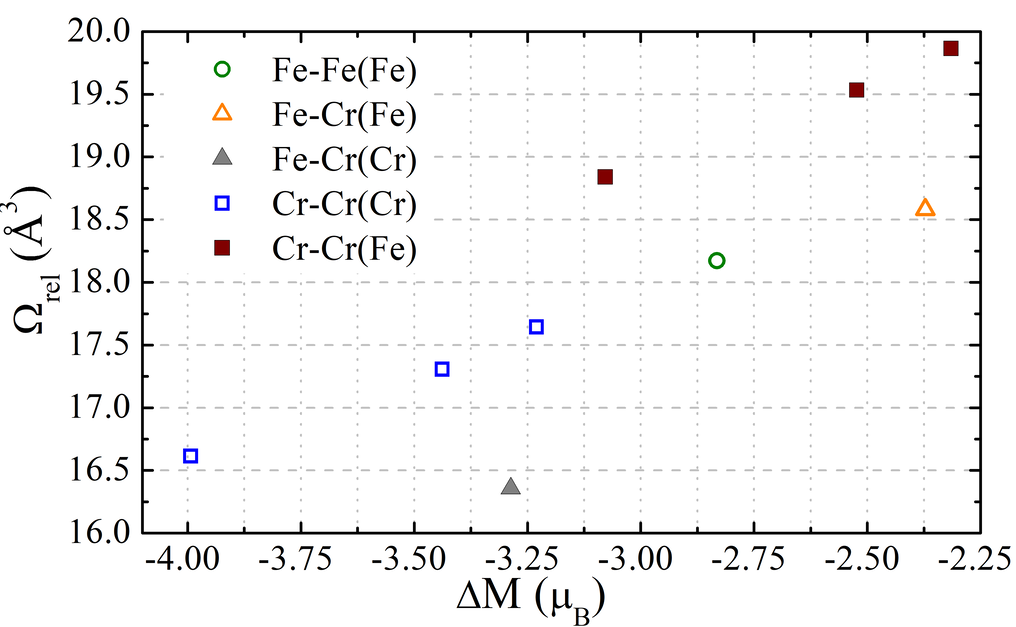}			  	
\caption{The relaxation volumes of dumbbells in bcc Fe matrix as a function of the change in the magnitude of the total magnetic moment in the supercell associated with the formation of a defect.
        \label{fig:Omega_vs_dM}}
\end{figure}

To understand the origin of differences between the relaxation volumes of dumbbells on Fe and Cr sites, the values of $\Omega_{rel}$ have been correlated with the variation of the magnitude of the magnetic moment of the supercell $\Delta M$ caused by the defect. Fig. \ref{fig:Omega_vs_dM} shows that the relaxation volumes of dumbbells are smaller when $\Delta M$ is more negative. In particular, $\Omega_{rel}$ of $\left\langle 110\right\rangle$ Fe-Cr dumbbell on a Fe site (18.581 \AA$^3$) is larger than that of a $\left\langle 110\right\rangle$ Fe-Fe dumbbell on a Fe site (18.171 \AA$^3$) since the sum of magnitudes of magnetic moments for the former structure is almost 0.5 $\mu_B$ larger. This suggests that magnetism is a significant factor affecting structural relaxation and hence relaxation volumes of defects in Fe-Cr alloys. The difference in magnetic properties between the structures containing Fe-Fe and Fe-Cr dumbbells is caused mainly by the differences in magnetic moments of atoms forming the dumbbells, which agrees with Refs. \cite{Nguyen-Manh2007,Becquart2018,Olsson2007}. 

In a $\left\langle 110\right\rangle$ Fe-Fe SIA dumbbell, the magnetic moments of Fe atoms are small (-0.207 $\mu_B$) and ordered antiferromagnetically with respect to other Fe atoms. In a $\left\langle 110\right\rangle$ Fe-Cr dumbbell the magnetic moment of Fe is larger (0.326 $\mu_B$) and ordered ferromagnetically with respect to other Fe moments and antiferromagnetically with respect to the moment of the Cr atom in the dumbbell, which has a notably larger magnitude of magnetic moment (-0.946 $\mu_B$). The magnetic moments of atoms in Fe-Fe and Fe-Cr dumbbells in bcc Fe are in agreement with the values given in Refs. \cite{Messina2020,Olsson2007}. The magnetic moments of Cr atoms in a $\left\langle 11\xi\right\rangle$ Cr-Cr dumbbell are -0.347 $\mu_B$, and both are aligned antiferromagnetically with respect to the magnetic moments of Fe atoms. 

\subsection{Elastic dipole tensors and relaxation volumes of point defects in random Fe-Cr alloys}

\begin{figure*}
\centering
      \begin{minipage}{.50\textwidth}
	    \centering
	    a)\includegraphics[width=.95\linewidth]{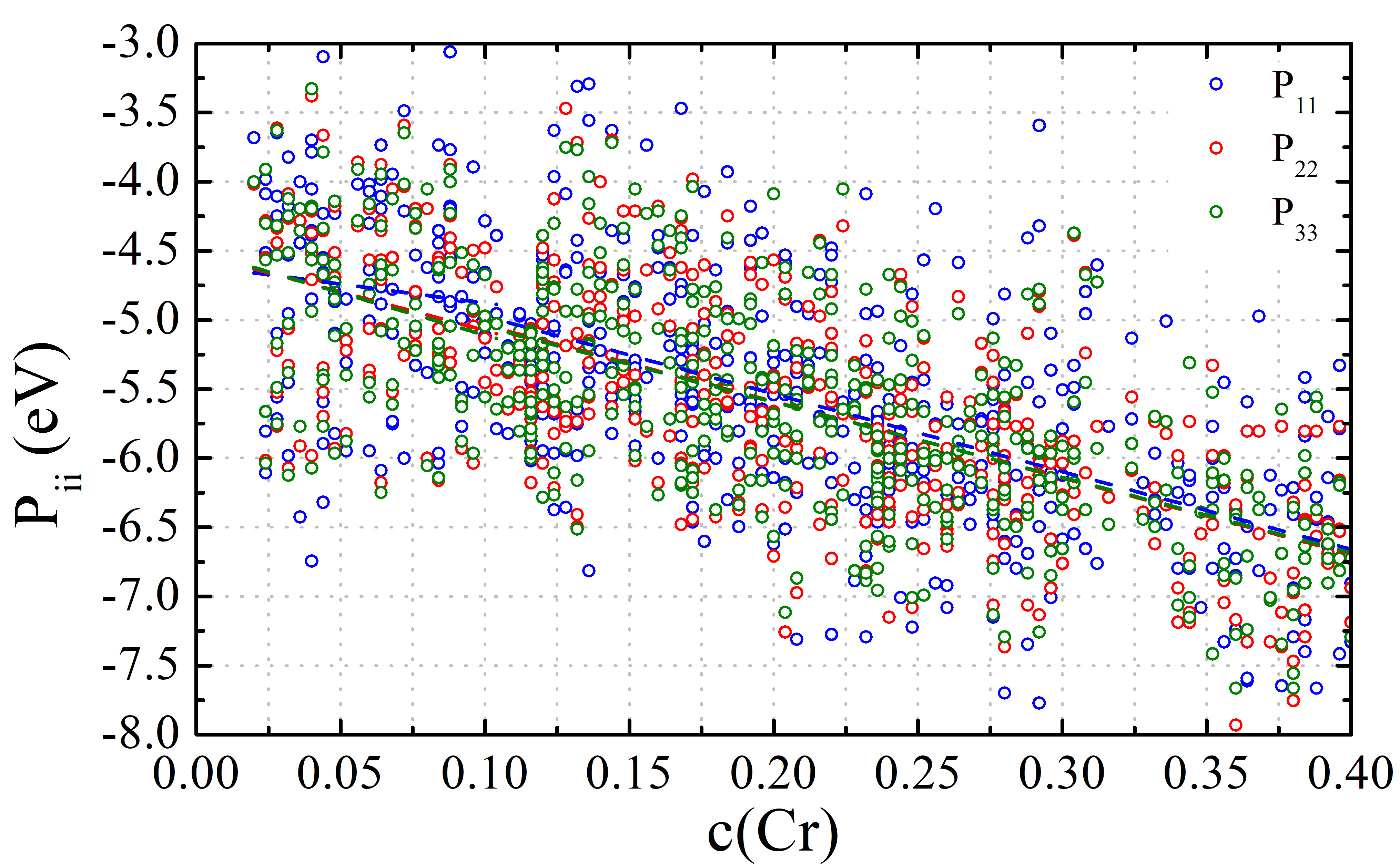}			  	
	\end{minipage}%
	\begin{minipage}{.50\textwidth}
	  	\centering
	  	b)\includegraphics[width=.95\linewidth]{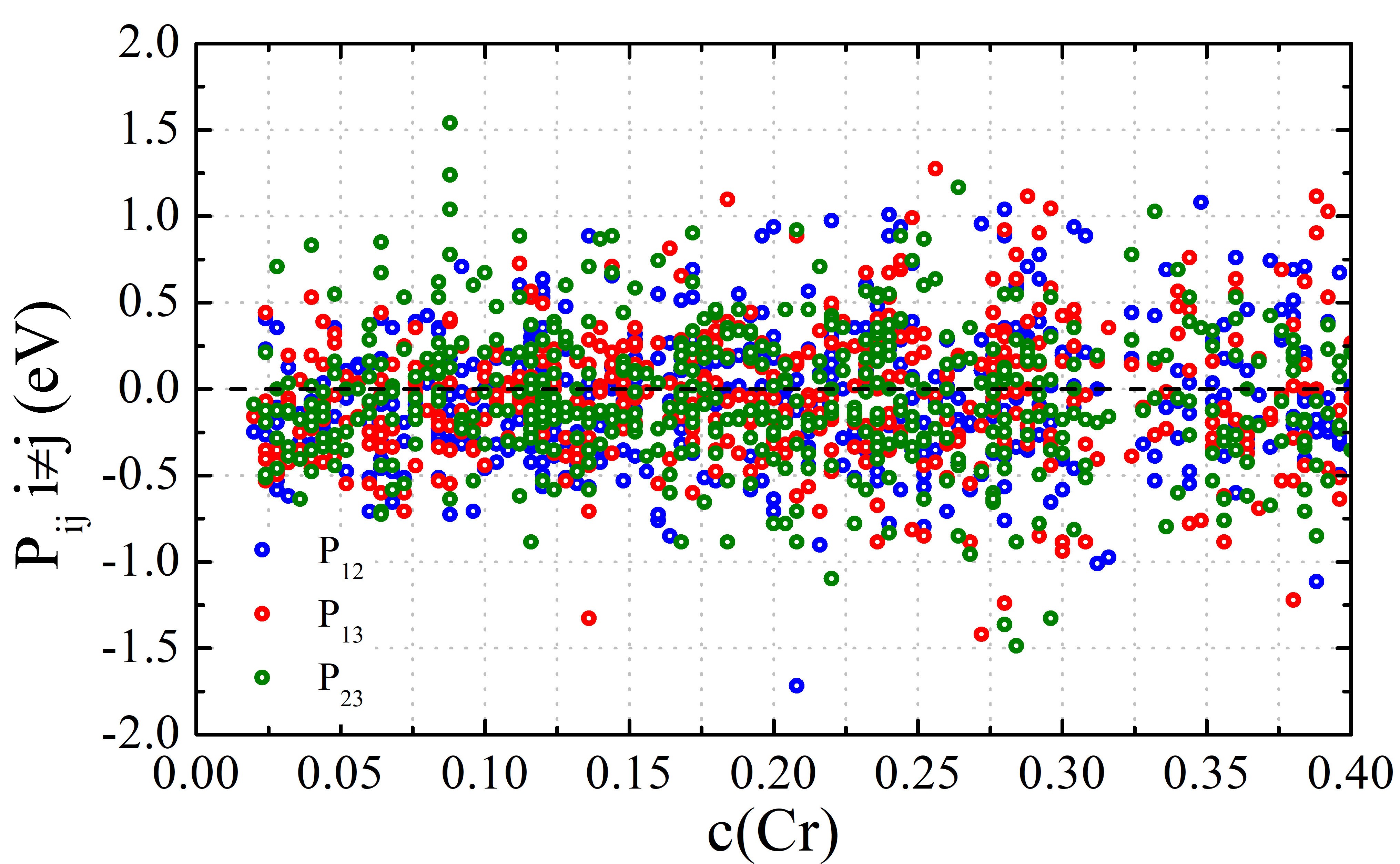}
	\end{minipage}%
\caption{(a) diagonal and (b) off-diagonal elements of elastic dipole tensor for vacancies on Fe and Cr sites in random Fe-Cr alloy structures.
        \label{fig:dipole_tensors_vac}}
\end{figure*}

\begin{figure*}
\centering
      \begin{minipage}{.50\textwidth}
	    \centering
	    a)\includegraphics[width=.95\linewidth]{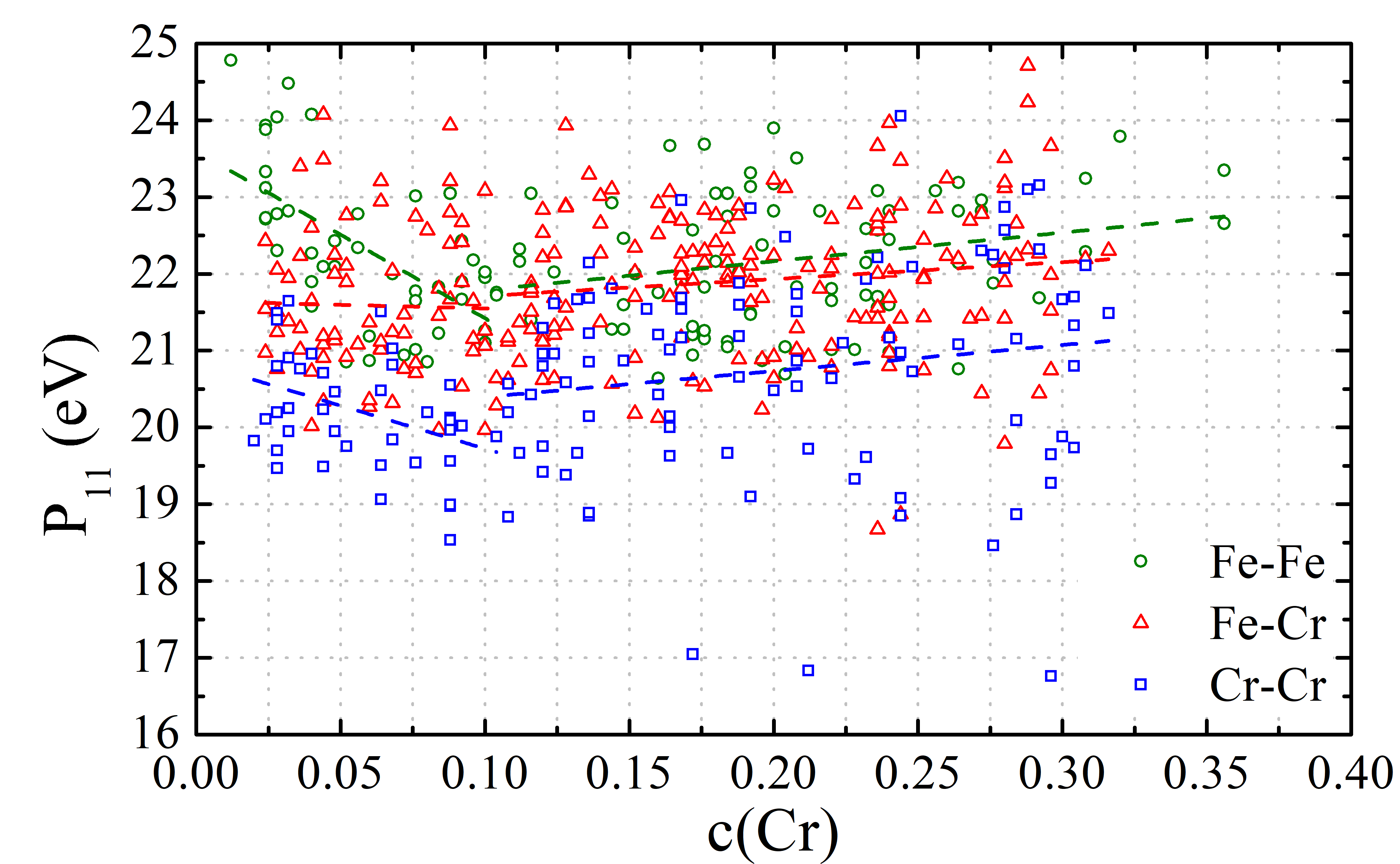}			  	
	\end{minipage}%
	\begin{minipage}{.50\textwidth}
	  	\centering
	  	b)\includegraphics[width=.95\linewidth]{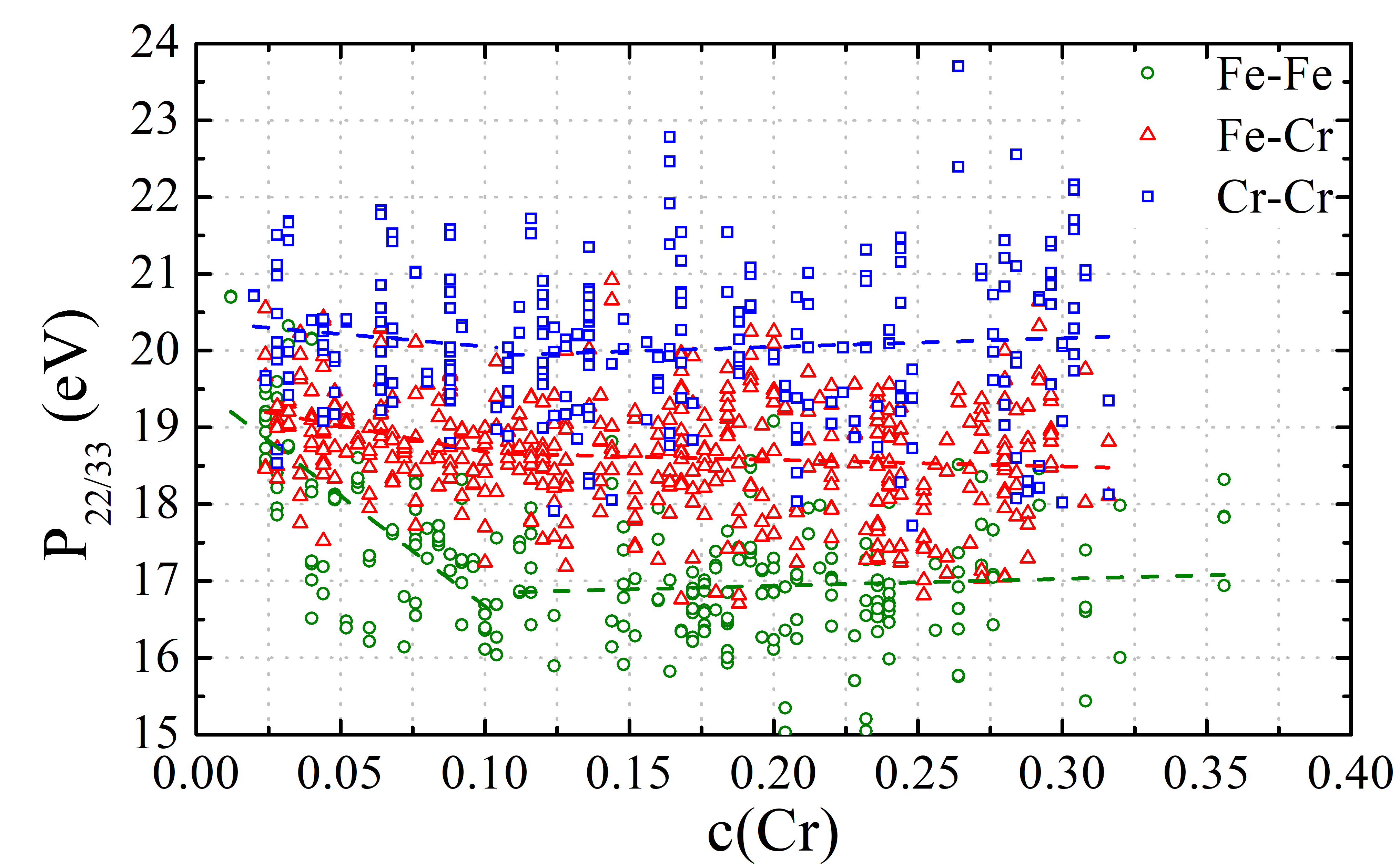}
	\end{minipage}%
	\newline
      \begin{minipage}{.50\textwidth}
	    \centering
	    c)\includegraphics[width=.95\linewidth]{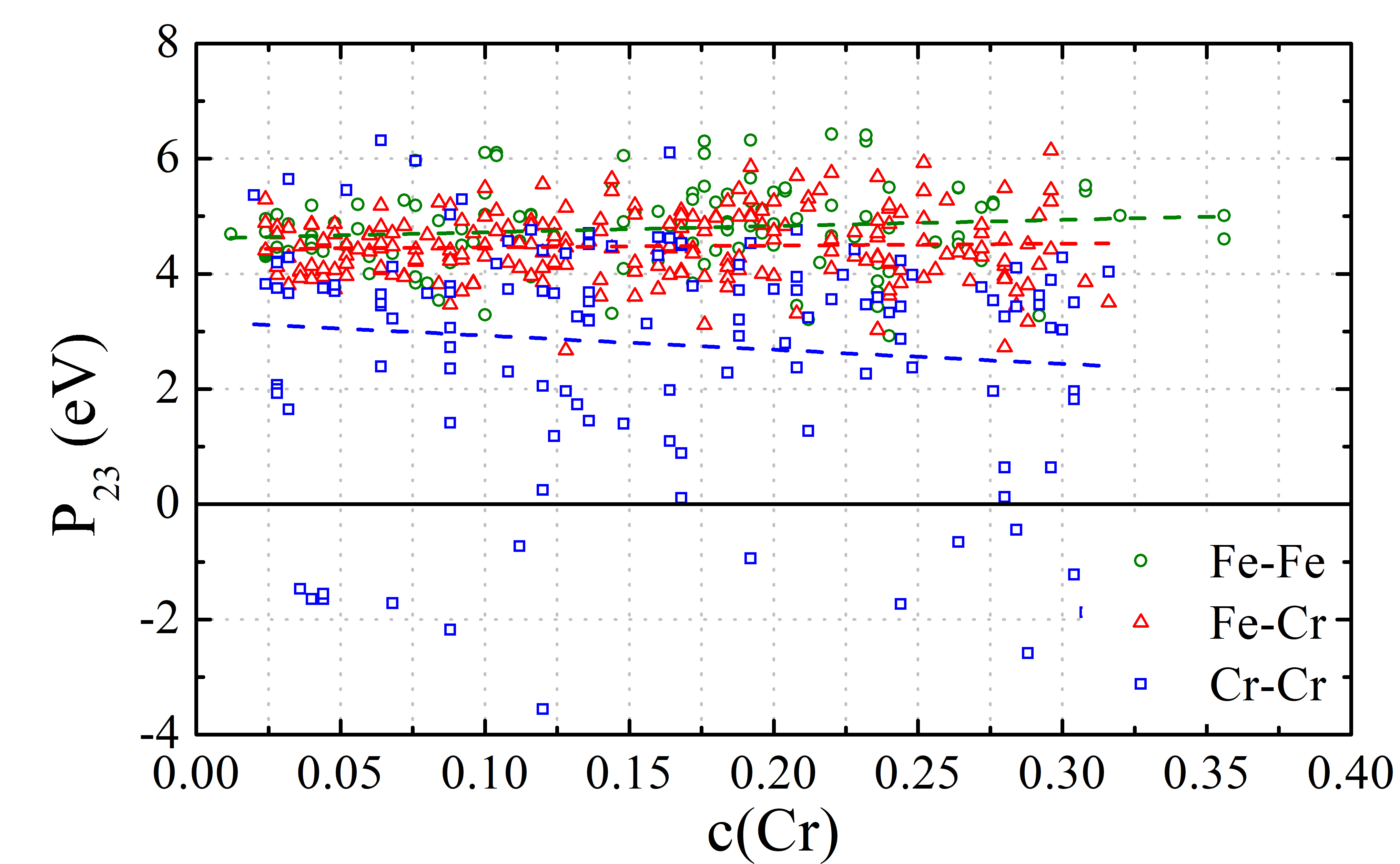}			  	
	\end{minipage}%
	\begin{minipage}{.50\textwidth}
	  	\centering
	  	d)\includegraphics[width=.95\linewidth]{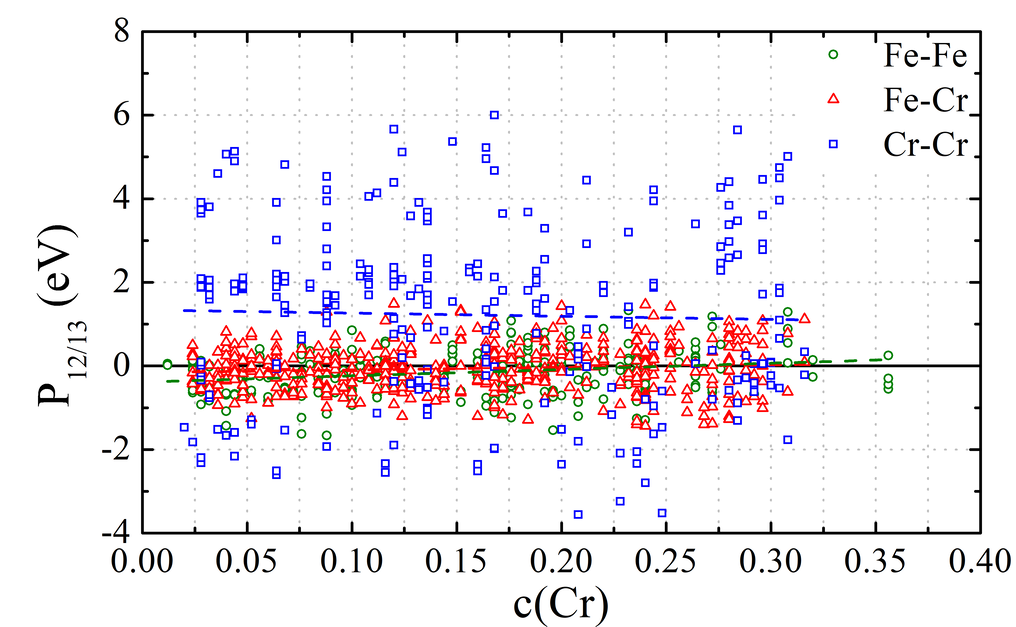}
	\end{minipage}%
\caption{Elements of elastic dipole tensor (a) $P_{11}$, (b) $P_{22}$ and $P_{33}$, (c) $P_{23}$, and (d) $P_{12}$ and $P_{13}$, computed for Fe-Fe, Fe-Cr and Cr-Cr dumbbells in random Fe-Cr alloys.
        \label{fig:dipole_tensors_dumbbells}}
\end{figure*}

In random Fe-Cr alloys, the elements of $P_{ij}$ and $\Omega_{ij}$ depend not only on the type of the defect but also on the atomic configuration of Cr and Fe in its local environment. Due to the random choice of positions of Cr atoms, all the elements of $P_{ij}$ and $\Omega_{ij}$ of defects differ from each other and are non-zero, even for vacancies - where in pure metals, because of cubic symmetry, we find that $P_{11}=P_{22}=P_{33}$ and $P_{12}=P_{23}=P_{31}=0$. Figs. \ref{fig:dipole_tensors_vac} and \ref{fig:dipole_tensors_dumbbells} show that the values of $P_{ij}$ for vacancies and dumbbells are fairly scattered. However, similarly to the data for defects in pure metals, there are notable identifiable trends that we discuss below. 

For vacancies, the magnitudes of $P_{11}$, $P_{22}$ and $P_{33}$ are notably larger than those of $P_{12}$, $P_{23}$ and $P_{31}$, and the mean values of the latter ones are very close to zero, see Fig. \ref{fig:dipole_tensors_vac}. This is expected, and is consistent with the argument given in Ref. \cite{Dudarev2018a} that averaging over configurations generally gives rise to the isotropic form of defect dipole and relaxation volume tensors. For dumbbells, as in the case of pure metals, values of $P_{22}$ and $P_{33}$ are usually similar, whereas $P_{11}$ can be either smaller or larger than $P_{22}$ and $P_{33}$, cf. Figs. \ref{fig:dipole_tensors_dumbbells}a and \ref{fig:dipole_tensors_dumbbells}b. 

\begin{figure*}
\centering
      \begin{minipage}{.50\textwidth}
	    \centering
	    a)\includegraphics[width=.95\linewidth]{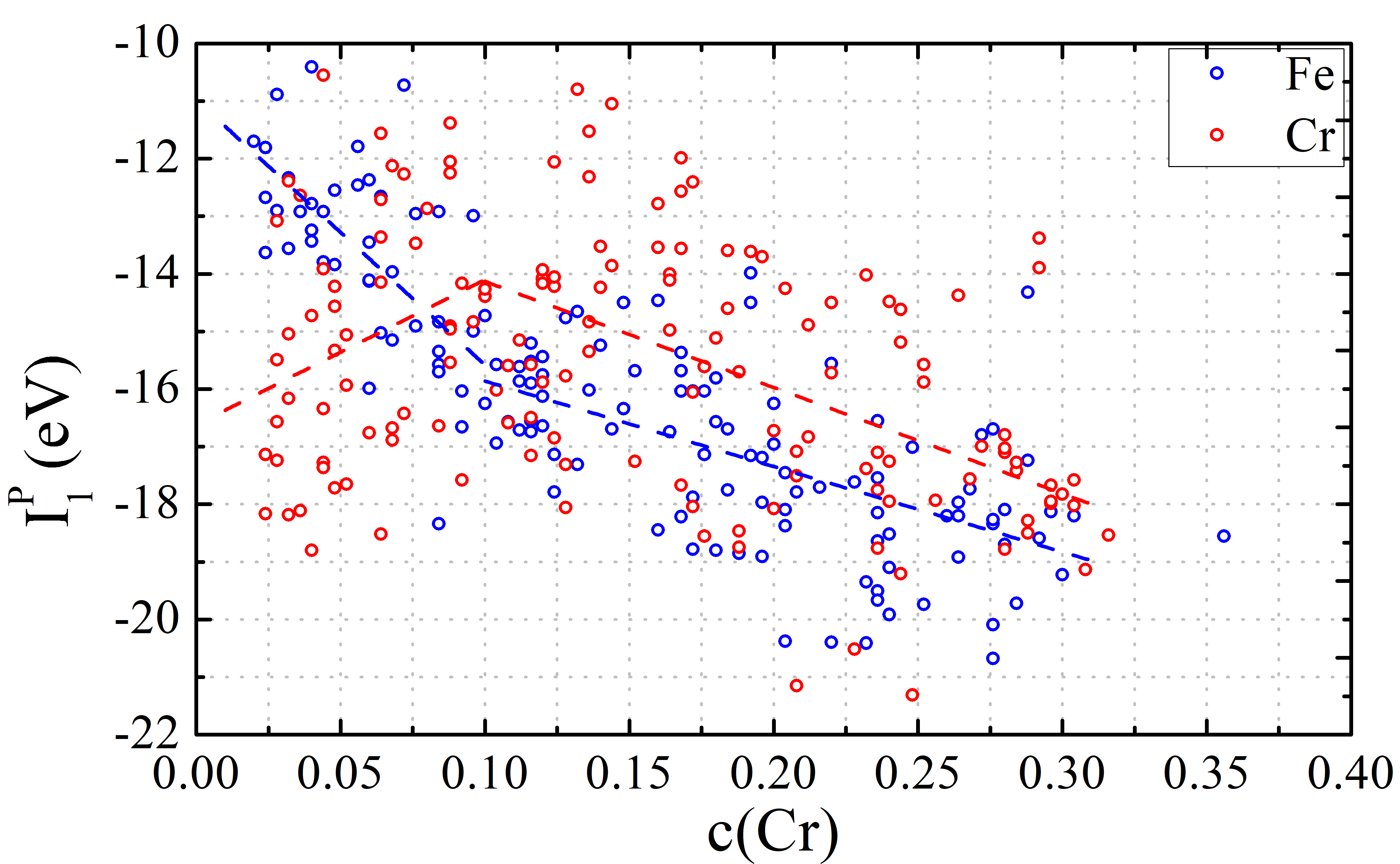}			  	
	\end{minipage}%
	\begin{minipage}{.50\textwidth}
	  	\centering
	  	b)\includegraphics[width=.95\linewidth]{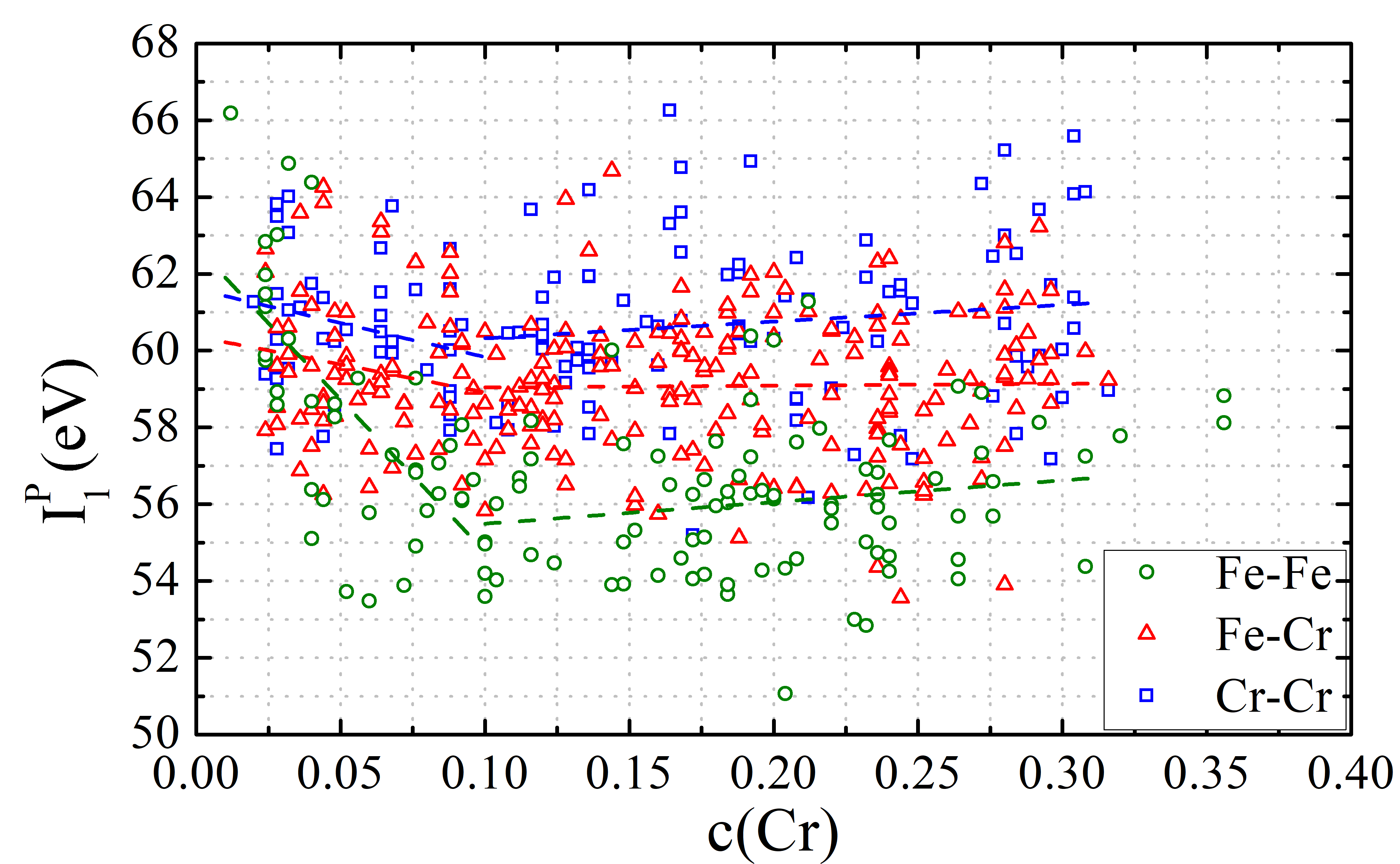}
	\end{minipage}%
	\newline
      \begin{minipage}{.50\textwidth}
	    \centering
	    c)\includegraphics[width=.95\linewidth]{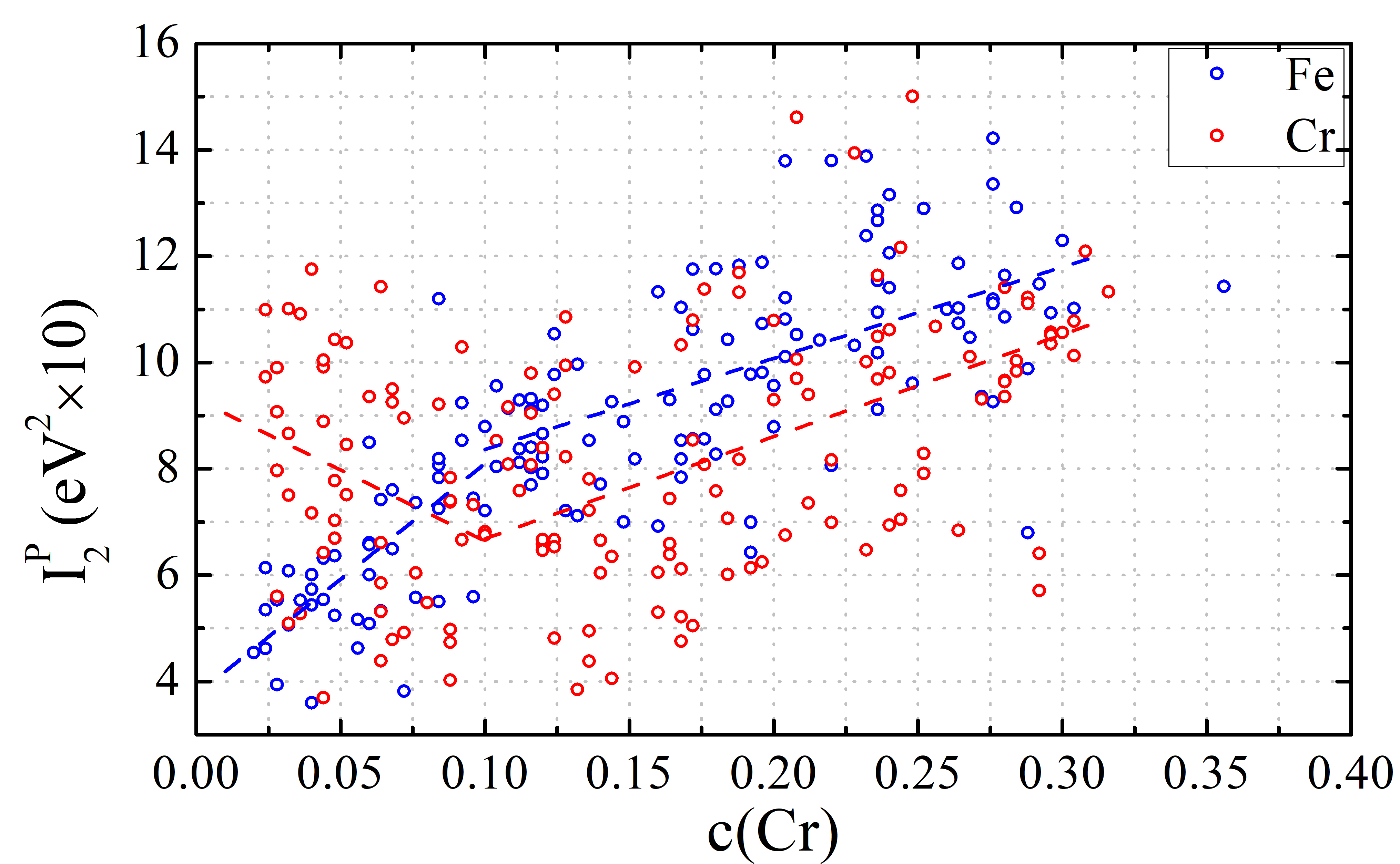}			  	
	\end{minipage}%
	\begin{minipage}{.50\textwidth}
	  	\centering
	  	d)\includegraphics[width=.95\linewidth]{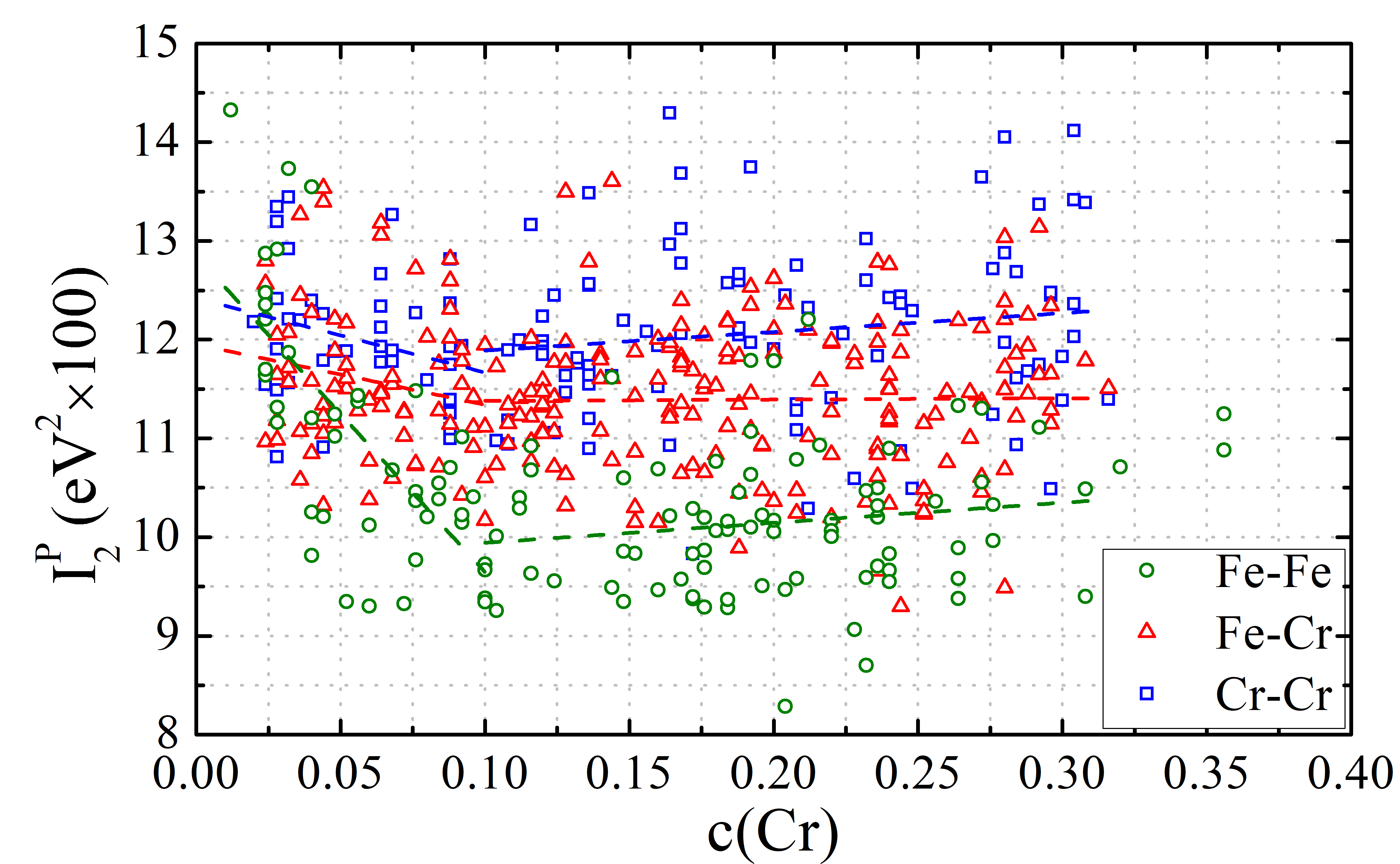}
	\end{minipage}%
	\newline
      \begin{minipage}{.50\textwidth}
	    \centering
	    e)\includegraphics[width=.95\linewidth]{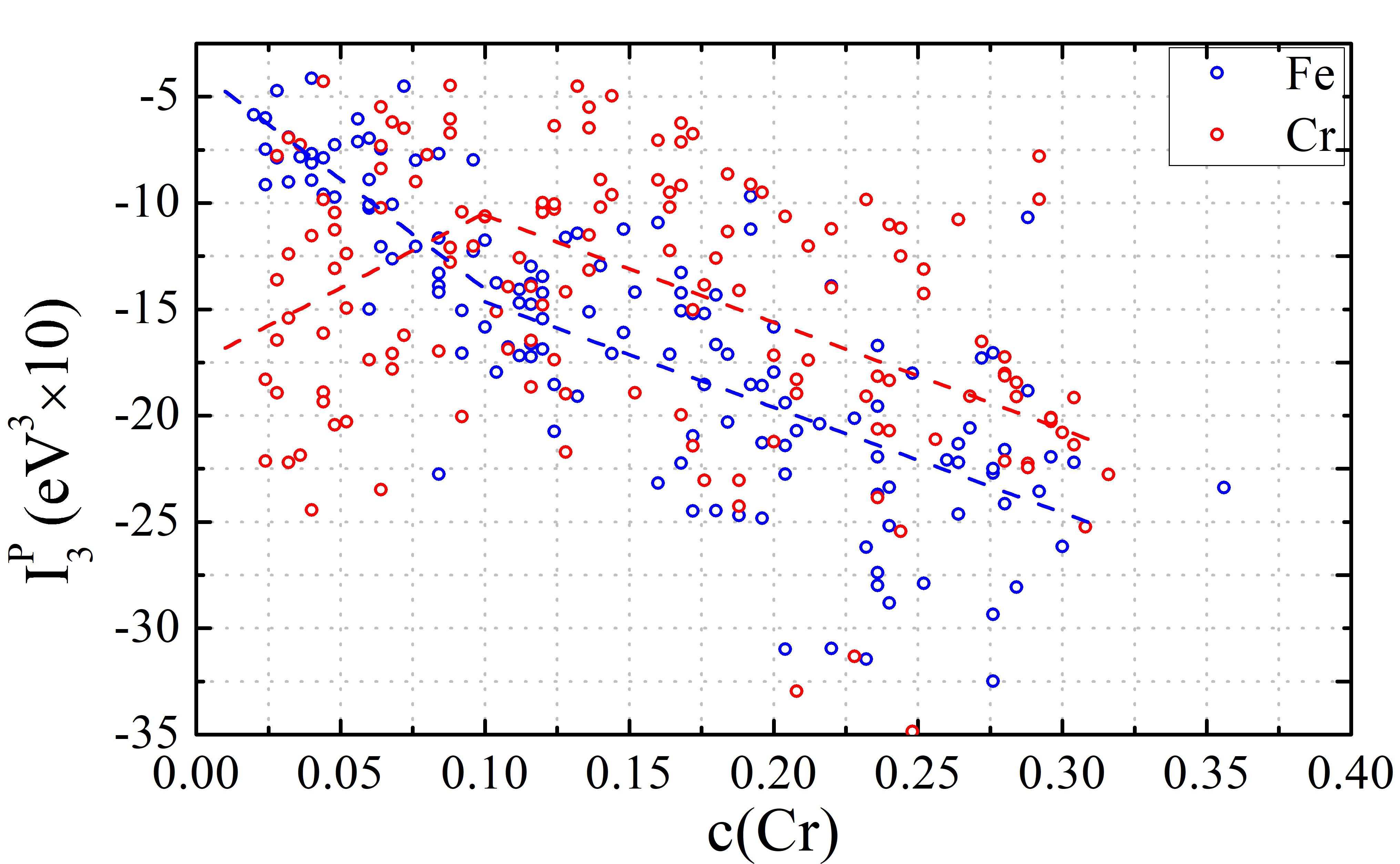}			  	
	\end{minipage}%
	\begin{minipage}{.50\textwidth}
	  	\centering
	  	f)\includegraphics[width=.95\linewidth]{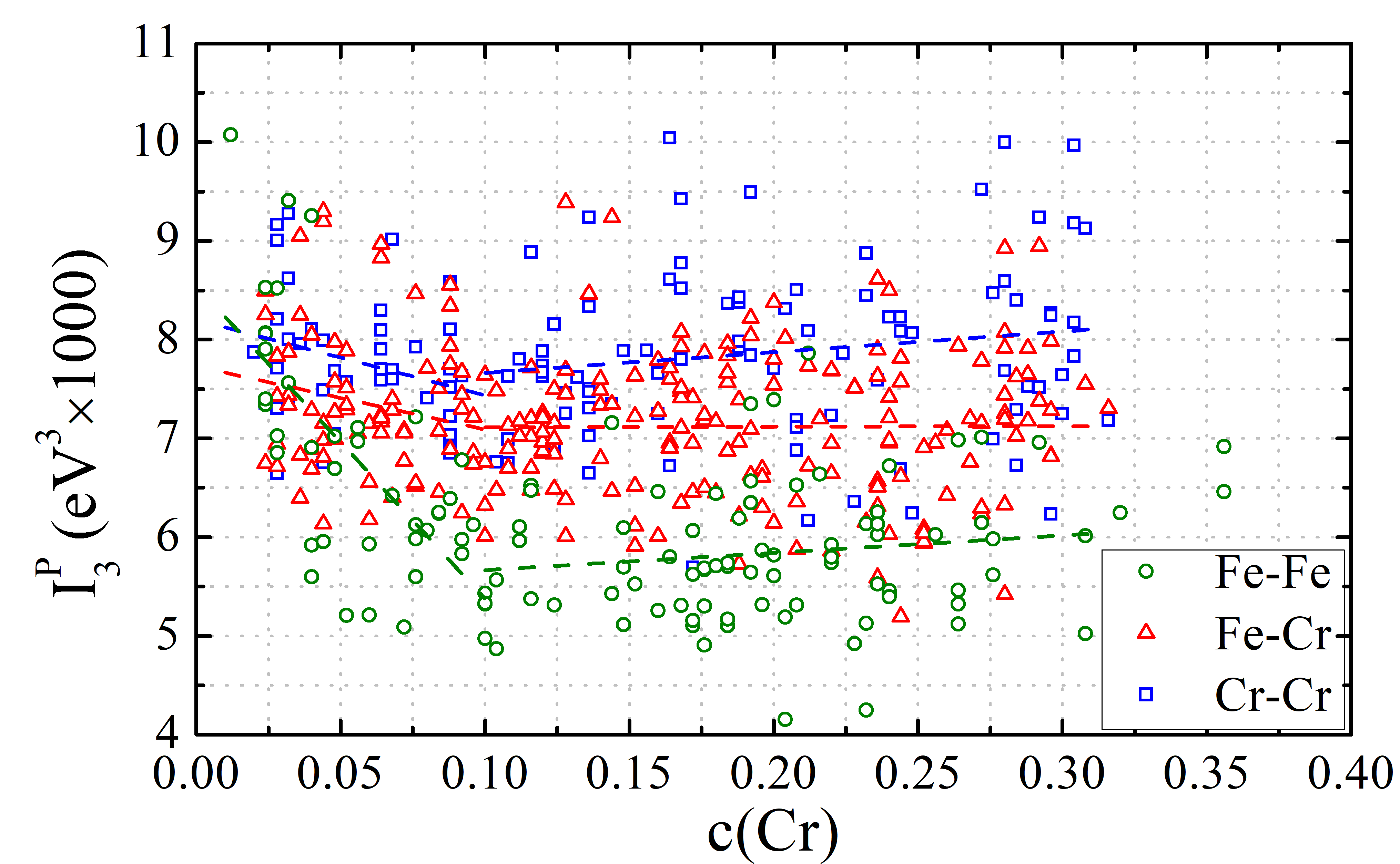}
	\end{minipage}%
\caption{Invariants of elastic dipole tensors (a,b) $I^P_{1}$, (c,d) $I^P_{2}$ (e,f) $I^P_{3}$ computed for (a,c,e) vacancies on Fe and Cr sites and (b,d,f) Fe-Fe, Fe-Cr and Cr-Cr dumbbells in random Fe-Cr alloys.
        \label{fig:tensor_invariants}}
\end{figure*}

Similarly to pure bcc Fe and Cr, the dipole tensors of $\left\langle 110\right\rangle$ Fe-Fe and Fe-Cr dumbbells are characterised by a significantly larger value of $P_{11}$ in comparison with $P_{22}$ and $P_{33}$ over the entire range of alloy compositions considered here. On the other hand, Cr-Cr dumbbells are characterised by notably smaller $P_{11}$ values, and larger $P_{22}$ and $P_{33}$ values, than Fe-Fe and Fe-Cr dumbbells. As a result, Cr-Cr dumbbells have the $P_{11}/P_{22}$ ratio much closer to unity than Fe-Fe and Fe-Cr dumbbells. 

The fact that the values of $P_{11}$, $P_{22}$ and $P_{33}$ for Cr-Cr dumbbells are similar does not mean that the elastic field of these defects is isotropic. The dipole tensor of every dumbbell defect has large off-diagonal terms. The main  difference between Fe-Fe, Fe-Cr and Cr-Cr dumbbells is that the two former ones have only one visibly non-zero off-diagonal $P_{ij}$ element, namely $P_{23}$, and the mean values of $P_{12}$ and $P_{31}$ are close to zero, whereas the latter one often has all the off-diagonal elements that are large. The values of these off-diagonal elements for Cr-Cr dumbbells also fluctuate stronger than those for Fe-Fe and Fe-Cr dumbbells.
This effect may be related to the fact that the direction of a Cr-Cr dumbbell is not necessarily close to $\left\langle 110\right\rangle$ as it is the case for  Fe-Fe and Fe-Cr dumbbells. For example, the most stable Cr-Cr dumbbell in pure Cr is symmetry broken \cite{Ma2019b} and its orientation is close to $\left\langle 11\xi\right\rangle$. Orientations of Cr-Cr dumbbells in random Fe-Cr alloys will be discussed in Section IV.A.

To understand the changes exhibited by $P_{ij}$ as a function of Cr content, we computed the trends shown in Figs. \ref{fig:dipole_tensors_vac} and \ref{fig:dipole_tensors_dumbbells}. For vacancies, $P_{11}$, $P_{22}$ and $P_{33}$ decrease as a function of Cr concentration. At low Cr concentration, these values approximately approach the value observed for a vacancy in pure bcc Fe. Equations for the trend lines are given in Table V in Appendix.

The data ranges for Fe-Fe, Fe-Cr and Cr-Cr dumbbells are divided into two categories: those corresponding to alloy compositions below and above 10\% at. Cr. The trend lines for these two concentration ranges may be significantly different. For example, the mean value of $P_{11}$  for dumbbells in alloys with Cr concentration lower than 10\% at. Cr decreases with Cr content whereas for larger Cr concentrations it increases. In the low Cr concentration limit, the steepest and slightest slopes are observed for the Fe-Fe and Fe-Cr dumbbells, respectively. At a low Cr concentration, $P_{11}$ is close to the value found for these defects formed on a Cr site in bcc Fe matrix. The mean values of $P_{22}$ and $P_{33}$ for Fe-Cr and Cr-Cr dumbbells are almost constant over the range of concentrations studied here, whereas for Fe-Fe, they decrease notably as a function of Cr content up to the Cr concentration close to approx. 10\% at.

\begin{figure*}
\centering
      \begin{minipage}{.50\textwidth}
	    \centering
	    a)\includegraphics[width=.95\linewidth]{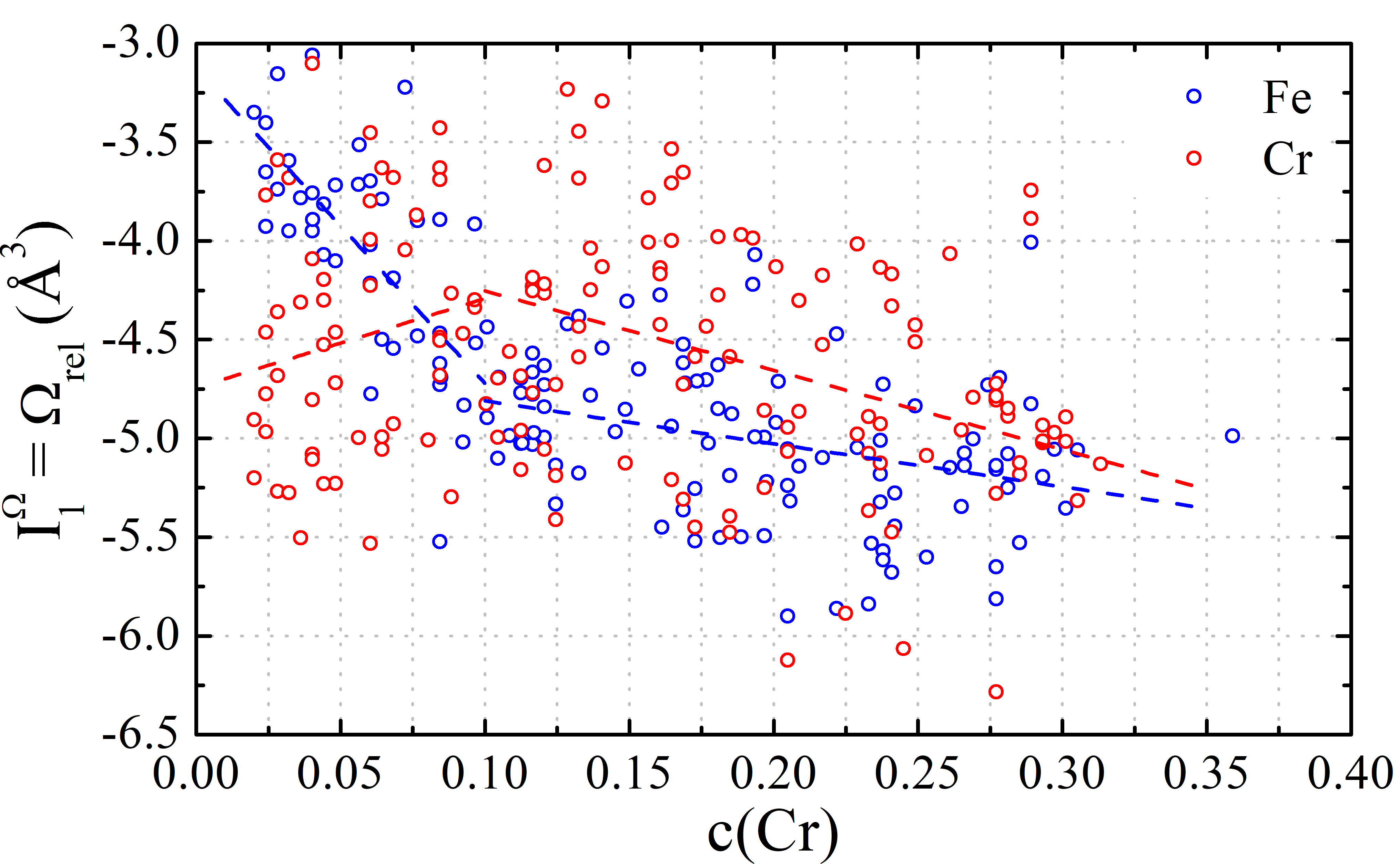}			  	
	\end{minipage}%
	\begin{minipage}{.50\textwidth}
	  	\centering
	  	b)\includegraphics[width=.95\linewidth]{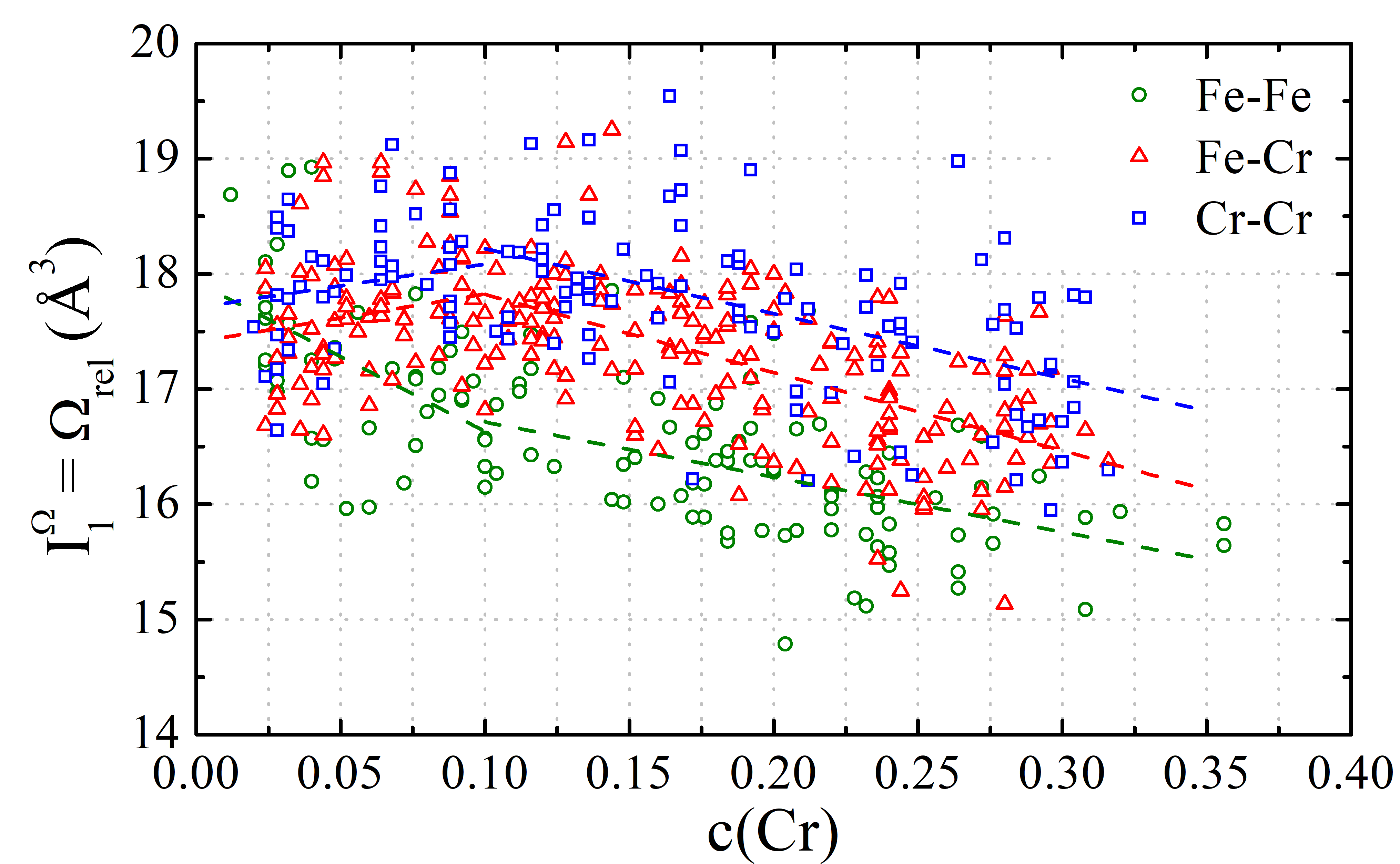}
	\end{minipage}%
	\newline
      \begin{minipage}{.50\textwidth}
	    \centering
	    c)\includegraphics[width=.95\linewidth]{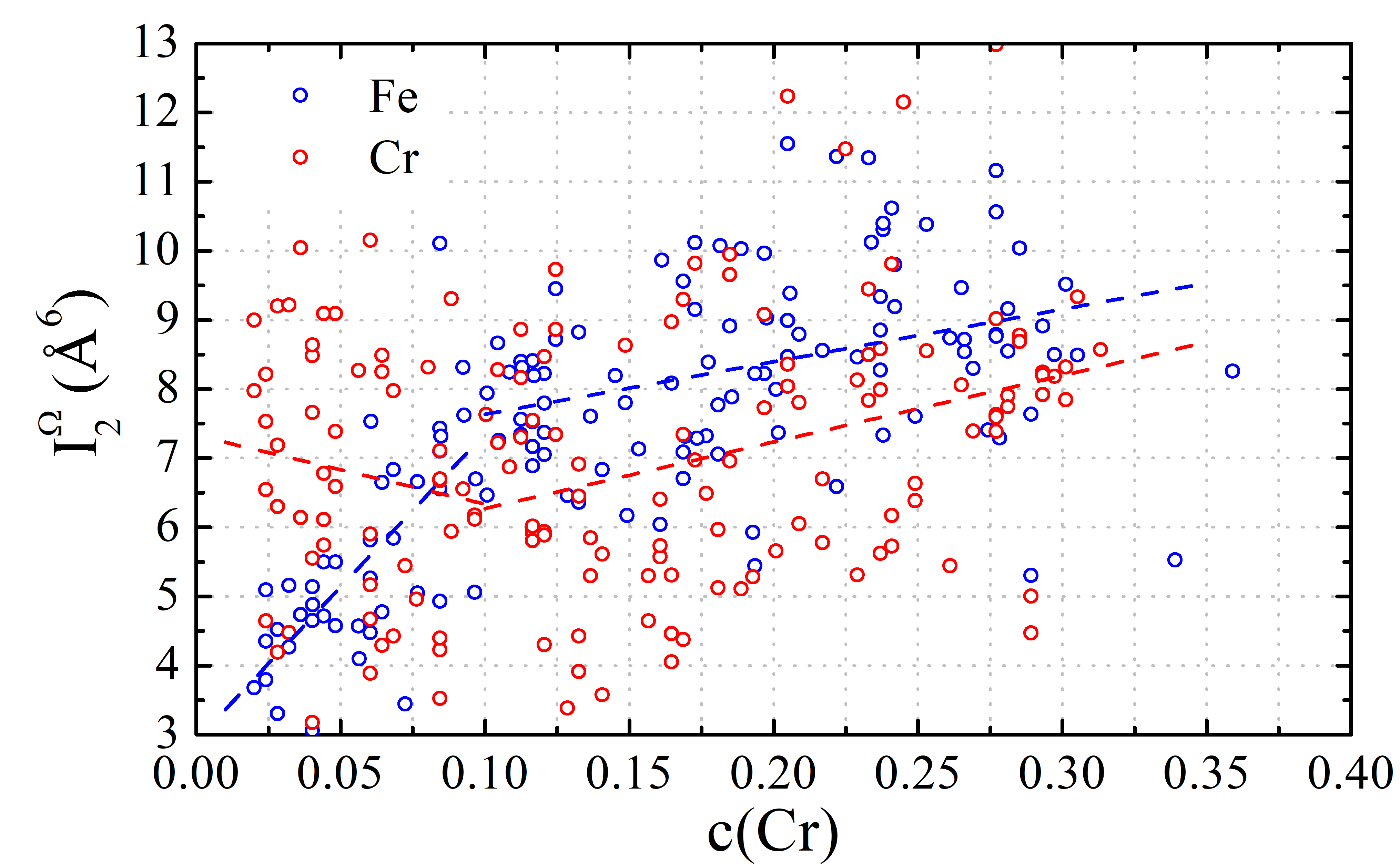}			  	
	\end{minipage}%
	\begin{minipage}{.50\textwidth}
	  	\centering
	  	d)\includegraphics[width=.95\linewidth]{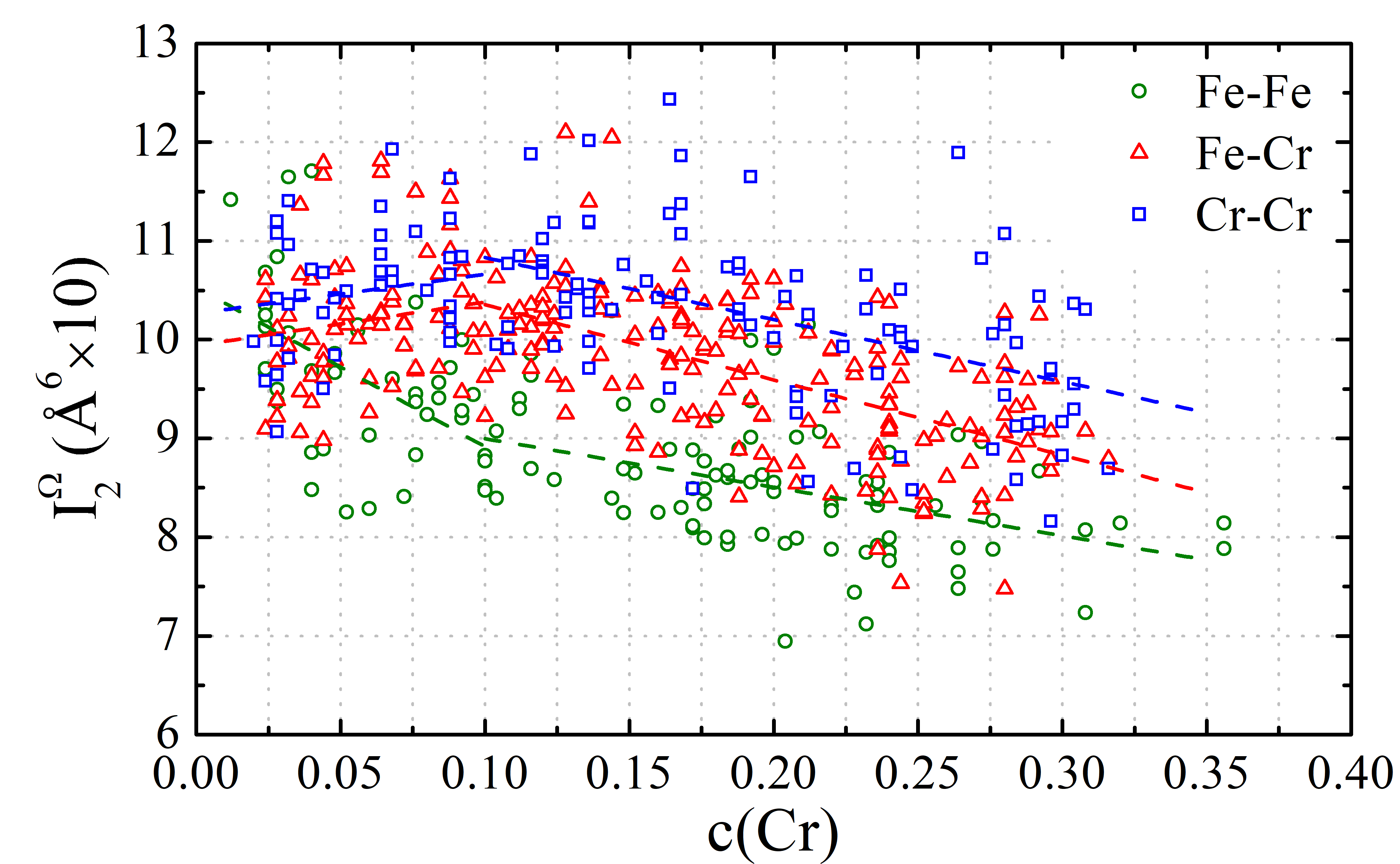}
	\end{minipage}%
	\newline
      \begin{minipage}{.50\textwidth}
	    \centering
	    e)\includegraphics[width=.95\linewidth]{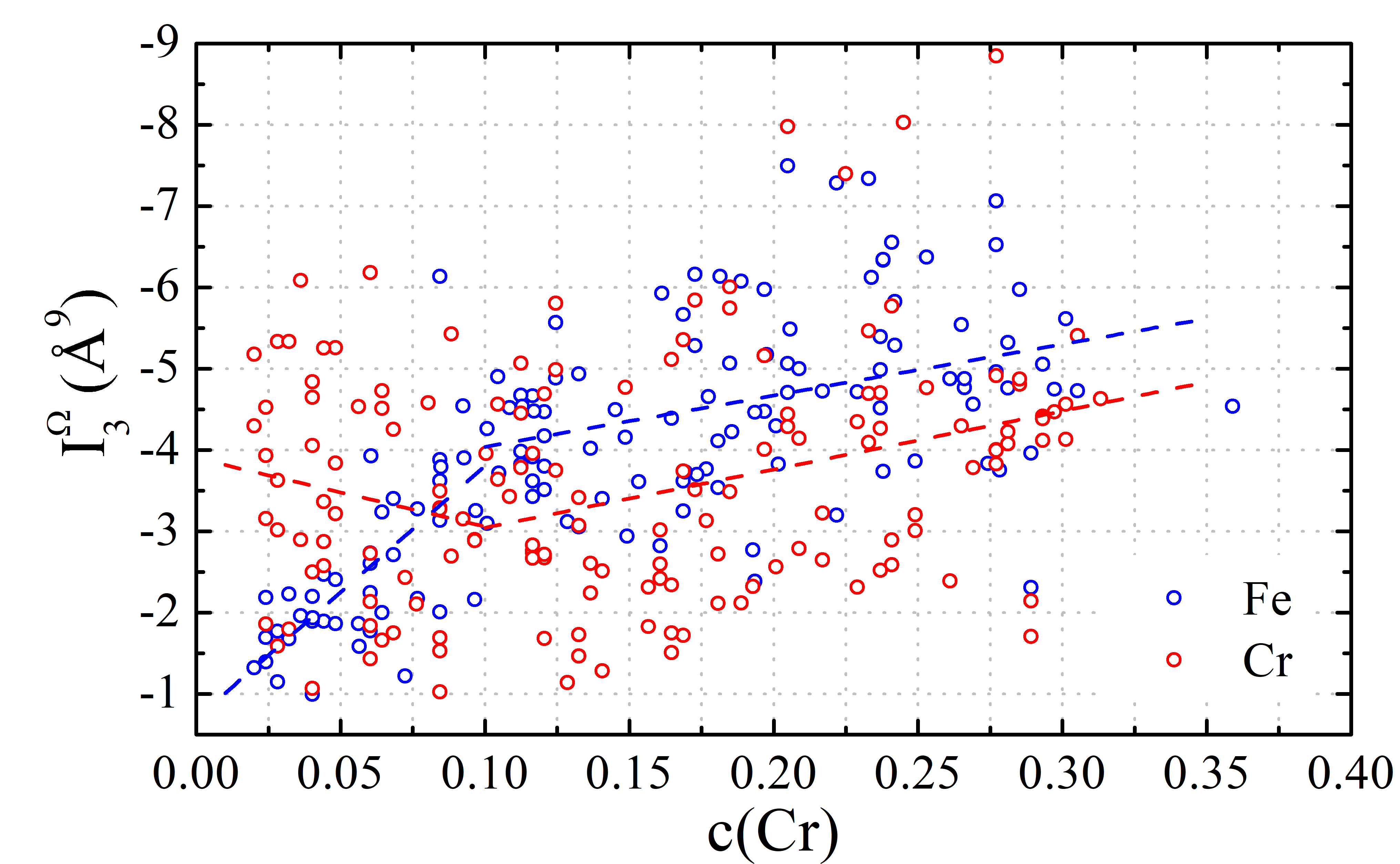}			  	
	\end{minipage}%
	\begin{minipage}{.50\textwidth}
	  	\centering
	  	f)\includegraphics[width=.95\linewidth]{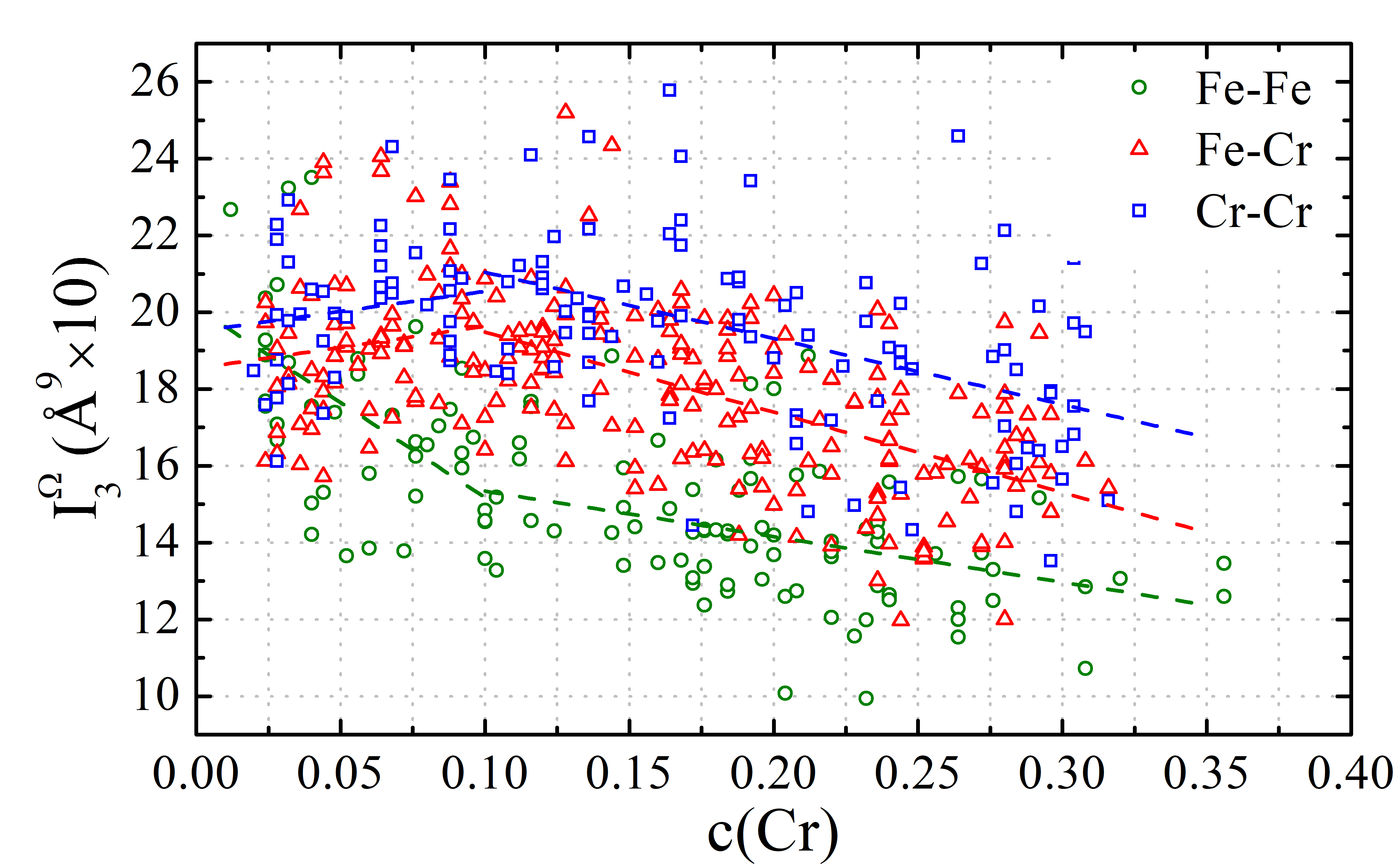}
	\end{minipage}%
\caption{Invariants of relaxation volume tensor: (a,b) $I^{\Omega}_{1}$, (c,d) $I^{\Omega}_{2}$, (e,f) $I^{\Omega}_{3}$ computed for (a,c,e) vacancies on Fe and Cr sites and (b,d,f) Fe-Fe, Fe-Cr, and Cr-Cr dumbbells in random Fe-Cr alloys}\label{fig:omega_tensor_invariants}
\end{figure*}

To characterise elastic dipole and relaxation volume tensors of point defects in Fe-Cr alloys in a way that is independent of rotations of coordinates, we have computed invariants of the two tensors, see Figs. \ref{fig:tensor_invariants} and \ref{fig:omega_tensor_invariants}. Invariants of elastic dipole tensors $I^P_1$, $I^P_2$ and $I^P_3$ are given in the units of eV, eV$^2$ and eV$^3$, whereas the invariants of relaxation volume tensors $I^{\Omega}_1$, $I^{\Omega}_2$ and $I^{\Omega}_3$ are given in \AA$^3$, \AA$^6$ and \AA$^9$. The first invariant of $\Omega _{ij}$, which is the sum of its diagonal elements, is the relaxation volume of the defect. 

For vacancies in Fe-Cr alloys, see Figs. \ref{fig:tensor_invariants}a,c,e, the variation of $I^P_1$ and $I^P_3$ is similar despite the fact that they describe different quantities and are given in different units. Both of them are negative and decrease as a function of Cr content for Cr concentrations above 10\% at. Cr. In both cases, for alloys with Cr concentration below 10\% at., the behaviour of mean values of $I^P_1$ and $I^P_3$ differ depending on whether the vacancy is formed on a Fe or Cr site (see Figs. \ref{fig:Fe-Cr_dumb_vac_scheme}c and \ref{fig:Fe-Cr_dumb_vac_scheme}c). The most rapid decrease of mean values of $I^P_1$ and $I^P_3$ as a function of Cr content is observed for vacancies on a Fe site, whereas the value for a vacancy on a Cr site increases as a function of Cr content. As opposed to $I^P_1$ and $I^P_3$, $I^P_2$ is positive definite, still the variation is similar to that of absolute values of $I^P_1$ and $I^P_3$.

Figs. \ref{fig:tensor_invariants}b,d,f show that all the three invariants of Fe-Fe, Fe-Cr and Cr-Cr dumbbells are positive and exhibit generally similar behaviour despite the fact that they describe different quantities. With the exception of very low Cr alloys, mean values of tensor invariants are the largest for Cr-Cr dumbbells and smallest for Fe-Fe dumbbells. For alloys with Cr concentration above 10\% at. Cr, they increase slightly whereas below 10\% at. they decrease as a function of Cr concentration. The steepest slope is observed for Fe-Fe dumbbells, which have by far the lowest values of $I^P_1$, $I^P_2$ and $I^P_3$ for larger concentrations of Cr whereas for concentration below approx. 3\% at. Cr their values are larger than for Fe-Cr dumbbells. 

Invariants of relaxation volumes tensor ($I^{\Omega}_1$, $I^{\Omega}_2$ and $I^{\Omega}_3$) of vacancies and dumbbells in random Fe-Cr alloys are shown in Fig. \ref{fig:omega_tensor_invariants}. Generally, the behaviour of $I^{\Omega}_1$, $I^{\Omega}_2$ and $I^{\Omega}_3$ is similar to that of invariants of elastic dipole tensors, cf. Figs. \ref{fig:tensor_invariants} and \ref{fig:omega_tensor_invariants}. All the three invariants for dumbbells and $I^{\Omega}_2$ for vacancies are positive whereas the $I^{\Omega}_1$ and $I^{\Omega}_3$ for vacancies are negative. Similarly to the formation energies and elements of relaxation volume tensor, the data points are scattered. Even for similar concentrations, the difference between the smallest and the largest values of $I^{\Omega}_1$ can be up to approx. 2 \AA$^3$, see Figs. \ref{fig:omega_tensor_invariants}a and \ref{fig:omega_tensor_invariants}b.

Relaxation volumes of vacancies in random Fe-Cr alloys are in general more negative than the volume of a vacancy in pure bcc Fe. Even at low Cr concentration the mean value of $\Omega_{rel}$ of a vacancy is equal to -2.4 \AA$^3$ and is approx. 50\% more negative than $\Omega_{rel}$ for a vacancy in bcc Fe. Fig. \ref{fig:omega_tensor_invariants}a shows mean relaxation volumes of vacancies depending on the kind of the atom (Fe or Cr) replaced by the vacancy. The results are noticeable different below and above approx. 10\% at. Cr and therefore the trend lines are described more accurately using two linear fits, one below and another above 10\% at. Cr. The most rapid decrease of the mean relaxation volume as a function of Cr content is observed for vacancies on a Fe site at low Cr concentration, whereas $\Omega_{rel}$ of a vacancy on a Cr site increases as a function of Cr content. For Cr concentrations above 10\% at. Cr, $\Omega_{rel}$ of a vacancy on both sites decreases as a function of Cr content but the slope for a vacancy on a Cr site is steeper. 

Relaxation volumes of dumbbells are all positive, and magnitudes are much larger than those of vacancies (see Fig. \ref{fig:omega_tensor_invariants}b). Results for Fe-Fe, Fe-Cr and Cr-Cr dumbbells are different above and below approx. 10\% at. Cr. For Cr concentrations below 10\% at. Cr, the mean values of $\Omega_{rel}$  for Fe-Cr and Cr-Cr SIAs increase whereas for Fe-Fe decrease rapidly as a function of Cr content. As a result, the mean values of $\Omega_{rel}$ for Fe-Fe dumbbells are the largest at very small Cr concentrations (below approx. 2\% at. Cr) and the lowest for larger Cr concentrations. At a low Cr concentration, the mean values of $\Omega_{rel}$ for the three types of dumbbells are similar to the values computed for pure bcc Fe (results for Fe-Cr dumbbells in Fe-Cr alloys are closer to the values for a dumbbell formed on a Cr site than on a Fe site in bcc Fe matrix). For Cr concentrations above approx. 10\% Cr,  $\Omega_{rel}$ for all the three types of dumbbells decreases as a function of Cr content, which is in agreement with that $\Omega_{rel}$ for these dumbbells in bcc Cr is notably smaller than in bcc Fe. The slopes in each case are similar. Equations for the trend lines describing how the invariants $I^P_1$, $I^P_2$, $I^P_3$, $I^{\Omega}_1 = \Omega_{rel}$ , $I^{\Omega}_2$ and $I^{\Omega}_3$ computed for point defects vary as functions of Cr concentration, are given in Table VI in Appendix.

\begin{figure*}
\centering
      \begin{minipage}{.50\textwidth}
	    \centering
	    a)\includegraphics[width=.95\linewidth]{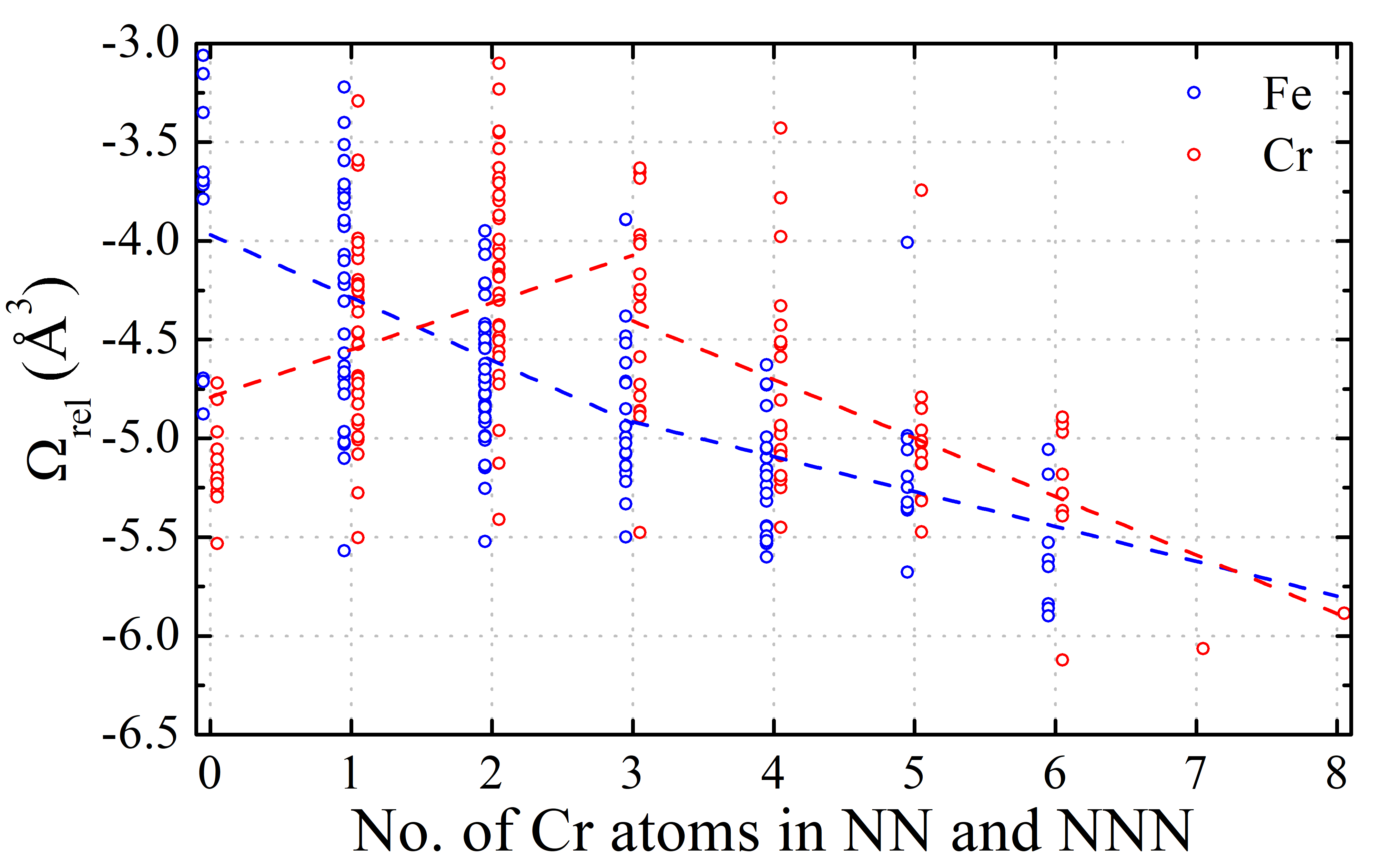}			  	
	\end{minipage}%
	\begin{minipage}{.50\textwidth}
	  	\centering
	  	b)\includegraphics[width=.95\linewidth]{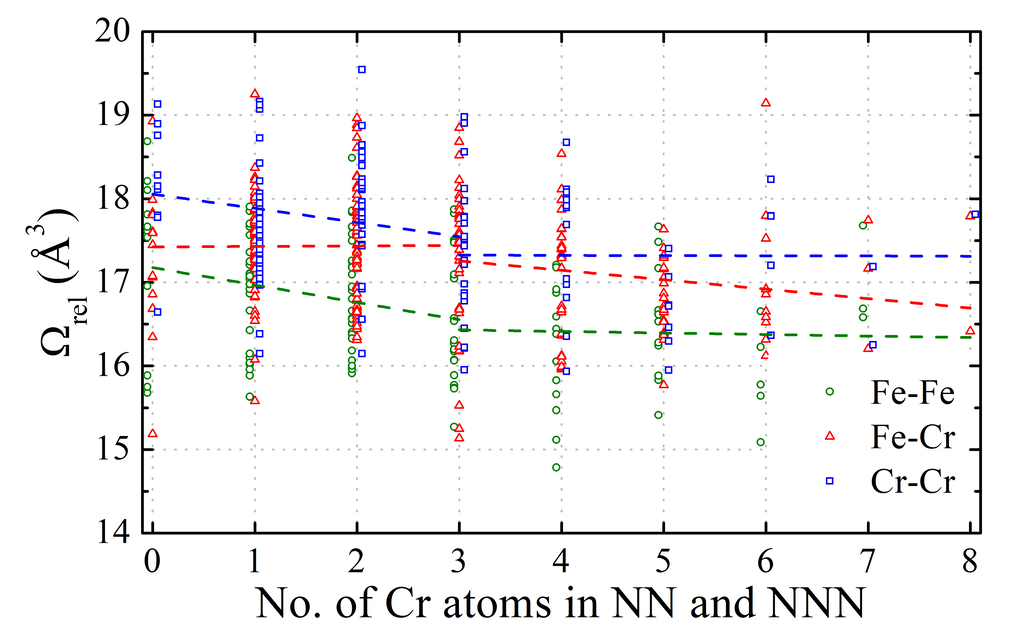}
	\end{minipage}%
	\newline
      \begin{minipage}{.50\textwidth}
	    \centering
	    c)\includegraphics[width=.95\linewidth]{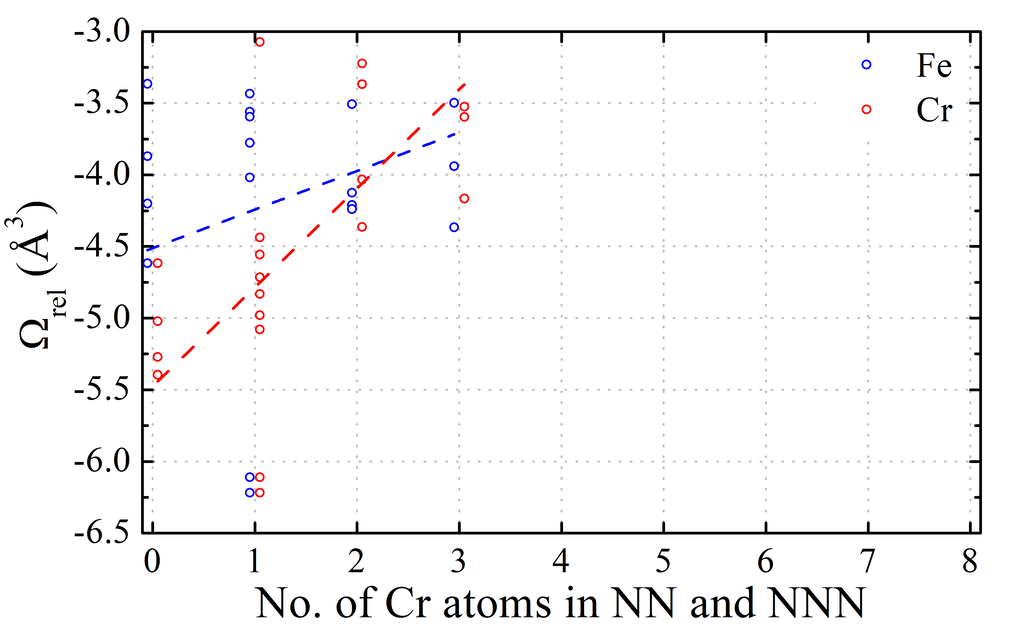}			  	
	\end{minipage}%
	\begin{minipage}{.50\textwidth}
	  	\centering
	  	d)\includegraphics[width=.95\linewidth]{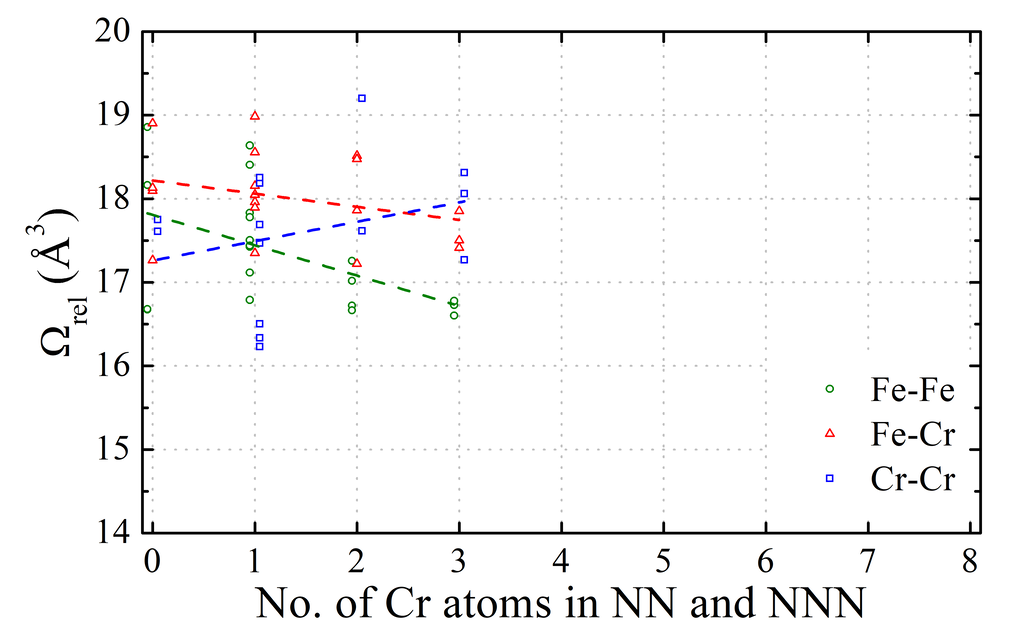}
	\end{minipage}%
\caption{Relaxation volumes of vacancies on Fe and Cr sites (a,c) and dumbbells (b,d) in random Fe-Cr alloys plotted over the entire range of compositions (a,b) and for the alloy containing 5\% at. Cr (c,d), as a function of the total number of Cr atoms in the NN and NNN coordination shells of a defect. Linear trends are indicated by dashed lines.  
        \label{fig:Relax_vol_NN}}
\end{figure*}

Comparison of relaxation volumes of vacancies and dumbbells in random Fe-Cr alloys for the entire range of concentrations and for the alloy with 5\% at. Cr  as a function of number of Cr atoms in the local environment of a defect is given in Fig. \ref{fig:Relax_vol_NN}. As in Fig. \ref{fig:form_en}, point defects in Fe-5\%Cr alloys are surrounded by up to 3 Cr atoms in NN and NNN. In order to be able to compare the results with those obtained for all the other Fe-Cr alloys, the latter ones have been divided into two regions: for point defects with $N_{Cr}^{def}$ smaller and larger than 3.

Results for the Fe-5\%Cr alloy and for all alloys show that the mean relaxation volume of a vacancy on a Fe site is larger than that on a Cr site for $N_{Cr}^{def}$ equal to 0 and 1, and smaller for the $N_{Cr}^{def}$ values 2 and 3. In the region with small number of Cr atoms in the local environment of a defect, $\Omega_{rel}$ on a Cr site decreases with increasing $N_{Cr}^{def}$ for both groups of alloys. However, $\Omega_{rel}$ on a Fe site, averaged over all the alloys, decreases as a function of $N_{Cr}^{def}$ whereas it slightly increases in the Fe-5\%Cr alloy. In the region with $N_{Cr}^{def}$ bigger than 3, $\Omega_{rel}$ decreases with increasing $N_{Cr}^{def}$ for vacancies bot on a Fe site and on a Cr site, which agrees with results presented as a function of Cr concentration, cf. Figs. \ref{fig:Relax_vol_NN}a and \ref{fig:omega_tensor_invariants}a.

The trends describing mean relaxation volumes of Fe-Fe and Fe-Cr dumbbells as functions of $N_{Cr}^{def}$ are generally similar in the Fe-5\%Cr alloy and in all the other alloys, however the mean values obtained for the Fe-5\%Cr alloy are approx. 0.5 \AA$^3$ larger, see Figs. \ref{fig:Relax_vol_NN}a and \ref{fig:Relax_vol_NN}c. Similarly to formation energies, the most notable difference between the groups of alloys is observed for Cr-Cr dumbbells -- the mean values for a Fe-5\%Cr alloy increase whereas those averaged over all the alloys decrease as a function of the number of Cr atoms in NN and NNN around a Cr-Cr dumbbell. The trends for the mean relaxation volumes of Fe-Fe and Cr-Cr dumbbells for $N_{Cr}^{def}$ larger than 3 are almost constant whereas those for Fe-Cr slightly decrease with the number of Cr atoms in the nearest neighbour shells. Equations for the trend lines describing mean relaxation volumes of point defect as functions of $N_{Cr}^{def}$ are given in Table VII of the Appendix.

\begin{figure*}
\centering
    \begin{minipage}{.50\textwidth}
	   \centering
	  a)\includegraphics[width=.95\linewidth]{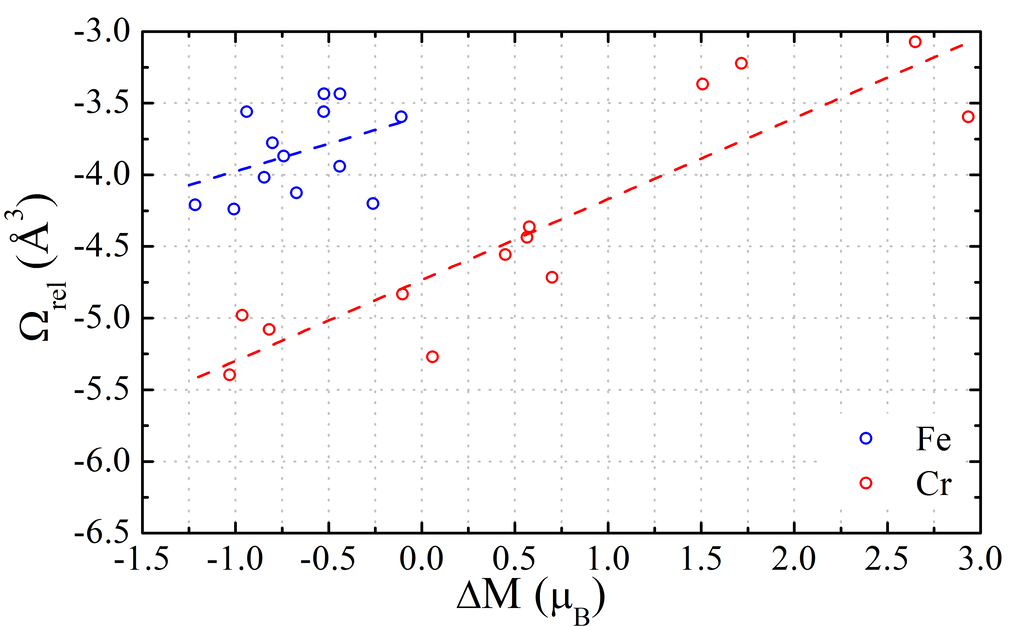}
	\end{minipage}%
	\begin{minipage}{.50\textwidth}
	  	\centering
	  	b)\includegraphics[width=.95\linewidth]{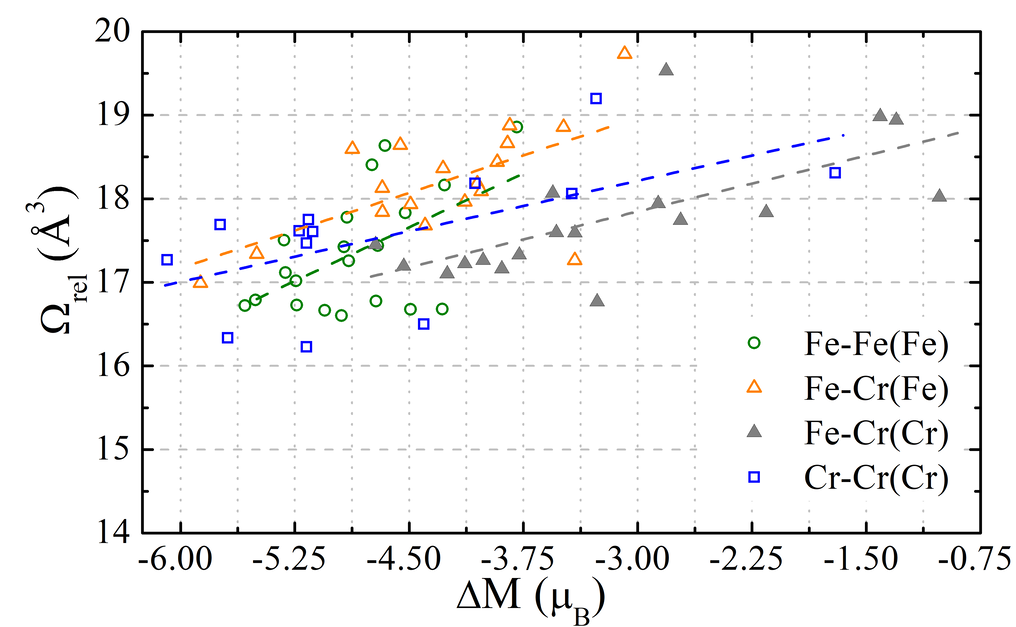}
	\end{minipage}
	\newline
	\begin{minipage}{.50\textwidth}
	  	\centering
	  	c)\includegraphics[width=.95\linewidth]{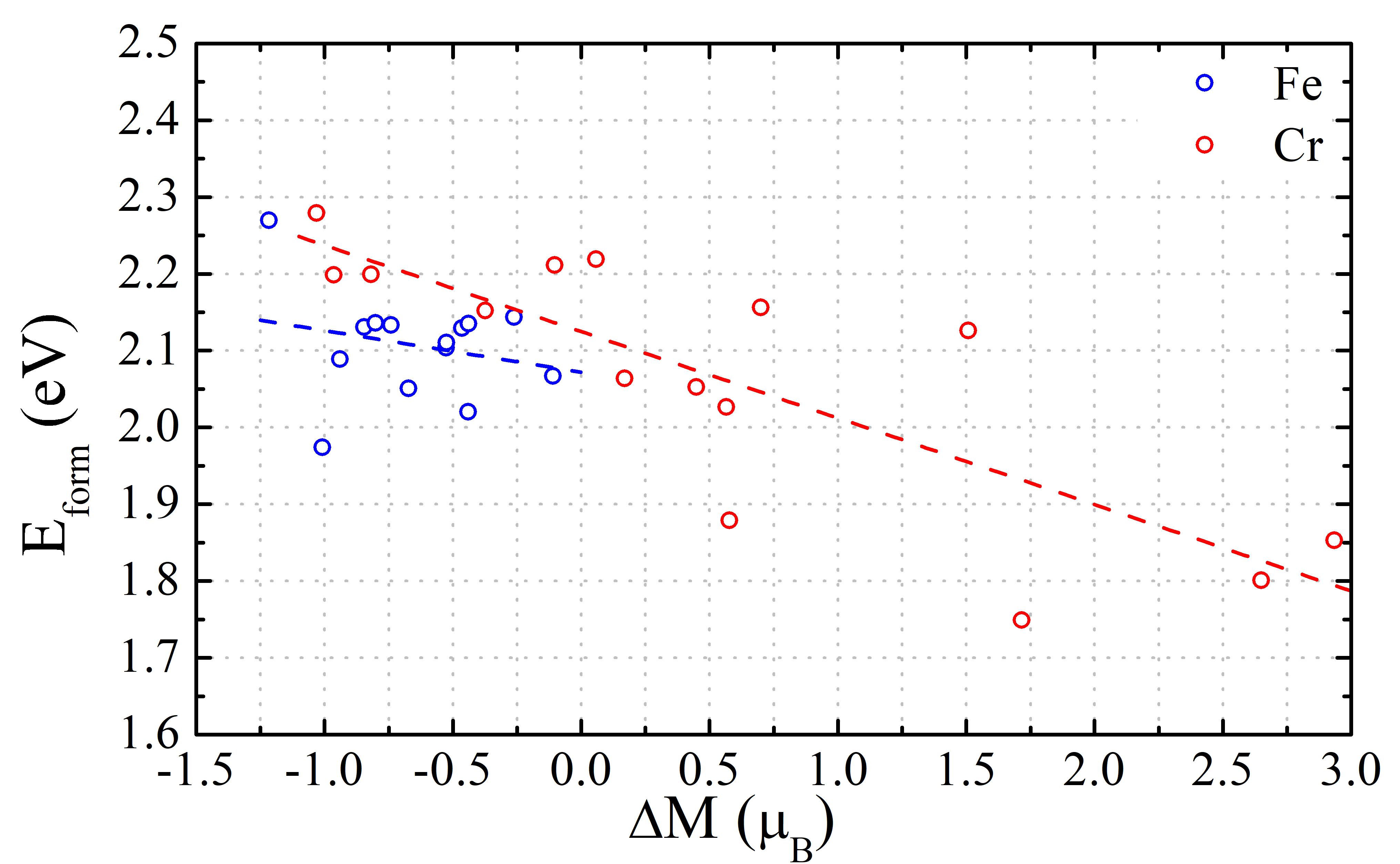}	  	
	\end{minipage}%
	\begin{minipage}{.50\textwidth}
	  	\centering
	  	d)\includegraphics[width=.95\linewidth]{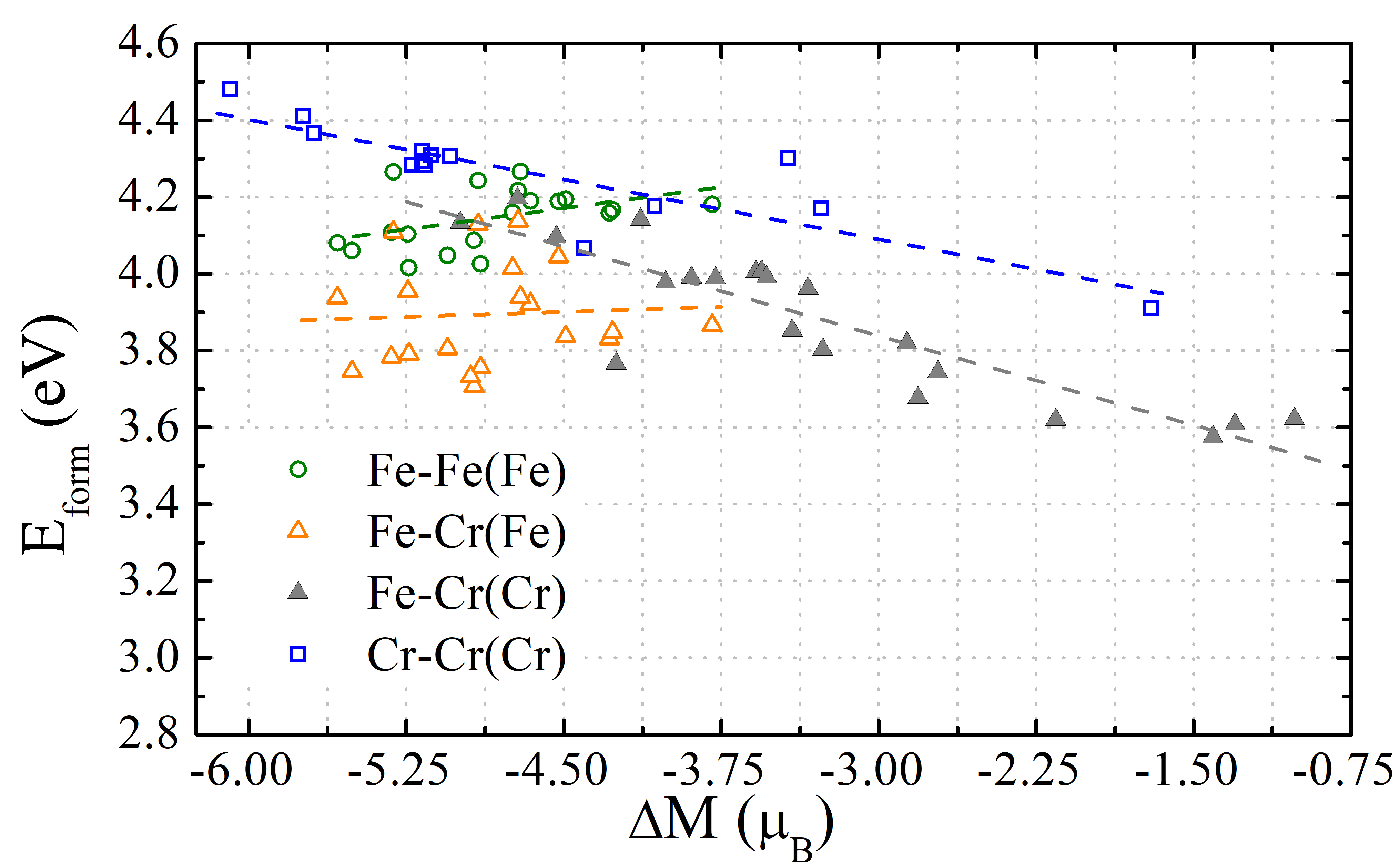}
	\end{minipage}%
	\newline
\caption{Relaxation volumes (a,b) and formation energies (c,d) of vacancies on Fe and Cr sites (a,c) and dumbbells (b,d) in random Fe-Cr alloys for the alloy with 5\% at. Cr as a function of variation of the  magnitude of the total magnetic moment in the supercell, caused by a defect. Linear trends are indicated by dashed lines.  
        \label{fig:Rel_Form_vs_dM}}
\end{figure*}

Similarly to dumbbells in Fe matrix, there is a correlation between the relaxation volume of a defect and the variation of the magnitude of the total magnetic moment in the supercell caused by a defect ($\Delta M$), see Figs. \ref{fig:Rel_Form_vs_dM}a-d. For vacancies and dumbbells, $\Omega_{rel}$ increases as a function of  $\Delta M$. As in bcc Fe matrix, Fe-Cr dumbbells on a Fe site in a Fe-5\%Cr alloy have larger magnitudes of magnetic moments and consequently larger relaxation volumes than Fe-Fe dumbbells on a Fe site. Slopes of trend lines for dumbbells indicate that the largest and smallest variations of $\Omega_{rel}$ with $\Delta M$ are observed for Fe-Fe and Cr-Cr dumbbells, respectively. Slopes of trend lines for vacancies and Fe-Cr do not change significantly depending on the lattice site where a defect is formed. Values of $\Omega_{rel}$ for defects formed on Cr sites are generally smaller.

Variation of magnitudes of magnetic moments associated with a defect also influences the formation energy of a defect. Fig. \ref{fig:Rel_Form_vs_dM}c shows that $E_{form}$ of vacancies decreases with increasing $\Delta M$. Comparing the results presented in Figs. \ref{fig:Rel_Form_vs_dM}a and \ref{fig:Rel_Form_vs_dM}c, we see a correlation between $\Omega_{rel}$ and $E_{form}$ of vacancies, indeed $E_{form}$ decreases as the absolute value of $\Omega_{rel}$ decreases. According to Fig. \ref{fig:Rel_Form_vs_dM}d, values of $E_{form}$ for Fe-Cr and Cr-Cr dumbbells on a Cr site decrease whereas those for Fe-Fe and Fe-Cr dumbbell on a Fe site slightly increase as a function of $\Delta M$. At the same time, a comparison of Figs. \ref{fig:Rel_Form_vs_dM}b and \ref{fig:Rel_Form_vs_dM}d does not show any clear correlation between $\Omega_{rel}$ and $E_{form}$ for dumbbells in Fe-Cr alloys.

\section{Discussion}
\subsection{The orientation of dumbbells}

\begin{figure*}
\centering
      \begin{minipage}{.50\textwidth}
	    \centering
	    a)\includegraphics[width=.95\linewidth]{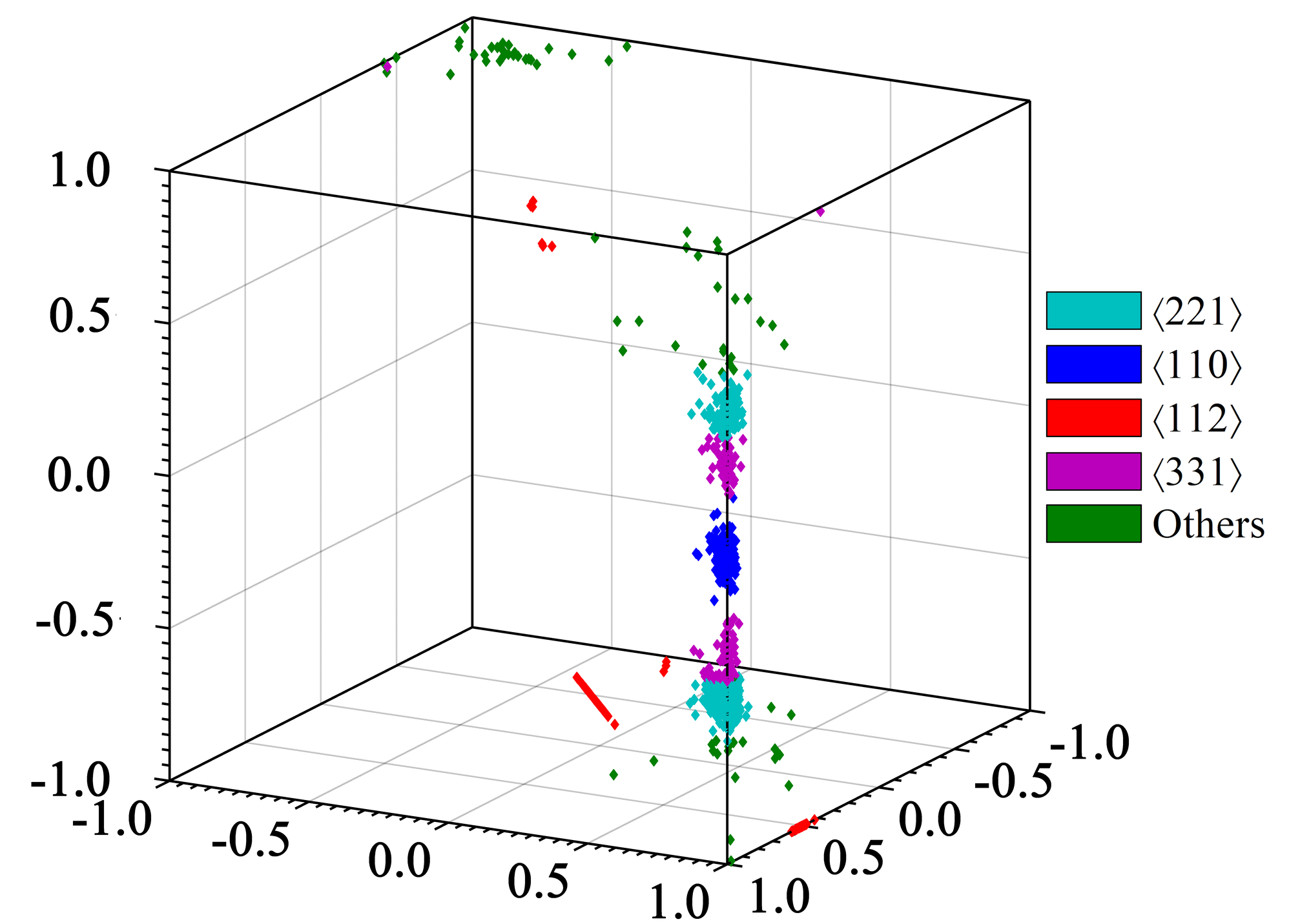}			  	
	\end{minipage}%
	\begin{minipage}{.50\textwidth}
  	    \centering
  	    b)\includegraphics[width=.95\linewidth]{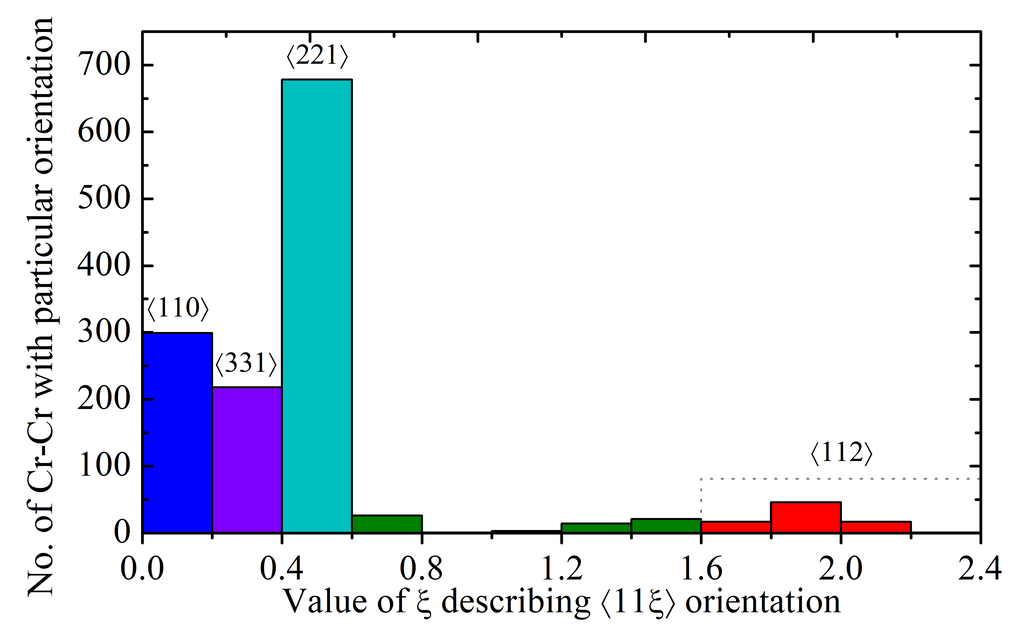}
	\end{minipage}%
    \newline
	\begin{minipage}{1.0\textwidth}
	  	\centering
	  	c)\includegraphics[width=.45\linewidth]{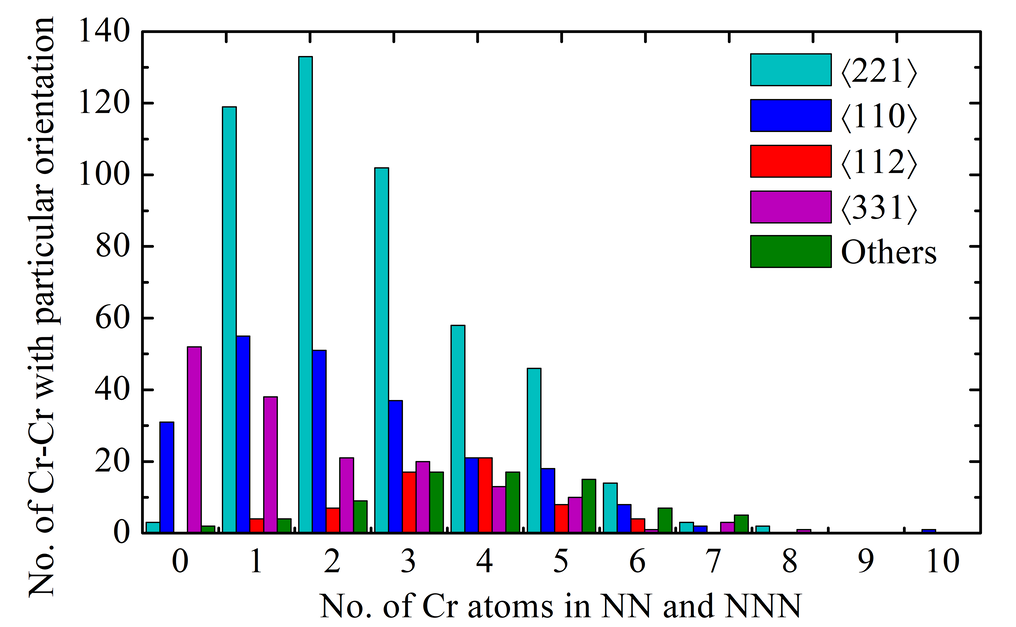}
	\end{minipage}%
	\newline
	\begin{minipage}{.25\textwidth}
	    \centering
	    d)\includegraphics[width=.9\linewidth]{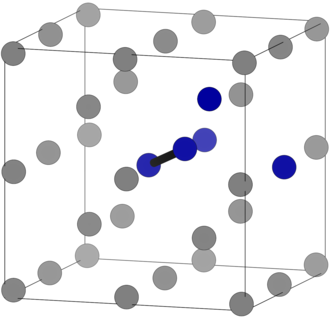}			  	
	\end{minipage}%
	\begin{minipage}{.25\textwidth}
	  	\centering
	  	e)\includegraphics[width=.9\linewidth]{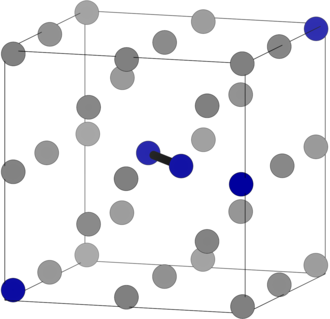}
	\end{minipage}%
	\begin{minipage}{.25\textwidth}
	    \centering
	    f)\includegraphics[width=.9\linewidth]{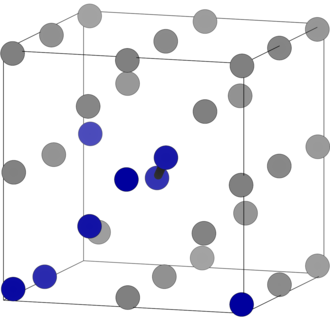}			  	
	\end{minipage}%
	\begin{minipage}{.25\textwidth}
	  	\centering
	  	g)\includegraphics[width=.9\linewidth]{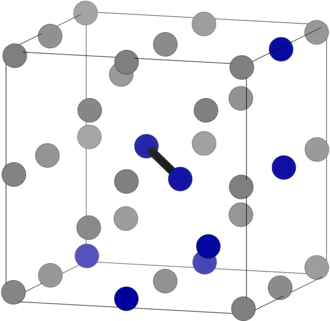}
	\end{minipage}%
\caption{(a) Schematic representation of Cr-Cr dumbbell orientations, (b) the number of $\left\langle 11\xi\right\rangle$ Cr-Cr dumbbells as a function of parameter $\xi$ and (c) the number of Cr-Cr dumbbells in a particular orientation as a function of the number of Cr atoms in the NN and NNN coordination shell of a defect. Examples of alloy configurations in the local environment of a Cr-Cr dumbbell adopting specific orientations: (d) $\left\langle 221\right\rangle$, (e) $\left\langle 110\right\rangle$, (f) $\left\langle 112\right\rangle$, (g) $\left\langle 311\right\rangle$. Fe and Cr atoms are shown by gray and blue spheres, respectively.
        \label{fig:dumbbell_directions}}
\end{figure*}

Orientations of SIA dumbbells, defined by the direction of the vector connecting the two central atoms forming a dumbbell defect, and explored in the calculations, are schematically shown in Fig. \ref{fig:dumbbell_directions}a. Both Fe-Fe and Fe-Cr dumbbells (corresponding to diamond symbols of lime colour) adopt a $\left\langle 110\right\rangle$ orientation after relaxation. This is similar to the orientation of a dumbbell defect in pure bcc Fe, where it adopts a $\left\langle 110\right\rangle$ orientation \cite{Domain2001,Fu2004,NguyenManh2006}. Variation of directions of Cr-Cr dumbbells is much larger, see Fig. \ref{fig:dumbbell_directions}a. In general, most orientations can be classified as an $\left\langle 11\xi\right\rangle$ orientation where $\xi$ spans the interval from 0 to 2.4. We have combined possible orientations of defects into five different groups, corresponding to different intervals of parameter $\xi$, namely $\left\langle 110\right\rangle$ ($0<\xi<0.2$), $\left\langle 331\right\rangle$ ($0.2<\xi<0.4$), $\left\langle 221\right\rangle$ ($0.4<\xi<0.6$), $\left\langle 112\right\rangle$ ($1.6<\xi<2.4$) orientations as well as others, see Fig. \ref{fig:dumbbell_directions}b. The number of dumbbells adopting a particular orientation as a function of the number of Cr atoms in the 1st and 2nd coordination shells around a defect is shown in Fig. \ref{fig:dumbbell_directions}c. Examples of alloy configurations in the local environment of a Cr-Cr dumbbell adopting a particular orientation are shown in Figs. \ref{fig:dumbbell_directions}d-g. The most common direction of a Cr-Cr dumbbell is $\left\langle 221\right\rangle$ (about 48.0\% of all Cr-Cr dumbbells), however this fraction decreases as the number of Cr atoms in the local environment of a dumbbell increases. The prevalence of the $\left\langle 221\right\rangle$ direction (indicated by the aquamarine colour in Fig. \ref{fig:dumbbell_directions}b) of Cr-Cr agrees with the earlier results by Klaver {\it et al.} \cite{Klaver2007}. For the configurations containing no Cr atoms in the 1st and 2nd coordination shells around a self-interstitial defect, the $\left\langle 331\right\rangle$ (purple) and $\left\langle 110\right\rangle$ (navy blue) orientations are more common (for example, $\left\langle 331\right\rangle$ and $\left\langle 110\right\rangle$ orientations represent 59.1\% and 35.2\% of all the directions of dumbbells that have no Cr atoms in their vicinity).  The occurrence of dumbbells with orientations $\left\langle 112\right\rangle$ (indicated by the red colour in Fig. \ref{fig:dumbbell_directions}b) as well as with orientations with higher crystallographic indices, the so-called ‘others’ (green), increases with the number of Cr atoms in the local environment of a defect.

\subsection{Comparison of the cell relaxation and stress methods}

\begin{figure*}
\centering
      \begin{minipage}{.50\textwidth}
	    \centering
	    a)\includegraphics[width=.95\linewidth]{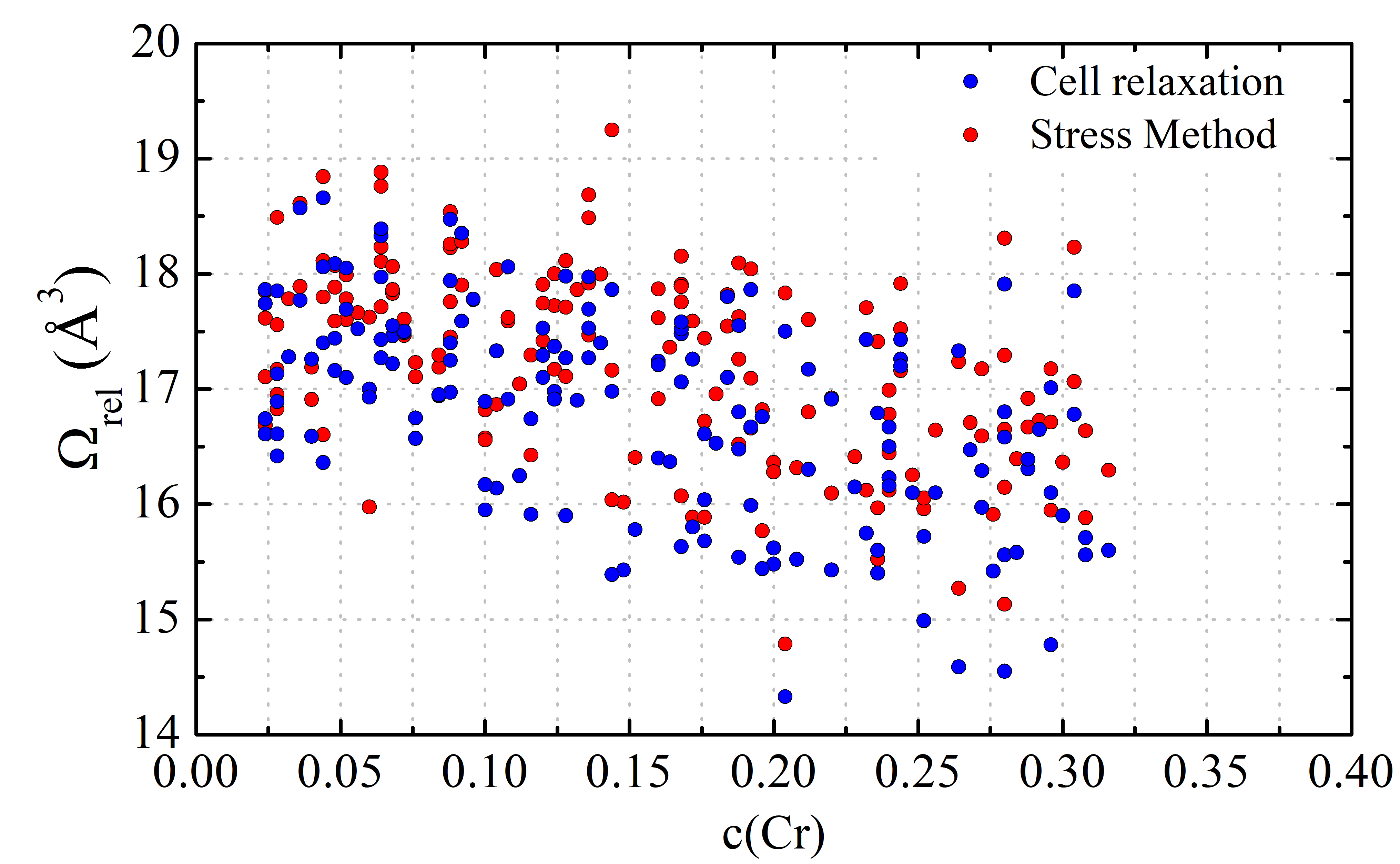}			  	
	\end{minipage}%
	\begin{minipage}{.50\textwidth}
	  	\centering
	  	b)\includegraphics[width=.95\linewidth]{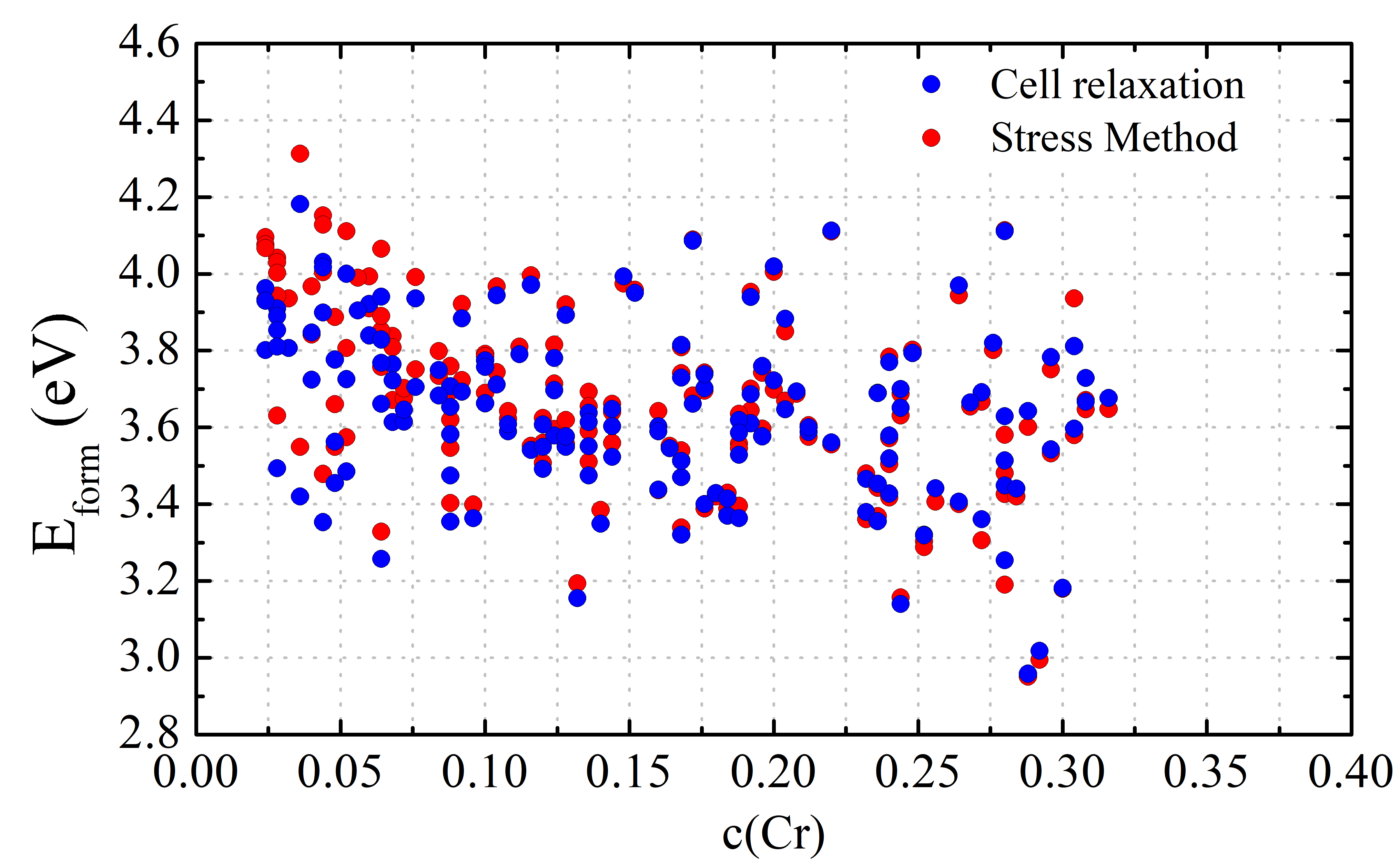}
	\end{minipage}%
	\newline
      \begin{minipage}{.50\textwidth}
	    \centering
	    c)\includegraphics[width=.95\linewidth]{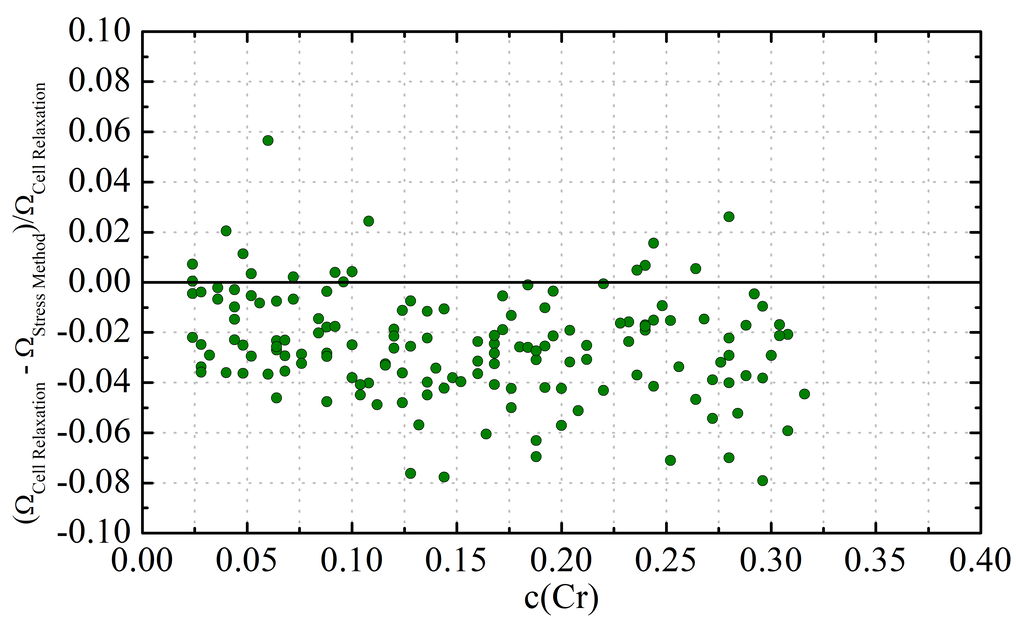}
	\end{minipage}%
	\begin{minipage}{.50\textwidth}
	  	\centering
	  	d)\includegraphics[width=.95\linewidth]{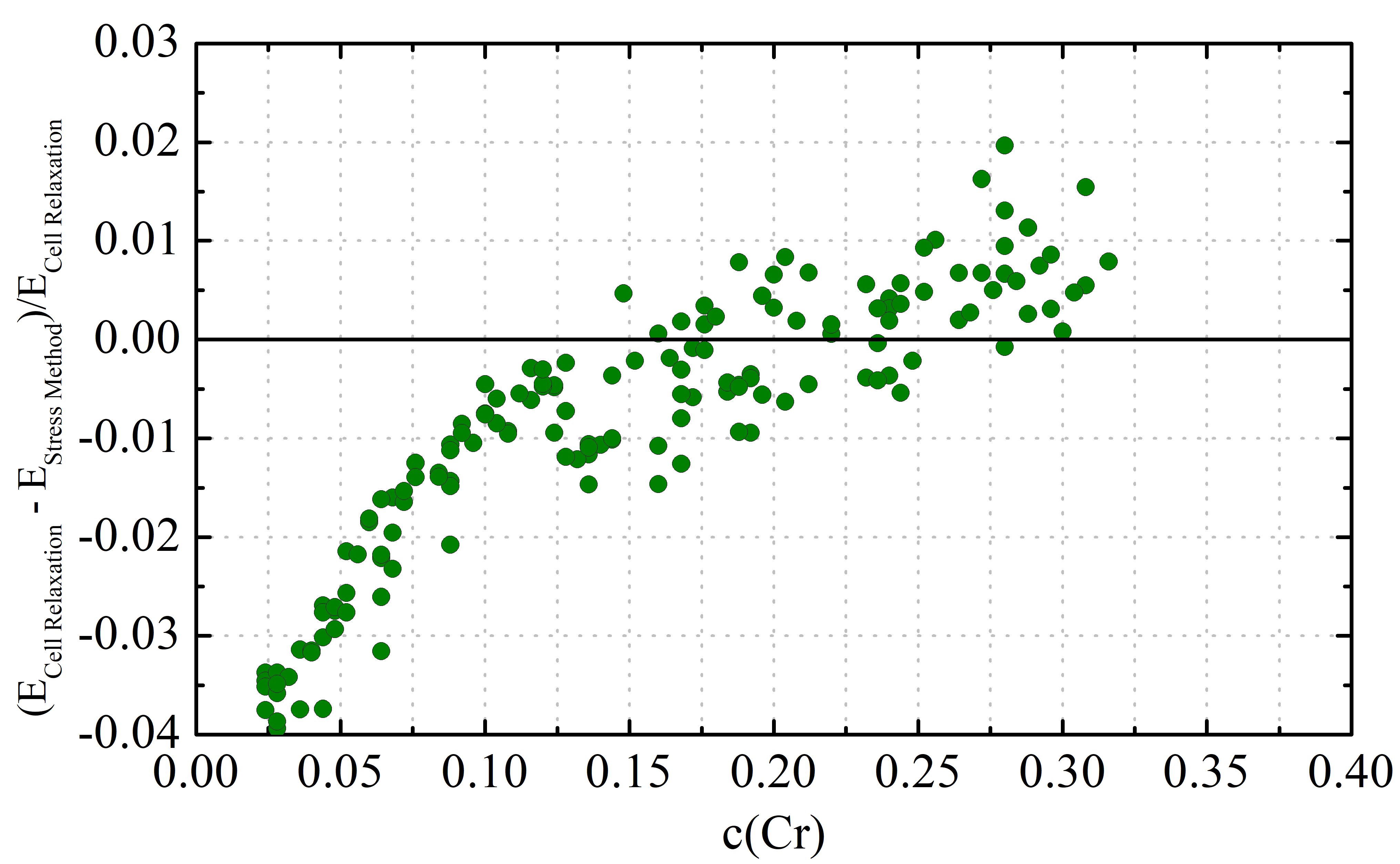}
	\end{minipage}%
\caption{Comparison of (a) relaxation volumes and (b) formation energies of SIA or Fe-Fe dumbbells evaluated using the stress and cell relaxation methods, and the relative difference between (c) relaxation volumes and (d) formation energies computed using both methods.
        \label{fig:Cell_relax_vs_stress_method}}
\end{figure*}

A comparison of results obtained using the stress method and full cell relaxation for 160 random Fe-Cr structures is shown in Fig. \ref{fig:Cell_relax_vs_stress_method}. Both approaches show that relaxation volumes and formation energies of dumbbells in random Fe-Cr alloys decrease with Cr content, see Fig. \ref{fig:Cell_relax_vs_stress_method}a and \ref{fig:Cell_relax_vs_stress_method}b. The stress method predicts somewhat larger values of relaxation volumes and formation energies than the cell relaxation method, exhibiting a correlation between $E_{form}$ and $\Omega_{rel}$. The relaxation volumes of defects computed using the stress method are on average 2.5\% larger, however there are a few structures where they are more than 5\% larger than the values derived using the cell relaxation method, see Fig. \ref{fig:Cell_relax_vs_stress_method}c.  Fig. \ref{fig:Cell_relax_vs_stress_method}d shows that the relative difference between formation energies of defects $E_{form}$ deduced using the stress and cell relaxation methods varies as a function of Cr content. Similarly to the majority of results given in sections above, the relative formation energy difference exhibits different behaviour in the two composition intervals, above and below 10\% at. Cr. Above 10\% at. Cr, the relative formation energy difference increases slowly as a function of Cr content. Values $E_{form}$ obtained using the stress method do not differ in general by more than 2\% in comparison with values computed using the full relaxation method, and the average relative formation energy difference in that interval of Cr concentrations is almost equal to zero. For Cr concentrations below 10\%, the overestimation of $E_{form}$ obtained using the stress method in caparison with that computed using the cell relaxation method increases towards low Cr content, reaching approximately $-4$\% for alloys containing approximately 3\% at. Cr.
It is worth noting that the elastic correction, implemented following Refs. \cite{Varvenne2013,Varvenne2017,Clouet2018,Dudarev2018,Ma2019a}, improves agreement between the results obtained using both methods.  Still, the use of elastic correction often proves insufficient, as it was found for defect clusters in Tungsten \cite{Hofmann2015,Mason2019}.

There are several reasons that might be responsible for the discrepancy. The lattice parameter used in the fixed volume calculations may influence the predicted relaxation volumes derived from the stress method. Here the calculations were performed using the lattice parameter of 2.831 \AA, whereas random Fe-Cr alloy structures can adopt the equilibrium lattice parameters up to 2.842 \AA. Also, the computed relaxation volumes may differ depending on the elastic constants used in the calculations. As was noted previously, the average elastic constants $\bar{C}_{11}$, $\bar{C}_{12}$, $\bar{C}_{44}$ are the interpolations derived from DFT calculations. Furthermore, slightly different convergence parameters were used in the calculations performed using the stress and cell relaxation methods. For example, the plane-wave energy cut-off for the fixed-volume calculations was 300 eV whereas for those with full cell relaxation was 400 eV.

Finally, we note that the values computed using the fixed cell volume method (the stress method) do not take into account non-elastic (non-harmonic) effects, which are implicitly included in the results obtained using the cell relaxation method. From the comparison of values of $E_{form}$ obtained using the stress method and the cell relaxation method, shown in Fig. \ref{fig:Cell_relax_vs_stress_method}d, it is reasonable to expect that the non-harmonic effects would play a particularly significant part in magnetic Fe-Cr alloys with low Cr concentration.

\section{Conclusions}

Concluding this study, we would like to highlight the clear benefits of an approach combining {\it ab initio} treatment of defects with auxilliary analysis based on elasticity. This has enabled quantifying the elastic effects of expansion and contraction of the lattice due to the fact that the atoms forming the alloy have different volumes. For example, we found that the volume of a substitutional Cr atom in bcc Fe lattice is approximately 18\% larger than the volume of a host Fe atom. At the same time,  the volume of a substitutional Fe atom in bcc Cr is 5\% smaller than the volume of a host Cr atom. We also found that elastic dipole and relaxation volume tensors of vacancies and self-interstitial atom (SIA) defects exhibit large fluctuations, with vacancies showing negative and SIA large positive relaxation volumes. Dipole tensors of vacancies are nearly isotropic across the entire alloy composition range. Fe-Fe and Fe-Cr self-interstitial atom dumbbells are more anisotropic than Cr-Cr dumbbells. Fluctuations of elastic dipole tensors of SIA defects are primarily associated with the variable orientation of defects. Statistical properties of tensors elastic dipole and relaxation volume tensors are analysed using their principal invariants, showing that properties of point defects differ significantly in alloys containing below and above 10\% at. Cr. The relaxation volume of a vacancy depends sensitively on whether it occupies a Fe or a Cr lattice site. The observed correlation between the elastic relaxation volumes and magnetic moments of defects suggests that magnetism is a significant factor influencing elastic fields of defects in Fe-Cr alloys.
These results also illustrate the significance of elastic relaxation effects in Fe-Cr alloys in the context of treatment of extended defects such as dislocation or grain boundaries, where elastic relaxation may affect segregation and diffusion of solute atoms in the alloy.

\section{Acknowledgements}

This work has been carried out within the framework of the EUROfusion Consortium and has received funding from the Euratom research and training programme 2014-2018 and 2019-2020 under Grant Agreement No. 633053, and by the M4F project under Grant Agreement No. 755039. The views and opinions expressed herein do not necessarily reflect those of the European Commission. We acknowledge funding by the RCUK Energy Programme (Grant No. EP/T012250/1). 
The work at WUT has been carried out as a part of an international project co-financed from the funds of the program of the Polish Minister of Science and Higher Education entitled "PMW" in 2019; Agreement No. 5018 / H2020-Euratom / 2019/2.
J.S.W. and D.N.M. acknowledge the support from high-performing computing facility MARCONI (Bologna, Italy) provided by EUROfusion. The simulations were also carried out with the support of the Poznan Supercomputing and Networking Center PCSS under grant No. 274. For the database of random Fe-Cr alloys, we acknowledge the use of resources on the EDF HPC Athos, Porthos and Gaia supercomputers, representing about 50 million CPU core hours.
We would like to thank P.-W. Ma, D. R. Mason and M.-C. Marinica for stimulating discussions.

\appendix*\section{\label{append}}

Tables \ref{tab:tab_eform}, \ref{tab:tab_elasticprop}, \ref{tab:tab_elasdip}, \ref{tab:tab_invariants} and \ref{tab:tab_relaxvol} contain equations for the trend lines that were shown in Figures in the main text.

\begin{table*}[]
\centering
\caption{Equations for the trend lines of chemical potentials ($\mu_{Fe/Cr}$), formation energies ($E_{form}$)of vacancies and dumbbells in random Fe-Cr alloys and formation energies as a function of change of magnitudes of magnetic moments ($\Delta{M}$) caused by a defect for the alloy with 5\% at. Cr.
\label{tab:tab_eform}}
\begin{ruledtabular}
\begin{tabular}{cccc}
{Parameter} & {\begin{tabular}[c]{@{}c@{}}Variation as a function of the variable\\in square brackets, denoted here as $x$\end{tabular}} & {Parameter} & {\begin{tabular}[c]{@{}c@{}}Variation as a function of the variable\\in square brackets, denoted here as $x$\end{tabular}} \\
\hline
\multicolumn{4}{c}{Chemical potentials} \\
{$\mu_{Fe}[c(Cr)]$} & {$y=-0.0088x-8.3193$} & $\mu_{Cr}[c(Cr)]$ & $y=\left\{ \begin{array}{ccl} 3.8323x-9.606 & \mbox{for} & x\leq0.1 \\-0.2328x-9.2251 & \mbox{for} & x>0.1\end{array}\right.$ \\
\hline
\multicolumn{4}{c}{Formation energies of vacancies} \\
{$E_{form}^{vac(Fe)}[c(Cr)]$} & {$y=-0.36439x+2.11763$} & {$E_{form}^{vac(Fe)}[N^{def}_{Cr}]$} & $y=\left\{ \begin{array}{ccl}-0.0294x+2.1291 & \mbox{for} & x\leq3 \\-0.0066x+2.055 & \mbox{for} & x>3 \end{array}\right.$ \\
{$E_{form}^{vac(Cr)}[c(Cr)]$} & $y=\left\{ \begin{array}{ccl} 1.09397x+2.02881 & \mbox{for} & x\leq0.1 \\0.13367x+2.05817 & \mbox{for} & x>0.1\end{array}\right.$ & {$E_{form}^{vac(Cr)}[N^{def}_{Cr}]$} & $y=\left\{ \begin{array}{ccl}-0.0613x+2.168 & \mbox{for} & x\leq3 \\0.0418x+1.9212 & \mbox{for} & x>3 \end{array}\right.$ \\
\hline
\multicolumn{4}{c}{Formation energies of dumbbells} \\
{$E_{form}^{Fe-Fe}[c(Cr)]$} & $y=\left\{ \begin{array}{ccl}-2.47363x+4.10369 & \mbox{for} & x\leq0.1 \\-2.10125x+4.17765 & \mbox{for} & x>0.1 \end{array}\right.$ & {$E_{form}^{Fe-Fe}[N^{def}_{Cr}]$} & $y=\left\{ \begin{array}{ccl}-0.0447x+3.929 & \mbox{for} & x\leq3 \\-0.0248x+3.8344 & \mbox{for} & x>3 \end{array}\right.$ \\
{$E_{form}^{Fe-Cr}[c(Cr)]$} & $y=\left\{ \begin{array}{ccl}-4.7541x+4.06472 & \mbox{for} & x\leq0.1 \\-0.34985x+3.63797 & \mbox{for} & x>0.1 \end{array}\right.$ & {$E_{form}^{Fe-Cr}[N^{def}_{Cr}]$} & $y=\left\{ \begin{array}{ccl}-0.0731x+3.78 & \mbox{for} & x\leq3 \\0.0125x+3.5232 & \mbox{for} & x>3 \end{array}\right.$ \\
{$E_{form}^{Cr-Cr}[c(Cr)]$} & $y=\left\{ \begin{array}{ccl}-6.92687x+4.37945 & \mbox{for} & x\leq0.1 \\-0.07179x+3.68609 & \mbox{for} & x>0.1 \end{array}\right.$ & {$E_{form}^{Cr-Cr}[N^{def}_{Cr}]$} & $y=\left\{ \begin{array}{ccl}-0.1225x+3.985 & \mbox{for} & x\leq3 \\0.0322x+3.5469 & \mbox{for} & x>3 \end{array}\right.$ \\
\hline
\multicolumn{4}{c}{{Formation energies of vacancies in alloy with 5\% at. Cr}}\\
{$E_{form}^{vac(Fe)}[N^{def}_{Cr}]$} & $y=-0.0394x+2.1451$ & {$E_{form}^{vac(Fe)}[\Delta{M}]$} & {$y=-0.054x+2.0721$} \\
{$E_{form}^{vac(Cr)}[N^{def}_{Cr}]$} & $y=-0.1438x+2.2508$ & {$E_{form}^{vac(Cr)}[\Delta{M}]$} & {$y=-0.1126x+2.1248$} \\
\hline
\multicolumn{4}{c}{{Formation energies of dumbbells in alloy with 5\% at. Cr}} \\
{$E_{form}^{Fe-Fe}[N^{def}_{Cr}]$} & $y=-0.0365x+4.1879$ & {$E_{form}^{Fe-Fe(Fe)}[\Delta{M}]$} & {$y=0.0733x+4.5006$} \\
 & & {$E_{form}^{Fe-Cr(Fe)}[\Delta{M}]$} & {$y=0.0176x+3.9803$} \\
\multirow{-2}{*}{{$E_{form}^{Fe-Cr}[N^{def}_{Cr}]$}} & \multirow{-2}{*}{{$y=-0.0504x+3.942$}} & {$E_{form}^{Fe-Cr(Cr)}[\Delta{M}]$} & {$y=-0.1552x+3.3731$} \\
{$E_{form}^{Cr-Cr}[N^{def}_{Cr}]$} & $y=-0.026x+4.3029$ & {$E_{form}^{Cr-Cr(Cr)}[\Delta{M}]$} & {$y=-0.104x+3.7778$} \\
\end{tabular}
\end{ruledtabular}
\end{table*}


\begin{table*}[]
\caption{Equations for the fitted curves of average lattice parameter ($a$), average elastic moduli ($\bar{C}_{11}$, $\bar{C}_{12}$, $\bar{C}_{44}$), bulk modulus ($B_{\mathrm{VRH}}$), shear modulus ($G_{\mathrm{VRH}}$), Young's modulus ($E_{\mathrm{VRH}}$), and Poisson's ratio ($\nu$) of fully relaxed random Fe-Cr alloy structures.
\label{tab:tab_elasticprop}}
\begin{ruledtabular}
\begin{tabular}{cccc}
{Parameter} & {\begin{tabular}[c]{@{}c@{}}Variation as a function of the variable\\in square brackets, denoted here as $x$\end{tabular}} & {Parameter} & {\begin{tabular}[c]{@{}c@{}}Variation as a function of the variable\\in square brackets, denoted here as $x$\end{tabular}} \\ 
\hline
{$a[c(Cr)]$} & {$y=1.2001x^3-0.8763x^2+0.1636x+2.8325$} & {$\bar{C}_{11}[c(Cr)]$} & {\begin{tabular}[c]{@{}c@{}}$y=10995x^4-8693.9x^3+2579.2x^2+$\\$-244.75x+270$\end{tabular}} \\
{$\bar{C}_{12}[c(Cr)]$} & {\begin{tabular}[c]{@{}c@{}}$y=9009.7x^4-9375.5x^3+3527.7x^2+$\\$-476.08x+154.66$\end{tabular}} & {$\bar{C}_{44}[c(Cr)]$} & {$y=1876.5x^3-1373.2x^2+309.7x+92.402$} \\
\hline
{$B_{\mathrm{VRH}}[c(Cr)]$} & {\begin{tabular}[c]{@{}c@{}}$y=9145.2x^4-8733.4x^3+3116.5x^2+$\\$-394.26x+193.38$\end{tabular}} & {$G_{\mathrm{VRH}}[c(Cr)]$} & {$y=1537.3x^3-1040.1x^2+223.56x+76.36$} \\
{$E_{\mathrm{VRH}}[c(Cr)]$} & {$y=3157x^3-2069.7x^2+450.37x+201.66$} & {$\nu[c(Cr)]$}&{\begin{tabular}[c]{@{}c@{}}$y=11.265x^4-13.4x^3+5.5182x^2+$\\$-0.8545x+0.3272$\end{tabular}} \\
\end{tabular}
\end{ruledtabular}
\end{table*}


\begin{table*}[]
\caption{Equations for the trend lines of diagonal ($P_{ii}$) elements of elastic dipole tensor for vacancies and elements of elastic dipole tensor ($P_{11}$, $P_{22/33}$, $P_{23}$, $P_{12/13}$) for dumbbells in random Fe-Cr alloy structures.
\label{tab:tab_elasdip}}
\begin{ruledtabular}
\begin{tabular}{cccc}
{Parameter}&{\begin{tabular}[c]{@{}c@{}}Variation as a function of the variable\\in square brackets, denoted here as $x$\end{tabular}} & {Parameter} &{\begin{tabular}[c]{@{}c@{}}Variation as a function of the variable\\in square brackets, denoted here as $x$\end{tabular}}\\ 
\hline
\multicolumn{4}{c}{{Elastic dipole tensors for vacancies}}\\
{$P_{11}[c(Cr)]$}& $y=\left\{ \begin{array}{ccl}-2.73089x-4.60421 & \mbox{for} & x\leq0.1 \\-5.62277x-4.41065 & \mbox{for} & x>0.1\end{array}\right.$ & {$P_{22}[c(Cr)]$} & $y=\left\{ \begin{array}{ccl}-5.55749x-4.51942 & \mbox{for} & x\leq0.1 \\-5.52298x-4.48534 & \mbox{for} & x>0.1\end{array}\right.$\\
{$P_{33}[c(Cr)]$}& $y=\left\{ \begin{array}{ccl}-6.11801x-4.49765 & \mbox{for} & x\leq0.1 \\-5.45409x-4.50325 & \mbox{for} & x>0.1\end{array}\right.$ & & \\
\hline
\multicolumn{4}{c}{{Elastic dipole tensors for dumbbells}}\\
{$P_{11}^{Fe-Fe}[c(Cr)]$}& $y=\left\{ \begin{array}{ccl}-21.79723x+23.60046 & \mbox{for} & x\leq0.1 \\3.73925x+21.41589 & \mbox{for} & x>0.1\end{array}\right.$ & {$P_{22/33}^{Fe-Fe}[c(Cr)]$}& $y=\left\{ \begin{array}{ccl}-28.79147x+19.55029 & \mbox{for} & x\leq0.1 \\0.91732x+16.75473 & \mbox{for} & x>0.1\end{array}\right.$\\
{$P_{11}^{Fe-Cr}[c(Cr)]$}& $y=\left\{ \begin{array}{ccl}-0.95036x+21.64462 & \mbox{for} & x\leq0.1 \\2.20376x+21.49155 & \mbox{for} & x>0.1\end{array}\right.$ & {$P_{22/33}^{Fe-Cr}[c(Cr)]$}& $y=\left\{ \begin{array}{ccl}-6.79236x+19.36198 & \mbox{for} & x\leq0.1 \\-0.85781x+18.75012 & \mbox{for} & x>0.1\end{array}\right.$\\
{$P_{11}^{Cr-Cr}[c(Cr)]$}& $y=\left\{ \begin{array}{ccl}-11.20776x+20.84527 & \mbox{for} & x\leq0.1 \\3.38226x+20.05774& \mbox{for} & x>0.1\end{array}\right.$ & {$P_{22/33}^{Cr-Cr}[c(Cr)]$}& $y=\left\{ \begin{array}{ccl}-3.24615x+20.37921 & \mbox{for} & x\leq0.1 \\1.13097x+19.82221& \mbox{for} & x>0.1\end{array}\right.$\\
{$P_{23}^{Fe-Fe}[c(Cr)]$}& {$y=1.07363x+4.61543$} & {$P_{12/13}^{Fe-Fe}[c(Cr)]$}& {$y=1.5062x-0.38669$}\\
{$P_{23}^{Fe-Cr}[c(Cr)]$}& {$y=0.32636x+4.42626$} & {$P_{12/13}^{Fe-Cr}[c(Cr)]$}& {$y=0.6145x-0.15704$}\\
{$P_{23}^{Cr-Cr}[c(Cr)]$}& {$y=-2.44943x+3.1727$} & {$P_{12/13}^{Cr-Cr}[c(Cr)]$}& {$y=-0.75586x+1.34013$}\\
\end{tabular}
\end{ruledtabular}
\end{table*}


\begin{table*}[]
\caption{Equations for the trend lines of invariants of elastic dipole tensor ($I^P_1$, $I^P_2$, $I^P_3$) and invariants of relaxation volume tensor ($I^{\Omega}_1$, $I^{\Omega}_2$, $I^{\Omega}_3$) computed for vacancies and dumbbells in random Fe-Cr alloy structures.
\label{tab:tab_invariants}}
\begin{ruledtabular}
\begin{tabular}{cccc}
{Parameter}&{\begin{tabular}[c]{@{}c@{}}Variation as a function of the variable\\in square brackets, denoted here as $x$\end{tabular}} & {Parameter}&{\begin{tabular}[c]{@{}c@{}}Variation as a function of the variable\\in square brackets, denoted here as $x$\end{tabular}}\\  
\hline
\multicolumn{2}{c}{{$I^P_1$ for vacancies}} & \multicolumn{2}{c}{{$I^{\Omega}_1=\Omega_{rel}$ for vacancies}} \\
{$I^{vac(Fe)}_1[c(Cr)]$} & $y=\left\{ \begin{array}{ccl}-46.0186x-10.9791 & \mbox{for} & x\leq0.1 \\-14.8308x-14.3788 & \mbox{for} & x>0.1\end{array}\right.$ & {$\Omega^{vac(Fe)}_{rel}[c(Cr)]$} & $y=\left\{ \begin{array}{ccl}-15.957x-3.126 & \mbox{for} & x\leq0.1 \\-2.175x-4.593 & \mbox{for} & x>0.1\end{array}\right.$ \\
{$I^{vac(Cr)}_1[c(Cr)]$} & $y=\left\{ \begin{array}{ccl}25.1814x-16.6189 & \mbox{for} & x\leq0.1 \\-18.4178x-12.2889 & \mbox{for} & x>0.1\end{array}\right.$ & {$\Omega^{vac(Cr)}_{rel}[c(Cr)]$} & $y=\left\{ \begin{array}{ccl}4.515x-4.743 & \mbox{for} & x\leq0.1 \\-4.021x-3.851 & \mbox{for} & x>0.1\end{array}\right.$\\
\hline
\multicolumn{2}{c}{{$I^P_2$ for vacancies}} & \multicolumn{2}{c}{{$I^{\Omega}_2$ for vacancies}}\\
{$I^{vac(Fe)}_2[c(Cr)]$} & $y=\left\{\begin{array}{ccl}434.5472x+37.5228&\mbox{for}&x\leq0.1\\171.4742x+66.4804&\mbox{for}&x>0.1\end{array}\right.$ & {$I^{vac(Fe)}_2[c(Cr)]$} & $y=\left\{ \begin{array}{ccl}44.369x+2.917 & \mbox{for} & x\leq0.1 \\7.6x+6.874 & \mbox{for} & x>0.1\end{array}\right.$ \\ 
{$I^{vac(Cr)}_2[c(Cr)]$} & $y=\left\{\begin{array}{ccl}-267.065x+93.1079 &\mbox{for} & x\leq0.1 \\ 192.0327x+47.634 & \mbox{for} & x>0.1\end{array}\right.$ & {$I^{vac(Cr)}_2[c(Cr)]$} & $y=\left\{ \begin{array}{ccl}-9.944x+7.331 & \mbox{for} & x\leq0.1 \\9.631x+5.307& \mbox{for} & x>0.1\end{array}\right.$ \\
\hline
\multicolumn{2}{c}{{$I^P_3$ for vacancies}} & \multicolumn{2}{c}{{$I^{\Omega}_3$ for vacancies}}\\
{$I^{vac(Fe)}_3[c(Cr)]$} & $y=\left\{ \begin{array}{ccl}-1032.122x-37.25 & \mbox{for} & x\leq0.1 \\-495.782x-96.919 & \mbox{for} & x>0.1\end{array}\right.$ & {$I^{vac(Fe)}_3[c(Cr)]$} & $y=\left\{ \begin{array}{ccl}-31.051x-0.699 & \mbox{for} & x\leq0.1 \\-6.308x-3.409 & \mbox{for} & x>0.1\end{array}\right.$\\
{$I^{vac(Cr)}_3[c(Cr)]$} & $y=\left\{ \begin{array}{ccl}705.724x-175.27 & \mbox{for} & x\leq0.1 \\-503.199x-55.482& \mbox{for} & x>0.1\end{array}\right.$ & {$I^{vac(Cr)}_3[c(Cr)]$} & $y=\left\{ \begin{array}{ccl}8.428x-3.901 & \mbox{for} & x\leq0.1 \\-7.175x-2.326& \mbox{for} & x>0.1\end{array}\right.$\\
\hline
\multicolumn{2}{c}{{$I^P_1$ for dumbbells}} & \multicolumn{2}{c}{{$I^{\Omega}_1=\Omega_{rel}$ for dumbbells}}\\
{$I^{Fe-Fe}_1[c(Cr)]$} & $y=\left\{ \begin{array}{ccl}-79.3802x+62.701 & \mbox{for} & x\leq0.1 \\5.6469x+54.927 & \mbox{for} & x>0.1\end{array}\right.$ & {$\Omega^{Fe-Fe}_{rel}[c(Cr)]$} & $y=\left\{ \begin{array}{ccl}-12.984x+17.93 & \mbox{for} & x\leq0.1 \\-4.776x+17.192 & \mbox{for} & x>0.1\end{array}\right.$\\
{$I^{Fe-Cr}_1[c(Cr)]$} & $y=\left\{ \begin{array}{ccl}-14.5351x+60.3686 & \mbox{for} & x\leq0.1 \\0.4882x+58.992& \mbox{for} & x>0.1\end{array}\right.$ & {$\Omega^{Fe-Cr}_{rel}[c(Cr)]$} & $y=\left\{ \begin{array}{ccl}4.171x+17.408 & \mbox{for} & x\leq0.1 \\-6.777x+18.498& \mbox{for} & x>0.1\end{array}\right.$\\
{$I^{Cr-Cr}_1[c(Cr)]$} & $y=\left\{ \begin{array}{ccl}-17.7001x+61.6037 & \mbox{for} & x\leq0.1 \\4.3449x+59.8876& \mbox{for} & x>0.1\end{array}\right.$ & {$\Omega^{Cr-Cr}_{rel}[c(Cr)]$} & $y=\left\{ \begin{array}{ccl}3.755x+17.708 & \mbox{for} & x\leq0.1 \\-5.618x+18.779& \mbox{for} & x>0.1\end{array}\right.$\\
\hline
\multicolumn{2}{c}{{$I^P_2$ for dumbbells}} & \multicolumn{2}{c}{{$I^{\Omega}_2$ for dumbbells}}\\
{$I^{Fe-Fe}_2[c(Cr)]$} & $y=\left\{\begin{array}{ccl} -3201.976x+1284.909 & \mbox{for} & x\leq0.1 \\ 203.521x+973.849 & \mbox{for} & x>0.1 \end{array}\right.$ & {$I^{Fe-Fe}_2[c(Cr)]$} & $y=\left\{ \begin{array}{ccl}-161.36x+105.29 & \mbox{for} & x\leq0.1 \\-49.119x+94.884 & \mbox{for} & x>0.1\end{array}\right.$\\
{$I^{Fe-Cr}_2[c(Cr)]$} & $y=\left\{\begin{array}{ccl} -611.626x+1195.304 &\mbox{for}&x\leq0.1\\13.143x+1136.725 & \mbox{for} & x>0.1\end{array}\right.$ & {$I^{Fe-Cr}_2[c(Cr)]$} & $y=\left\{ \begin{array}{ccl}44.119x+99.391 & \mbox{for} & x\leq0.1 \\-76.175x+111.16& \mbox{for} & x>0.1\end{array}\right.$\\
{$I^{Cr-Cr}_2[c(Cr)]$} & $y=\left\{\begin{array}{ccl} -749.875x+1241.88 & \mbox{for} & x\leq0.1\\190.091x+1169.838 & \mbox{for} & x>0.1\end{array}\right.$ & {$I^{Cr-Cr}_2[c(Cr)]$} & $y=\left\{ \begin{array}{ccl}39.69x+102.65 & \mbox{for} & x\leq0.1 \\-62.865x+114.61& \mbox{for} & x>0.1\end{array}\right.$\\
\hline
\multicolumn{2}{c}{{$I^P_3$ for dumbbells}} & \multicolumn{2}{c}{{$I^{\Omega}_3$ for dumbbells}} \\
{$I^{Fe-Fe}_3[c(Cr)]$}& $y=\left\{ \begin{array}{ccl}-31708.36x+8547.57 & \mbox{for} & x\leq0.1 \\1755.89x+5489.92 & \mbox{for} & x>0.1\end{array}\right.$ & {$I^{Fe-Fe}_3[c(Cr)]$} & $y=\left\{ \begin{array}{ccl}-494.76x+201.21 & \mbox{for} & x\leq0.1 \\-118.52x+165.24 & \mbox{for} & x>0.1\end{array}\right.$\\
{$I^{Fe-Cr}_3[c(Cr)]$}& $y=\left\{ \begin{array}{ccl}-6385.02x+7729.52 & \mbox{for} & x\leq0.1 \\52.26x+7107.83& \mbox{for} & x>0.1\end{array}\right.$ & {$I^{Fe-Cr}_3[c(Cr)]$} & $y=\left\{ \begin{array}{ccl}109.57x+185.38 & \mbox{for} & x\leq0.1 \\-208.8x+215.73& \mbox{for} & x>0.1\end{array}\right.$\\
{$I^{Cr-Cr}_3[c(Cr)]$}& $y=\left\{ \begin{array}{ccl}-7637.41x+8200.96 & \mbox{for} & x\leq0.1 \\2039.45x+7454.34& \mbox{for} & x>0.1\end{array}\right.$ & {$I^{Cr-Cr}_3[c(Cr)]$} & $y=\left\{ \begin{array}{ccl}104.53x+194.99 & \mbox{for} & x\leq0.1 \\-172.21x+227.64& \mbox{for} & x>0.1\end{array}\right.$\\
\end{tabular}
\end{ruledtabular}
\end{table*}


\begin{table*}[]
\caption{Equations for the trend lines of relaxation volumes ($\Omega_{rel}$) for vacancies and dumbbells in random Fe-Cr alloys as a function of the number of Cr atoms in the vicinity of a defect $N^{def}_{Cr}$ and as a function of change of magnitudes of magnetic moments ($\Delta{M}$) caused by a defect for the alloy with 5\% at. Cr.
\label{tab:tab_relaxvol}}
\begin{ruledtabular}
\begin{tabular}{cccc}
{Parameter}&{\begin{tabular}[c]{@{}c@{}}Variation as a function of the variable\\in square brackets, denoted here as $x$\end{tabular}} & {Parameter} &{\begin{tabular}[c]{@{}c@{}}Variation as a function of the variable\\in square brackets, denoted here as $x$\end{tabular}}\\  
\hline
\multicolumn{4}{c}{{$\Omega_{rel}$ for vacancies}}\\
{$\Omega^{vac(Fe)}_{rel}[N^{def}_{Cr}]$} & $y=\left\{ \begin{array}{ccl}-0.3185x-3.968 & \mbox{for} & x\leq3 \\-0.1766x-4.3859 & \mbox{for} & x>3\end{array}\right.$ & {$\Omega^{vac(Cr)}_{rel}[N^{def}_{Cr}]$} & $y=\left\{ \begin{array}{ccl}0.2393x-4.791 & \mbox{for} & x\leq3 \\-0.2965x-3.5152 & \mbox{for} & x>3\end{array}\right.$ \\
\hline
\multicolumn{4}{c}{{$\Omega_{rel}$ for vacancies in alloy with 5\% at. Cr}} \\
{$\Omega^{vac(Fe)}_{rel}[N^{def}_{Cr}]$} & $y=0.2662x-4.5148$ & {$\Omega^{vac(Fe)}_{rel}[\Delta{M}]$}&{$y=0.385x-359$} \\
{$\Omega^{vac(Cr)}_{rel}[N^{def}_{Cr}]$} & $y=0.6806x-5.4627$ & {$\Omega^{vac(Cr)}_{rel}[\Delta{M}]$}&{$y=0.5649x-4.7339$} \\
\hline
\multicolumn{4}{c}{{$\Omega_{rel}$ for dumbbells}}\\
{$\Omega^{Fe-Fe}_{rel}[N^{def}_{Cr}]$} & $y=\left\{ \begin{array}{ccl}-0.208x+17.176 & \mbox{for} & x\leq3 \\-0.018x+16.486 & \mbox{for} & x>3\end{array}\right.$ & {$\Omega^{Fe-Cr}_{rel}[N^{def}_{Cr}]$} & $y=\left\{ \begin{array}{ccl}0.006x+17.424 & \mbox{for} & x\leq3 \\-0.113x+17.595 & \mbox{for} & x>3\end{array}\right.$\\
{$\Omega^{Cr-Cr}_{rel}[N^{def}_{Cr}]$} & $y=\left\{ \begin{array}{ccl}-0.169x+18.055 & \mbox{for} & x\leq3 \\-0.002x+17.331 & \mbox{for} & x>3\end{array}\right.$ & & \\
\hline
\multicolumn{4}{c}{{$\Omega_{rel}$ for dumbbells in alloy with 5\% at. Cr}}\\
{$\Omega^{Fe-Fe}_{rel}[N^{def}_{Cr}]$} & $y=-0.3632x+17.782$ & {$\Omega^{Fe-Fe(Fe)}_{rel}[\Delta{M}]$}&{$y=0.858x+21.519$}\\
 & & {$\Omega^{Fe-Cr(Fe)}_{rel}[\Delta{M}]$} & {$y=0.5994x+20.767$}\\
\multirow{-2}{*}{{$\Omega^{Fe-Cr}_{rel}[N^{def}_{Cr}]$}} & \multirow{-2}{*}{$y=-0.1552x+18.204$} & {$\Omega^{Fe-Cr(Cr)}_{rel}[\Delta{M}]$}&{$y=0.4457x+19.184$}\\
{$\Omega^{Cr-Cr}_{rel}[N^{def}_{Cr}]$} & $y=0.2287x+17.265$ & {$\Omega^{Cr-Cr(Cr)}_{rel}[\Delta{M}]$}&{$y=0.4034x+19.425$}\\
\end{tabular}
\end{ruledtabular}
\end{table*}

\bibliography{FeCr_paper}

\providecommand{\noopsort}[1]{}\providecommand{\singleletter}[1]{#1}%
\begin{thebibliography}{85}%
\makeatletter
\providecommand \@ifxundefined [1]{%
 \@ifx{#1\undefined}
}%
\providecommand \@ifnum [1]{%
 \ifnum #1\expandafter \@firstoftwo
 \else \expandafter \@secondoftwo
 \fi
}%
\providecommand \@ifx [1]{%
 \ifx #1\expandafter \@firstoftwo
 \else \expandafter \@secondoftwo
 \fi
}%
\providecommand \natexlab [1]{#1}%
\providecommand \enquote  [1]{``#1''}%
\providecommand \bibnamefont  [1]{#1}%
\providecommand \bibfnamefont [1]{#1}%
\providecommand \citenamefont [1]{#1}%
\providecommand \href@noop [0]{\@secondoftwo}%
\providecommand \href [0]{\begingroup \@sanitize@url \@href}%
\providecommand \@href[1]{\@@startlink{#1}\@@href}%
\providecommand \@@href[1]{\endgroup#1\@@endlink}%
\providecommand \@sanitize@url [0]{\catcode `\\12\catcode `\$12\catcode
  `\&12\catcode `\#12\catcode `\^12\catcode `\_12\catcode `\%12\relax}%
\providecommand \@@startlink[1]{}%
\providecommand \@@endlink[0]{}%
\providecommand \url  [0]{\begingroup\@sanitize@url \@url }%
\providecommand \@url [1]{\endgroup\@href {#1}{\urlprefix }}%
\providecommand \urlprefix  [0]{URL }%
\providecommand \Eprint [0]{\href }%
\providecommand \doibase [0]{http://dx.doi.org/}%
\providecommand \selectlanguage [0]{\@gobble}%
\providecommand \bibinfo  [0]{\@secondoftwo}%
\providecommand \bibfield  [0]{\@secondoftwo}%
\providecommand \translation [1]{[#1]}%
\providecommand \BibitemOpen [0]{}%
\providecommand \bibitemStop [0]{}%
\providecommand \bibitemNoStop [0]{.\EOS\space}%
\providecommand \EOS [0]{\spacefactor3000\relax}%
\providecommand \BibitemShut  [1]{\csname bibitem#1\endcsname}%
\let\auto@bib@innerbib\@empty
\bibitem [{\citenamefont {Cai}\ and\ \citenamefont {Nix}(2016)}]{CaiNix}%
  \BibitemOpen
  \bibfield  {author} {\bibinfo {author} {\bibfnamefont {W.}~\bibnamefont
  {Cai}}\ and\ \bibinfo {author} {\bibfnamefont {W.~D.}\ \bibnamefont {Nix}},\
  }\href@noop {} {\emph {\bibinfo {title} {{Imperfections in Crystalline
  Solids}}}}\ (\bibinfo  {publisher} {Cambridge University Press},\ \bibinfo
  {address} {Cambridge, England, UK},\ \bibinfo {year} {2016})\BibitemShut
  {NoStop}%
\bibitem [{\citenamefont {Ruban}\ \emph {et~al.}(2008)\citenamefont {Ruban},
  \citenamefont {Korzhavyi},\ and\ \citenamefont {Johansson}}]{Ruban2008a}%
  \BibitemOpen
  \bibfield  {author} {\bibinfo {author} {\bibfnamefont {A.~V.}\ \bibnamefont
  {Ruban}}, \bibinfo {author} {\bibfnamefont {P.~A.}\ \bibnamefont
  {Korzhavyi}}, \ and\ \bibinfo {author} {\bibfnamefont {B.}~\bibnamefont
  {Johansson}},\ }\href {\doibase 10.1103/PhysRevB.77.094436} {\bibfield
  {journal} {\bibinfo  {journal} {Phys. Rev. B}\ }\textbf {\bibinfo {volume}
  {77}},\ \bibinfo {pages} {094436} (\bibinfo {year} {2008})}\BibitemShut
  {NoStop}%
\bibitem [{\citenamefont {Wr{\'{o}}bel}\ \emph {et~al.}(2015)\citenamefont
  {Wr{\'{o}}bel}, \citenamefont {Nguyen-Manh}, \citenamefont {Lavrentiev},
  \citenamefont {Muzyk},\ and\ \citenamefont {Dudarev}}]{Wrobel2015}%
  \BibitemOpen
  \bibfield  {author} {\bibinfo {author} {\bibfnamefont {J.~S.}\ \bibnamefont
  {Wr{\'{o}}bel}}, \bibinfo {author} {\bibfnamefont {D.}~\bibnamefont
  {Nguyen-Manh}}, \bibinfo {author} {\bibfnamefont {M.~Y.}\ \bibnamefont
  {Lavrentiev}}, \bibinfo {author} {\bibfnamefont {M.}~\bibnamefont {Muzyk}}, \
  and\ \bibinfo {author} {\bibfnamefont {S.~L.}\ \bibnamefont {Dudarev}},\
  }\href {\doibase 10.1103/PhysRevB.91.024108} {\bibfield  {journal} {\bibinfo
  {journal} {Phys. Rev. B}\ }\textbf {\bibinfo {volume} {91}},\ \bibinfo
  {pages} {024108} (\bibinfo {year} {2015})}\BibitemShut {NoStop}%
\bibitem [{\citenamefont {Lavrentiev}\ \emph {et~al.}(2007)\citenamefont
  {Lavrentiev}, \citenamefont {Drautz}, \citenamefont {Nguyen-Manh},
  \citenamefont {Klaver},\ and\ \citenamefont {Dudarev}}]{Lavrentiev2007}%
  \BibitemOpen
  \bibfield  {author} {\bibinfo {author} {\bibfnamefont {M.~Y.}\ \bibnamefont
  {Lavrentiev}}, \bibinfo {author} {\bibfnamefont {R.}~\bibnamefont {Drautz}},
  \bibinfo {author} {\bibfnamefont {D.}~\bibnamefont {Nguyen-Manh}}, \bibinfo
  {author} {\bibfnamefont {T.~P.~C.}\ \bibnamefont {Klaver}}, \ and\ \bibinfo
  {author} {\bibfnamefont {S.~L.}\ \bibnamefont {Dudarev}},\ }\href {\doibase
  10.1103/PhysRevB.75.014208} {\bibfield  {journal} {\bibinfo  {journal} {Phys.
  Rev. B}\ }\textbf {\bibinfo {volume} {75}},\ \bibinfo {pages} {014208}
  (\bibinfo {year} {2007})}\BibitemShut {NoStop}%
\bibitem [{\citenamefont {Wr{\'{o}}bel}\ \emph
  {et~al.}(2017{\natexlab{a}})\citenamefont {Wr{\'{o}}bel}, \citenamefont
  {Nguyen-Manh}, \citenamefont {Kurzyd{\l}owski},\ and\ \citenamefont
  {Dudarev}}]{Wrobel2017}%
  \BibitemOpen
  \bibfield  {author} {\bibinfo {author} {\bibfnamefont {J.~S.}\ \bibnamefont
  {Wr{\'{o}}bel}}, \bibinfo {author} {\bibfnamefont {D.}~\bibnamefont
  {Nguyen-Manh}}, \bibinfo {author} {\bibfnamefont {K.~J.}\ \bibnamefont
  {Kurzyd{\l}owski}}, \ and\ \bibinfo {author} {\bibfnamefont {S.~L.}\
  \bibnamefont {Dudarev}},\ }\href
  {http://stacks.iop.org/0953-8984/29/i=14/a=145403?key=crossref.64d730b8e8d600fbd485cff23cdaa579}
  {\bibfield  {journal} {\bibinfo  {journal} {J. Phys. Condens. Matter}\
  }\textbf {\bibinfo {volume} {29}},\ \bibinfo {pages} {145403} (\bibinfo
  {year} {2017}{\natexlab{a}})}\BibitemShut {NoStop}%
\bibitem [{\citenamefont {Fernandez-Caballero}\ \emph
  {et~al.}(2017)\citenamefont {Fernandez-Caballero}, \citenamefont
  {Wr{\'{o}}bel}, \citenamefont {Mummery},\ and\ \citenamefont
  {Nguyen-Manh}}]{Fernandez-Caballero2017}%
  \BibitemOpen
  \bibfield  {author} {\bibinfo {author} {\bibfnamefont {A.}~\bibnamefont
  {Fernandez-Caballero}}, \bibinfo {author} {\bibfnamefont {J.~S.}\
  \bibnamefont {Wr{\'{o}}bel}}, \bibinfo {author} {\bibfnamefont
  {P.}~\bibnamefont {Mummery}}, \ and\ \bibinfo {author} {\bibfnamefont
  {D.}~\bibnamefont {Nguyen-Manh}},\ }\href@noop {} {\bibfield  {journal}
  {\bibinfo  {journal} {J. Phase Equilib. Diff.}\ }\textbf {\bibinfo {volume}
  {38}},\ \bibinfo {pages} {391} (\bibinfo {year} {2017})}\BibitemShut
  {NoStop}%
\bibitem [{\citenamefont {Fedorov}\ \emph {et~al.}(2020)\citenamefont
  {Fedorov}, \citenamefont {Wr{\'{o}}bel}, \citenamefont
  {Fern{\'{a}}ndez-Caballero}, \citenamefont {Kurzyd{\l}owski},\ and\
  \citenamefont {Nguyen-Manh}}]{Fedorov2020}%
  \BibitemOpen
  \bibfield  {author} {\bibinfo {author} {\bibfnamefont {M.}~\bibnamefont
  {Fedorov}}, \bibinfo {author} {\bibfnamefont {J.~S.}\ \bibnamefont
  {Wr{\'{o}}bel}}, \bibinfo {author} {\bibfnamefont {A.}~\bibnamefont
  {Fern{\'{a}}ndez-Caballero}}, \bibinfo {author} {\bibfnamefont {K.~J.}\
  \bibnamefont {Kurzyd{\l}owski}}, \ and\ \bibinfo {author} {\bibfnamefont
  {D.}~\bibnamefont {Nguyen-Manh}},\ }\href {\doibase
  10.1103/PhysRevB.101.174416} {\bibfield  {journal} {\bibinfo  {journal}
  {Phys. Rev. B}\ }\textbf {\bibinfo {volume} {101}},\ \bibinfo {pages}
  {174416} (\bibinfo {year} {2020})}\BibitemShut {NoStop}%
\bibitem [{\citenamefont {Leibfried}\ and\ \citenamefont
  {Breuer}(1978)}]{Leibfried}%
  \BibitemOpen
  \bibfield  {author} {\bibinfo {author} {\bibfnamefont {G.}~\bibnamefont
  {Leibfried}}\ and\ \bibinfo {author} {\bibfnamefont {N.}~\bibnamefont
  {Breuer}},\ }\href@noop {} {\emph {\bibinfo {title} {{Point Defects in Metals
  I: Introduction to the Theory}}}}\ (\bibinfo  {publisher} {Springer-Verlag},\
  \bibinfo {address} {Berlin, Germany},\ \bibinfo {year} {1978})\BibitemShut
  {NoStop}%
\bibitem [{\citenamefont {Bacon}\ \emph {et~al.}(1979)\citenamefont {Bacon},
  \citenamefont {Barnett},\ and\ \citenamefont {Scattergood}}]{Bacon1979}%
  \BibitemOpen
  \bibfield  {author} {\bibinfo {author} {\bibfnamefont {D.~J.}\ \bibnamefont
  {Bacon}}, \bibinfo {author} {\bibfnamefont {D.~M.}\ \bibnamefont {Barnett}},
  \ and\ \bibinfo {author} {\bibfnamefont {R.~O.}\ \bibnamefont
  {Scattergood}},\ }\href@noop {} {\bibfield  {journal} {\bibinfo  {journal}
  {Prog. Mater. Sci.}\ }\textbf {\bibinfo {volume} {23}},\ \bibinfo {pages}
  {51} (\bibinfo {year} {1979})}\BibitemShut {NoStop}%
\bibitem [{\citenamefont {Freysoldt}\ \emph {et~al.}(2014)\citenamefont
  {Freysoldt}, \citenamefont {Grabowski}, \citenamefont {Hickel}, \citenamefont
  {Neugebauer}, \citenamefont {Kresse}, \citenamefont {Janotti},\ and\
  \citenamefont {{Van de Walle}}}]{Freysoldt2014}%
  \BibitemOpen
  \bibfield  {author} {\bibinfo {author} {\bibfnamefont {C.}~\bibnamefont
  {Freysoldt}}, \bibinfo {author} {\bibfnamefont {B.}~\bibnamefont
  {Grabowski}}, \bibinfo {author} {\bibfnamefont {T.}~\bibnamefont {Hickel}},
  \bibinfo {author} {\bibfnamefont {J.}~\bibnamefont {Neugebauer}}, \bibinfo
  {author} {\bibfnamefont {G.}~\bibnamefont {Kresse}}, \bibinfo {author}
  {\bibfnamefont {A.}~\bibnamefont {Janotti}}, \ and\ \bibinfo {author}
  {\bibfnamefont {C.~G.}\ \bibnamefont {{Van de Walle}}},\ }\href {\doibase
  10.1103/RevModPhys.86.253} {\bibfield  {journal} {\bibinfo  {journal} {Rev.
  Mod. Phys.}\ }\textbf {\bibinfo {volume} {86}},\ \bibinfo {pages} {253}
  (\bibinfo {year} {2014})}\BibitemShut {NoStop}%
\bibitem [{\citenamefont {Clouet}\ \emph {et~al.}(2008)\citenamefont {Clouet},
  \citenamefont {Garruchet}, \citenamefont {Nguyen}, \citenamefont {Perez},\
  and\ \citenamefont {Becquart}}]{Clouet2008}%
  \BibitemOpen
  \bibfield  {author} {\bibinfo {author} {\bibfnamefont {E.}~\bibnamefont
  {Clouet}}, \bibinfo {author} {\bibfnamefont {S.}~\bibnamefont {Garruchet}},
  \bibinfo {author} {\bibfnamefont {H.}~\bibnamefont {Nguyen}}, \bibinfo
  {author} {\bibfnamefont {M.}~\bibnamefont {Perez}}, \ and\ \bibinfo {author}
  {\bibfnamefont {C.~S.}\ \bibnamefont {Becquart}},\ }\href {\doibase
  10.1016/j.actamat.2008.03.024} {\bibfield  {journal} {\bibinfo  {journal}
  {Acta Mater.}\ }\textbf {\bibinfo {volume} {56}},\ \bibinfo {pages} {3450}
  (\bibinfo {year} {2008})}\BibitemShut {NoStop}%
\bibitem [{\citenamefont {Varvenne}\ \emph {et~al.}(2013)\citenamefont
  {Varvenne}, \citenamefont {Bruneval}, \citenamefont {Marinica},\ and\
  \citenamefont {Clouet}}]{Varvenne2013}%
  \BibitemOpen
  \bibfield  {author} {\bibinfo {author} {\bibfnamefont {C.}~\bibnamefont
  {Varvenne}}, \bibinfo {author} {\bibfnamefont {F.}~\bibnamefont {Bruneval}},
  \bibinfo {author} {\bibfnamefont {M.-C.}\ \bibnamefont {Marinica}}, \ and\
  \bibinfo {author} {\bibfnamefont {E.}~\bibnamefont {Clouet}},\ }\href
  {\doibase 10.1103/PhysRevB.88.134102} {\bibfield  {journal} {\bibinfo
  {journal} {Phys. Rev. B}\ }\textbf {\bibinfo {volume} {88}},\ \bibinfo
  {pages} {134102} (\bibinfo {year} {2013})}\BibitemShut {NoStop}%
\bibitem [{\citenamefont {Varvenne}\ and\ \citenamefont
  {Clouet}(2017)}]{Varvenne2017}%
  \BibitemOpen
  \bibfield  {author} {\bibinfo {author} {\bibfnamefont {C.}~\bibnamefont
  {Varvenne}}\ and\ \bibinfo {author} {\bibfnamefont {E.}~\bibnamefont
  {Clouet}},\ }\href {\doibase 10.1103/PhysRevB.96.224103} {\bibfield
  {journal} {\bibinfo  {journal} {Phys. Rev. B}\ }\textbf {\bibinfo {volume}
  {96}},\ \bibinfo {pages} {224103} (\bibinfo {year} {2017})}\BibitemShut
  {NoStop}%
\bibitem [{\citenamefont {Dudarev}\ and\ \citenamefont
  {Ma}(2018)}]{Dudarev2018}%
  \BibitemOpen
  \bibfield  {author} {\bibinfo {author} {\bibfnamefont {S.~L.}\ \bibnamefont
  {Dudarev}}\ and\ \bibinfo {author} {\bibfnamefont {P.-W.}\ \bibnamefont
  {Ma}},\ }\href {\doibase 10.1103/PhysRevMaterials.2.033602} {\bibfield
  {journal} {\bibinfo  {journal} {Phys. Rev. Mater.}\ }\textbf {\bibinfo
  {volume} {2}},\ \bibinfo {pages} {033602} (\bibinfo {year}
  {2018})}\BibitemShut {NoStop}%
\bibitem [{\citenamefont {Ma}\ and\ \citenamefont
  {Dudarev}(2019{\natexlab{a}})}]{Ma2019}%
  \BibitemOpen
  \bibfield  {author} {\bibinfo {author} {\bibfnamefont {P.-W.}\ \bibnamefont
  {Ma}}\ and\ \bibinfo {author} {\bibfnamefont {S.~L.}\ \bibnamefont
  {Dudarev}},\ }\href {\doibase 10.1103/PhysRevMaterials.3.013605} {\bibfield
  {journal} {\bibinfo  {journal} {Phys. Rev. Mater.}\ }\textbf {\bibinfo
  {volume} {3}},\ \bibinfo {pages} {013605} (\bibinfo {year}
  {2019}{\natexlab{a}})}\BibitemShut {NoStop}%
\bibitem [{\citenamefont {Ma}\ and\ \citenamefont
  {Dudarev}(2019{\natexlab{b}})}]{Ma2019a}%
  \BibitemOpen
  \bibfield  {author} {\bibinfo {author} {\bibfnamefont {P.-W.}\ \bibnamefont
  {Ma}}\ and\ \bibinfo {author} {\bibfnamefont {S.~L.}\ \bibnamefont
  {Dudarev}},\ }\href {\doibase 10.1103/PhysRevMaterials.3.063601} {\bibfield
  {journal} {\bibinfo  {journal} {Phys. Rev. Mater.}\ }\textbf {\bibinfo
  {volume} {3}},\ \bibinfo {pages} {063601} (\bibinfo {year}
  {2019}{\natexlab{b}})}\BibitemShut {NoStop}%
\bibitem [{\citenamefont {Ma}\ and\ \citenamefont
  {Dudarev}(2019{\natexlab{c}})}]{Ma2019b}%
  \BibitemOpen
  \bibfield  {author} {\bibinfo {author} {\bibfnamefont {P.-W.}\ \bibnamefont
  {Ma}}\ and\ \bibinfo {author} {\bibfnamefont {S.~L.}\ \bibnamefont
  {Dudarev}},\ }\href {\doibase 10.1103/PhysRevMaterials.3.043606} {\bibfield
  {journal} {\bibinfo  {journal} {Phys. Rev. Mater.}\ }\textbf {\bibinfo
  {volume} {3}},\ \bibinfo {pages} {043606} (\bibinfo {year}
  {2019}{\natexlab{c}})}\BibitemShut {NoStop}%
\bibitem [{\citenamefont {Domain}\ and\ \citenamefont
  {Becquart}(2001)}]{Domain2001}%
  \BibitemOpen
  \bibfield  {author} {\bibinfo {author} {\bibfnamefont {C.}~\bibnamefont
  {Domain}}\ and\ \bibinfo {author} {\bibfnamefont {C.~S.}\ \bibnamefont
  {Becquart}},\ }\href {\doibase 10.1103/PhysRevB.65.024103} {\bibfield
  {journal} {\bibinfo  {journal} {Phys. Rev. B}\ }\textbf {\bibinfo {volume}
  {65}},\ \bibinfo {pages} {024103} (\bibinfo {year} {2001})}\BibitemShut
  {NoStop}%
\bibitem [{\citenamefont {Dudarev}\ \emph {et~al.}(2010)\citenamefont
  {Dudarev}, \citenamefont {Gilbert}, \citenamefont {Arakawa}, \citenamefont
  {Mori}, \citenamefont {Yao}, \citenamefont {Jenkins},\ and\ \citenamefont
  {Derlet}}]{Dudarev2010}%
  \BibitemOpen
  \bibfield  {author} {\bibinfo {author} {\bibfnamefont {S.~L.}\ \bibnamefont
  {Dudarev}}, \bibinfo {author} {\bibfnamefont {M.~R.}\ \bibnamefont
  {Gilbert}}, \bibinfo {author} {\bibfnamefont {K.}~\bibnamefont {Arakawa}},
  \bibinfo {author} {\bibfnamefont {H.}~\bibnamefont {Mori}}, \bibinfo {author}
  {\bibfnamefont {Z.}~\bibnamefont {Yao}}, \bibinfo {author} {\bibfnamefont
  {M.~L.}\ \bibnamefont {Jenkins}}, \ and\ \bibinfo {author} {\bibfnamefont
  {P.~M.}\ \bibnamefont {Derlet}},\ }\href {\doibase
  10.1103/PhysRevB.81.224107} {\bibfield  {journal} {\bibinfo  {journal} {Phys.
  Rev. B}\ }\textbf {\bibinfo {volume} {81}},\ \bibinfo {pages} {224107}
  (\bibinfo {year} {2010})}\BibitemShut {NoStop}%
\bibitem [{\citenamefont {Dudarev}\ \emph {et~al.}(2018)\citenamefont
  {Dudarev}, \citenamefont {Mason}, \citenamefont {Tarleton}, \citenamefont
  {Ma},\ and\ \citenamefont {Sand}}]{Dudarev2018a}%
  \BibitemOpen
  \bibfield  {author} {\bibinfo {author} {\bibfnamefont {S.~L.}\ \bibnamefont
  {Dudarev}}, \bibinfo {author} {\bibfnamefont {D.~R.}\ \bibnamefont {Mason}},
  \bibinfo {author} {\bibfnamefont {E.}~\bibnamefont {Tarleton}}, \bibinfo
  {author} {\bibfnamefont {P.-W.}\ \bibnamefont {Ma}}, \ and\ \bibinfo {author}
  {\bibfnamefont {A.~E.}\ \bibnamefont {Sand}},\ }\href {\doibase
  10.1088/1741-4326/aadb48} {\bibfield  {journal} {\bibinfo  {journal} {Nucl.
  Fusion}\ }\textbf {\bibinfo {volume} {58}},\ \bibinfo {pages} {126002}
  (\bibinfo {year} {2018})}\BibitemShut {NoStop}%
\bibitem [{\citenamefont {Clouet}\ \emph {et~al.}(2018)\citenamefont {Clouet},
  \citenamefont {Varvenne},\ and\ \citenamefont {Jourdan}}]{Clouet2018}%
  \BibitemOpen
  \bibfield  {author} {\bibinfo {author} {\bibfnamefont {E.}~\bibnamefont
  {Clouet}}, \bibinfo {author} {\bibfnamefont {C.}~\bibnamefont {Varvenne}}, \
  and\ \bibinfo {author} {\bibfnamefont {T.}~\bibnamefont {Jourdan}},\ }\href
  {\doibase 10.1016/j.commatsci.2018.01.053} {\bibfield  {journal} {\bibinfo
  {journal} {Comput. Mater. Sci.}\ }\textbf {\bibinfo {volume} {147}},\
  \bibinfo {pages} {49} (\bibinfo {year} {2018})}\BibitemShut {NoStop}%
\bibitem [{\citenamefont {Wr{\'{o}}bel}\ \emph
  {et~al.}(2017{\natexlab{b}})\citenamefont {Wr{\'{o}}bel}, \citenamefont
  {Nguyen-Manh}, \citenamefont {Dudarev},\ and\ \citenamefont
  {Kurzyd{\l}owski}}]{Wrobel2017a}%
  \BibitemOpen
  \bibfield  {author} {\bibinfo {author} {\bibfnamefont {J.~S.}\ \bibnamefont
  {Wr{\'{o}}bel}}, \bibinfo {author} {\bibfnamefont {D.}~\bibnamefont
  {Nguyen-Manh}}, \bibinfo {author} {\bibfnamefont {S.~L.}\ \bibnamefont
  {Dudarev}}, \ and\ \bibinfo {author} {\bibfnamefont {K.~J.}\ \bibnamefont
  {Kurzyd{\l}owski}},\ }\href {\doibase 10.1016/j.nimb.2016.10.024} {\bibfield
  {journal} {\bibinfo  {journal} {Nucl. Instr. Meth. Phys. Res. B}\ }\textbf
  {\bibinfo {volume} {393}},\ \bibinfo {pages} {126} (\bibinfo {year}
  {2017}{\natexlab{b}})}\BibitemShut {NoStop}%
\bibitem [{\citenamefont {Samin}\ \emph {et~al.}(2019)\citenamefont {Samin},
  \citenamefont {Andersson}, \citenamefont {Holby},\ and\ \citenamefont
  {Uberuaga}}]{Samin2019}%
  \BibitemOpen
  \bibfield  {author} {\bibinfo {author} {\bibfnamefont {A.~J.}\ \bibnamefont
  {Samin}}, \bibinfo {author} {\bibfnamefont {D.~A.}\ \bibnamefont
  {Andersson}}, \bibinfo {author} {\bibfnamefont {E.~F.}\ \bibnamefont
  {Holby}}, \ and\ \bibinfo {author} {\bibfnamefont {B.~P.}\ \bibnamefont
  {Uberuaga}},\ }\href {\doibase 10.1103/PhysRevB.99.174202} {\bibfield
  {journal} {\bibinfo  {journal} {Phys. Rev. B}\ }\textbf {\bibinfo {volume}
  {99}},\ \bibinfo {pages} {174202} (\bibinfo {year} {2019})}\BibitemShut
  {NoStop}%
\bibitem [{\citenamefont {Klaver}\ \emph {et~al.}(2006)\citenamefont {Klaver},
  \citenamefont {Drautz},\ and\ \citenamefont {Finnis}}]{Klaver2006}%
  \BibitemOpen
  \bibfield  {author} {\bibinfo {author} {\bibfnamefont {T.~P.~C.}\
  \bibnamefont {Klaver}}, \bibinfo {author} {\bibfnamefont {R.}~\bibnamefont
  {Drautz}}, \ and\ \bibinfo {author} {\bibfnamefont {M.~W.}\ \bibnamefont
  {Finnis}},\ }\href {\doibase 10.1103/PhysRevB.74.094435} {\bibfield
  {journal} {\bibinfo  {journal} {Phys. Rev. B}\ }\textbf {\bibinfo {volume}
  {74}},\ \bibinfo {pages} {094435} (\bibinfo {year} {2006})}\BibitemShut
  {NoStop}%
\bibitem [{\citenamefont {Klaver}\ \emph {et~al.}(2007)\citenamefont {Klaver},
  \citenamefont {Olsson},\ and\ \citenamefont {Finnis}}]{Klaver2007}%
  \BibitemOpen
  \bibfield  {author} {\bibinfo {author} {\bibfnamefont {T.~P.~C.}\
  \bibnamefont {Klaver}}, \bibinfo {author} {\bibfnamefont {P.}~\bibnamefont
  {Olsson}}, \ and\ \bibinfo {author} {\bibfnamefont {M.~W.}\ \bibnamefont
  {Finnis}},\ }\href {\doibase 10.1103/PhysRevB.76.214110} {\bibfield
  {journal} {\bibinfo  {journal} {Phys. Rev. B}\ }\textbf {\bibinfo {volume}
  {76}},\ \bibinfo {pages} {214110} (\bibinfo {year} {2007})}\BibitemShut
  {NoStop}%
\bibitem [{\citenamefont {Lavrentiev}\ \emph {et~al.}(2009)\citenamefont
  {Lavrentiev}, \citenamefont {Dudarev},\ and\ \citenamefont
  {Nguyen-Manh}}]{Lavrentiev2009}%
  \BibitemOpen
  \bibfield  {author} {\bibinfo {author} {\bibfnamefont {M.~Y.}\ \bibnamefont
  {Lavrentiev}}, \bibinfo {author} {\bibfnamefont {S.~L.}\ \bibnamefont
  {Dudarev}}, \ and\ \bibinfo {author} {\bibfnamefont {D.}~\bibnamefont
  {Nguyen-Manh}},\ }\href {\doibase 10.1016/j.jnucmat.2008.12.052} {\bibfield
  {journal} {\bibinfo  {journal} {J. Nucl. Mater.}\ }\textbf {\bibinfo {volume}
  {386-388}},\ \bibinfo {pages} {22} (\bibinfo {year} {2009})}\BibitemShut
  {NoStop}%
\bibitem [{\citenamefont {Lavrentiev}\ \emph {et~al.}(2011)\citenamefont
  {Lavrentiev}, \citenamefont {{Nguyen Manh}},\ and\ \citenamefont
  {Dudarev}}]{Lavrentiev2011}%
  \BibitemOpen
  \bibfield  {author} {\bibinfo {author} {\bibfnamefont {M.~Y.}\ \bibnamefont
  {Lavrentiev}}, \bibinfo {author} {\bibfnamefont {D.}~\bibnamefont {{Nguyen
  Manh}}}, \ and\ \bibinfo {author} {\bibfnamefont {S.~L.}\ \bibnamefont
  {Dudarev}},\ }\href {\doibase 10.4028/www.scientific.net/SSP.172-174.1002}
  {\bibfield  {journal} {\bibinfo  {journal} {Solid State Phenom.}\ }\textbf
  {\bibinfo {volume} {172-174}},\ \bibinfo {pages} {1002} (\bibinfo {year}
  {2011})}\BibitemShut {NoStop}%
\bibitem [{\citenamefont {Nguyen-Manh}\ \emph {et~al.}(2007)\citenamefont
  {Nguyen-Manh}, \citenamefont {Lavrentiev},\ and\ \citenamefont
  {Dudarev}}]{Nguyen-Manh2007}%
  \BibitemOpen
  \bibfield  {author} {\bibinfo {author} {\bibfnamefont {D.}~\bibnamefont
  {Nguyen-Manh}}, \bibinfo {author} {\bibfnamefont {M.~Y.}\ \bibnamefont
  {Lavrentiev}}, \ and\ \bibinfo {author} {\bibfnamefont {S.~L.}\ \bibnamefont
  {Dudarev}},\ }\href {\doibase 10.1007/s10820-007-9079-4} {\bibfield
  {journal} {\bibinfo  {journal} {J. Comput.-Aided Mater. Des.}\ }\textbf
  {\bibinfo {volume} {14}},\ \bibinfo {pages} {159} (\bibinfo {year}
  {2007})}\BibitemShut {NoStop}%
\bibitem [{\citenamefont {Nguyen-Manh}\ \emph {et~al.}(2008)\citenamefont
  {Nguyen-Manh}, \citenamefont {Lavrentiev},\ and\ \citenamefont
  {Dudarev}}]{Nguyen-Manh2008}%
  \BibitemOpen
  \bibfield  {author} {\bibinfo {author} {\bibfnamefont {D.}~\bibnamefont
  {Nguyen-Manh}}, \bibinfo {author} {\bibfnamefont {M.~Y.}\ \bibnamefont
  {Lavrentiev}}, \ and\ \bibinfo {author} {\bibfnamefont {S.~L.}\ \bibnamefont
  {Dudarev}},\ }\href {\doibase 10.1016/j.crhy.2007.10.011} {\bibfield
  {journal} {\bibinfo  {journal} {C. R. Physique}\ }\textbf {\bibinfo {volume}
  {9}},\ \bibinfo {pages} {379} (\bibinfo {year} {2008})}\BibitemShut {NoStop}%
\bibitem [{\citenamefont {Nguyen-Manh}\ \emph {et~al.}(2012)\citenamefont
  {Nguyen-Manh}, \citenamefont {Lavrentiev}, \citenamefont {Muzyk},\ and\
  \citenamefont {Dudarev}}]{Nguyen-Manh2012}%
  \BibitemOpen
  \bibfield  {author} {\bibinfo {author} {\bibfnamefont {D.}~\bibnamefont
  {Nguyen-Manh}}, \bibinfo {author} {\bibfnamefont {M.~Y.}\ \bibnamefont
  {Lavrentiev}}, \bibinfo {author} {\bibfnamefont {M.}~\bibnamefont {Muzyk}}, \
  and\ \bibinfo {author} {\bibfnamefont {S.~L.}\ \bibnamefont {Dudarev}},\
  }\href {\doibase 10.1007/s10853-012-6657-y} {\bibfield  {journal} {\bibinfo
  {journal} {J. Mater. Sci.}\ }\textbf {\bibinfo {volume} {47}},\ \bibinfo
  {pages} {7385} (\bibinfo {year} {2012})}\BibitemShut {NoStop}%
\bibitem [{\citenamefont {Olsson}\ \emph {et~al.}(2003)\citenamefont {Olsson},
  \citenamefont {Abrikosov}, \citenamefont {Vitos},\ and\ \citenamefont
  {Wallenius}}]{Olsson2003}%
  \BibitemOpen
  \bibfield  {author} {\bibinfo {author} {\bibfnamefont {P.}~\bibnamefont
  {Olsson}}, \bibinfo {author} {\bibfnamefont {I.~A.}\ \bibnamefont
  {Abrikosov}}, \bibinfo {author} {\bibfnamefont {L.}~\bibnamefont {Vitos}}, \
  and\ \bibinfo {author} {\bibfnamefont {J.}~\bibnamefont {Wallenius}},\ }\href
  {\doibase 10.1016/S0022-3115(03)00207-1} {\bibfield  {journal} {\bibinfo
  {journal} {J. Nucl. Mater.}\ }\textbf {\bibinfo {volume} {321}},\ \bibinfo
  {pages} {84} (\bibinfo {year} {2003})}\BibitemShut {NoStop}%
\bibitem [{\citenamefont {Olsson}\ \emph {et~al.}(2006)\citenamefont {Olsson},
  \citenamefont {Abrikosov},\ and\ \citenamefont {Wallenius}}]{Olsson2006}%
  \BibitemOpen
  \bibfield  {author} {\bibinfo {author} {\bibfnamefont {P.}~\bibnamefont
  {Olsson}}, \bibinfo {author} {\bibfnamefont {I.~A.}\ \bibnamefont
  {Abrikosov}}, \ and\ \bibinfo {author} {\bibfnamefont {J.}~\bibnamefont
  {Wallenius}},\ }\href {\doibase 10.1103/PhysRevB.73.104416} {\bibfield
  {journal} {\bibinfo  {journal} {Phys. Rev. B}\ }\textbf {\bibinfo {volume}
  {73}},\ \bibinfo {pages} {104416} (\bibinfo {year} {2006})}\BibitemShut
  {NoStop}%
\bibitem [{\citenamefont {Olsson}\ \emph {et~al.}(2007)\citenamefont {Olsson},
  \citenamefont {Domain},\ and\ \citenamefont {Wallenius}}]{Olsson2007}%
  \BibitemOpen
  \bibfield  {author} {\bibinfo {author} {\bibfnamefont {P.}~\bibnamefont
  {Olsson}}, \bibinfo {author} {\bibfnamefont {C.}~\bibnamefont {Domain}}, \
  and\ \bibinfo {author} {\bibfnamefont {J.}~\bibnamefont {Wallenius}},\ }\href
  {\doibase 10.1103/PhysRevB.75.014110} {\bibfield  {journal} {\bibinfo
  {journal} {Phys. Rev. B}\ }\textbf {\bibinfo {volume} {75}},\ \bibinfo
  {pages} {014110} (\bibinfo {year} {2007})}\BibitemShut {NoStop}%
\bibitem [{\citenamefont {Senninger}\ \emph {et~al.}(2016)\citenamefont
  {Senninger}, \citenamefont {Soisson}, \citenamefont {Mart\'{i}nez},
  \citenamefont {Nastar}, \citenamefont {Fu},\ and\ \citenamefont
  {Br\'{e}chet}}]{Senninger2016}%
  \BibitemOpen
  \bibfield  {author} {\bibinfo {author} {\bibfnamefont {O.}~\bibnamefont
  {Senninger}}, \bibinfo {author} {\bibfnamefont {F.}~\bibnamefont {Soisson}},
  \bibinfo {author} {\bibfnamefont {E.}~\bibnamefont {Mart\'{i}nez}}, \bibinfo
  {author} {\bibfnamefont {M.}~\bibnamefont {Nastar}}, \bibinfo {author}
  {\bibfnamefont {C.-C.}\ \bibnamefont {Fu}}, \ and\ \bibinfo {author}
  {\bibfnamefont {Y.}~\bibnamefont {Br\'{e}chet}},\ }\href {\doibase
  10.1016/j.actamat.2015.09.058} {\bibfield  {journal} {\bibinfo  {journal}
  {Acta Mater.}\ }\textbf {\bibinfo {volume} {103}},\ \bibinfo {pages} {1}
  (\bibinfo {year} {2016})}\BibitemShut {NoStop}%
\bibitem [{\citenamefont {Mirebeau}\ and\ \citenamefont
  {Parette}(2010)}]{Mirebeau2010}%
  \BibitemOpen
  \bibfield  {author} {\bibinfo {author} {\bibfnamefont {I.}~\bibnamefont
  {Mirebeau}}\ and\ \bibinfo {author} {\bibfnamefont {G.}~\bibnamefont
  {Parette}},\ }\href {\doibase 10.1103/PhysRevB.82.104203} {\bibfield
  {journal} {\bibinfo  {journal} {Phys. Rev. B}\ }\textbf {\bibinfo {volume}
  {82}},\ \bibinfo {pages} {104203} (\bibinfo {year} {2010})}\BibitemShut
  {NoStop}%
\bibitem [{\citenamefont {Hardie}\ \emph {et~al.}(2013)\citenamefont {Hardie},
  \citenamefont {Williams}, \citenamefont {Xu},\ and\ \citenamefont
  {Roberts}}]{Hardie2013}%
  \BibitemOpen
  \bibfield  {author} {\bibinfo {author} {\bibfnamefont {C.~D.}\ \bibnamefont
  {Hardie}}, \bibinfo {author} {\bibfnamefont {C.~A.}\ \bibnamefont
  {Williams}}, \bibinfo {author} {\bibfnamefont {S.}~\bibnamefont {Xu}}, \ and\
  \bibinfo {author} {\bibfnamefont {S.~G.}\ \bibnamefont {Roberts}},\ }\href
  {\doibase 10.1016/j.jnucmat.2013.03.052} {\bibfield  {journal} {\bibinfo
  {journal} {J. Nucl. Mater.}\ }\textbf {\bibinfo {volume} {439}},\ \bibinfo
  {pages} {33} (\bibinfo {year} {2013})}\BibitemShut {NoStop}%
\bibitem [{\citenamefont {Porollo}\ \emph {et~al.}(1998)\citenamefont
  {Porollo}, \citenamefont {Dvoriashin}, \citenamefont {Vorobyev},\ and\
  \citenamefont {Konobeev}}]{Porollo1998}%
  \BibitemOpen
  \bibfield  {author} {\bibinfo {author} {\bibfnamefont {S.}~\bibnamefont
  {Porollo}}, \bibinfo {author} {\bibfnamefont {A.}~\bibnamefont {Dvoriashin}},
  \bibinfo {author} {\bibfnamefont {A.}~\bibnamefont {Vorobyev}}, \ and\
  \bibinfo {author} {\bibfnamefont {Y.}~\bibnamefont {Konobeev}},\ }\href
  {\doibase 10.1016/S0022-3115(98)00043-9} {\bibfield  {journal} {\bibinfo
  {journal} {J. Nucl. Mater.}\ }\textbf {\bibinfo {volume} {256}},\ \bibinfo
  {pages} {247} (\bibinfo {year} {1998})}\BibitemShut {NoStop}%
\bibitem [{\citenamefont {Kuksenko}\ \emph {et~al.}(2011)\citenamefont
  {Kuksenko}, \citenamefont {Pareige}, \citenamefont {Genevois}, \citenamefont
  {Cuvilly}, \citenamefont {Roussel},\ and\ \citenamefont
  {Pareige}}]{Kuksenko2011}%
  \BibitemOpen
  \bibfield  {author} {\bibinfo {author} {\bibfnamefont {V.}~\bibnamefont
  {Kuksenko}}, \bibinfo {author} {\bibfnamefont {C.}~\bibnamefont {Pareige}},
  \bibinfo {author} {\bibfnamefont {C.}~\bibnamefont {Genevois}}, \bibinfo
  {author} {\bibfnamefont {F.}~\bibnamefont {Cuvilly}}, \bibinfo {author}
  {\bibfnamefont {M.}~\bibnamefont {Roussel}}, \ and\ \bibinfo {author}
  {\bibfnamefont {P.}~\bibnamefont {Pareige}},\ }\href {\doibase
  10.1016/j.jnucmat.2011.05.042} {\bibfield  {journal} {\bibinfo  {journal} {J.
  Nucl. Mater.}\ }\textbf {\bibinfo {volume} {415}},\ \bibinfo {pages} {61}
  (\bibinfo {year} {2011})}\BibitemShut {NoStop}%
\bibitem [{\citenamefont {Lavrentiev}\ \emph {et~al.}(2018)\citenamefont
  {Lavrentiev}, \citenamefont {Nguyen-Manh},\ and\ \citenamefont
  {Dudarev}}]{Lavrentiev2018}%
  \BibitemOpen
  \bibfield  {author} {\bibinfo {author} {\bibfnamefont {M.~Y.}\ \bibnamefont
  {Lavrentiev}}, \bibinfo {author} {\bibfnamefont {D.}~\bibnamefont
  {Nguyen-Manh}}, \ and\ \bibinfo {author} {\bibfnamefont {S.~L.}\ \bibnamefont
  {Dudarev}},\ }\href {\doibase 10.1016/j.jnucmat.2017.10.038} {\bibfield
  {journal} {\bibinfo  {journal} {J. Nucl. Mater.}\ }\textbf {\bibinfo {volume}
  {499}},\ \bibinfo {pages} {613} (\bibinfo {year} {2018})}\BibitemShut
  {NoStop}%
\bibitem [{\citenamefont {Klaver}\ \emph {et~al.}(2016)\citenamefont {Klaver},
  \citenamefont {del Rio}, \citenamefont {Bonny}, \citenamefont {Eich},\ and\
  \citenamefont {Caro}}]{Klaver2016}%
  \BibitemOpen
  \bibfield  {author} {\bibinfo {author} {\bibfnamefont {T.~P.~C.}\
  \bibnamefont {Klaver}}, \bibinfo {author} {\bibfnamefont {E.}~\bibnamefont
  {del Rio}}, \bibinfo {author} {\bibfnamefont {G.}~\bibnamefont {Bonny}},
  \bibinfo {author} {\bibfnamefont {S.~M.}\ \bibnamefont {Eich}}, \ and\
  \bibinfo {author} {\bibfnamefont {A.}~\bibnamefont {Caro}},\ }\href {\doibase
  10.1016/j.commatsci.2016.04.033} {\bibfield  {journal} {\bibinfo  {journal}
  {Comput. Mater. Sci.}\ }\textbf {\bibinfo {volume} {121}},\ \bibinfo {pages}
  {204} (\bibinfo {year} {2016})}\BibitemShut {NoStop}%
\bibitem [{\citenamefont {Becquart}\ \emph {et~al.}(2018)\citenamefont
  {Becquart}, \citenamefont {{Ngayam Happy}}, \citenamefont {Olsson},\ and\
  \citenamefont {Domain}}]{Becquart2018}%
  \BibitemOpen
  \bibfield  {author} {\bibinfo {author} {\bibfnamefont {C.~S.}\ \bibnamefont
  {Becquart}}, \bibinfo {author} {\bibfnamefont {R.}~\bibnamefont {{Ngayam
  Happy}}}, \bibinfo {author} {\bibfnamefont {P.}~\bibnamefont {Olsson}}, \
  and\ \bibinfo {author} {\bibfnamefont {C.}~\bibnamefont {Domain}},\ }\href
  {\doibase 10.1016/j.jnucmat.2017.12.022} {\bibfield  {journal} {\bibinfo
  {journal} {J. Nucl. Mater.}\ }\textbf {\bibinfo {volume} {500}},\ \bibinfo
  {pages} {92} (\bibinfo {year} {2018})}\BibitemShut {NoStop}%
\bibitem [{\citenamefont {Hofmann}\ \emph {et~al.}(2015)\citenamefont
  {Hofmann}, \citenamefont {Nguyen-Manh}, \citenamefont {Gilbert},
  \citenamefont {Beck}, \citenamefont {Eliason}, \citenamefont {Maznev},
  \citenamefont {Liu}, \citenamefont {Armstrong}, \citenamefont {Nelson},\ and\
  \citenamefont {Dudarev}}]{Hofmann2015}%
  \BibitemOpen
  \bibfield  {author} {\bibinfo {author} {\bibfnamefont {F.}~\bibnamefont
  {Hofmann}}, \bibinfo {author} {\bibfnamefont {D.}~\bibnamefont
  {Nguyen-Manh}}, \bibinfo {author} {\bibfnamefont {M.~R.}\ \bibnamefont
  {Gilbert}}, \bibinfo {author} {\bibfnamefont {C.~E.}\ \bibnamefont {Beck}},
  \bibinfo {author} {\bibfnamefont {J.~K.}\ \bibnamefont {Eliason}}, \bibinfo
  {author} {\bibfnamefont {A.~A.}\ \bibnamefont {Maznev}}, \bibinfo {author}
  {\bibfnamefont {W.}~\bibnamefont {Liu}}, \bibinfo {author} {\bibfnamefont
  {D.~E.~J.}\ \bibnamefont {Armstrong}}, \bibinfo {author} {\bibfnamefont
  {K.~A.}\ \bibnamefont {Nelson}}, \ and\ \bibinfo {author} {\bibfnamefont
  {S.~L.}\ \bibnamefont {Dudarev}},\ }\href {\doibase
  10.1016/j.actamat.2015.01.055} {\bibfield  {journal} {\bibinfo  {journal}
  {Acta Mater.}\ }\textbf {\bibinfo {volume} {89}},\ \bibinfo {pages} {352}
  (\bibinfo {year} {2015})}\BibitemShut {NoStop}%
\bibitem [{\citenamefont {Mason}\ \emph {et~al.}(2019)\citenamefont {Mason},
  \citenamefont {Nguyen-Manh}, \citenamefont {Marinica}, \citenamefont
  {Alexander}, \citenamefont {Sand},\ and\ \citenamefont
  {Dudarev}}]{Mason2019}%
  \BibitemOpen
  \bibfield  {author} {\bibinfo {author} {\bibfnamefont {D.~R.}\ \bibnamefont
  {Mason}}, \bibinfo {author} {\bibfnamefont {D.}~\bibnamefont {Nguyen-Manh}},
  \bibinfo {author} {\bibfnamefont {M.-C.}\ \bibnamefont {Marinica}}, \bibinfo
  {author} {\bibfnamefont {R.}~\bibnamefont {Alexander}}, \bibinfo {author}
  {\bibfnamefont {A.~E.}\ \bibnamefont {Sand}}, \ and\ \bibinfo {author}
  {\bibfnamefont {S.~L.}\ \bibnamefont {Dudarev}},\ }\href {\doibase
  10.1063.1.5098452} {\bibfield  {journal} {\bibinfo  {journal} {J. Appl.
  Phys.}\ }\textbf {\bibinfo {volume} {126}},\ \bibinfo {pages} {075112}
  (\bibinfo {year} {2019})}\BibitemShut {NoStop}%
\bibitem [{\citenamefont {Ma}\ and\ \citenamefont
  {Dudarev}(2019{\natexlab{d}})}]{CALANIE}%
  \BibitemOpen
  \bibfield  {author} {\bibinfo {author} {\bibfnamefont {P.-W.}\ \bibnamefont
  {Ma}}\ and\ \bibinfo {author} {\bibfnamefont {S.~L.}\ \bibnamefont
  {Dudarev}},\ }\href {\doibase https://doi.org/10.1016/j.cpc.2019.107130}
  {\bibfield  {journal} {\bibinfo  {journal} {Comput. Phys. Commun.}\ }\textbf
  {\bibinfo {volume} {252}},\ \bibinfo {pages} {107130} (\bibinfo {year}
  {2019}{\natexlab{d}})}\BibitemShut {NoStop}%
\bibitem [{\citenamefont {Mura}(1987)}]{Mura1987}%
  \BibitemOpen
  \bibfield  {author} {\bibinfo {author} {\bibfnamefont {T.}~\bibnamefont
  {Mura}},\ }\href {\doibase 10.1007/978-94-009-3489-4} {\emph {\bibinfo
  {title} {{Micromechanics of Defects in Solids}}}},\ \bibinfo {edition} {2nd}\
  ed.\ (\bibinfo  {publisher} {Kluwer Academic Publishers},\ \bibinfo {address}
  {Dordrecht},\ \bibinfo {year} {1987})\BibitemShut {NoStop}%
\bibitem [{\citenamefont {Heald}\ and\ \citenamefont
  {Speight}(1975)}]{HealdSpeight1975}%
  \BibitemOpen
  \bibfield  {author} {\bibinfo {author} {\bibfnamefont {P.~T.}\ \bibnamefont
  {Heald}}\ and\ \bibinfo {author} {\bibfnamefont {M.~V.}\ \bibnamefont
  {Speight}},\ }\href@noop {} {\bibfield  {journal} {\bibinfo  {journal} {Acta
  Metall.}\ }\textbf {\bibinfo {volume} {23}},\ \bibinfo {pages} {1389 }
  (\bibinfo {year} {1975})}\BibitemShut {NoStop}%
\bibitem [{\citenamefont {Nye}(1985)}]{Nye1985}%
  \BibitemOpen
  \bibfield  {author} {\bibinfo {author} {\bibfnamefont {J.~F.}\ \bibnamefont
  {Nye}},\ }\href@noop {} {\emph {\bibinfo {title} {Physical Properties Of
  Crystals}}},\ Oxford Science Publications\ (\bibinfo  {publisher} {Clarendon
  Press},\ \bibinfo {address} {Oxford},\ \bibinfo {year} {1985})\BibitemShut
  {NoStop}%
\bibitem [{\citenamefont {Yamamoto}\ \emph {et~al.}(2008)\citenamefont
  {Yamamoto}, \citenamefont {Kitamura},\ and\ \citenamefont
  {Ogata}}]{Yamamoto2008}%
  \BibitemOpen
  \bibfield  {author} {\bibinfo {author} {\bibfnamefont {M.}~\bibnamefont
  {Yamamoto}}, \bibinfo {author} {\bibfnamefont {T.}~\bibnamefont {Kitamura}},
  \ and\ \bibinfo {author} {\bibfnamefont {T.}~\bibnamefont {Ogata}},\ }\href
  {\doibase 10.1016/j.engfracmech.2007.01.015} {\bibfield  {journal} {\bibinfo
  {journal} {Eng. Fract. Mech.}\ }\textbf {\bibinfo {volume} {75}},\ \bibinfo
  {pages} {779} (\bibinfo {year} {2008})}\BibitemShut {NoStop}%
\bibitem [{\citenamefont {Yang}\ and\ \citenamefont
  {Volinsky}(2008)}]{Yang2008}%
  \BibitemOpen
  \bibfield  {author} {\bibinfo {author} {\bibfnamefont {B.}~\bibnamefont
  {Yang}}\ and\ \bibinfo {author} {\bibfnamefont {A.}~\bibnamefont
  {Volinsky}},\ }\href {\doibase 10.1016/j.engfracmech.2007.12.008} {\bibfield
  {journal} {\bibinfo  {journal} {Eng. Fract. Mech.}\ }\textbf {\bibinfo
  {volume} {75}},\ \bibinfo {pages} {3121} (\bibinfo {year}
  {2008})}\BibitemShut {NoStop}%
\bibitem [{\citenamefont {Eshelby}(1955)}]{Eshelby1955}%
  \BibitemOpen
  \bibfield  {author} {\bibinfo {author} {\bibfnamefont {J.~D.}\ \bibnamefont
  {Eshelby}},\ }\href@noop {} {\bibfield  {journal} {\bibinfo  {journal} {Acta
  Metall.}\ }\textbf {\bibinfo {volume} {3}},\ \bibinfo {pages} {487} (\bibinfo
  {year} {1955})}\BibitemShut {NoStop}%
\bibitem [{\citenamefont {Lie}\ and\ \citenamefont {Koehler}(1968)}]{Lie1968}%
  \BibitemOpen
  \bibfield  {author} {\bibinfo {author} {\bibfnamefont {K.~H.~C.}\
  \bibnamefont {Lie}}\ and\ \bibinfo {author} {\bibfnamefont {J.~S.}\
  \bibnamefont {Koehler}},\ }\href {\doibase {10.1080/00018736800101326}}
  {\bibfield  {journal} {\bibinfo  {journal} {Adv. Phys.}\ }\textbf {\bibinfo
  {volume} {17}},\ \bibinfo {pages} {421} (\bibinfo {year} {1968})}\BibitemShut
  {NoStop}%
\bibitem [{\citenamefont {Hudson}\ \emph {et~al.}(2005)\citenamefont {Hudson},
  \citenamefont {Dudarev}, \citenamefont {Caturla},\ and\ \citenamefont
  {Sutton}}]{Hudson2005}%
  \BibitemOpen
  \bibfield  {author} {\bibinfo {author} {\bibfnamefont {T.~S.}\ \bibnamefont
  {Hudson}}, \bibinfo {author} {\bibfnamefont {S.~L.}\ \bibnamefont {Dudarev}},
  \bibinfo {author} {\bibfnamefont {M.~J.}\ \bibnamefont {Caturla}}, \ and\
  \bibinfo {author} {\bibfnamefont {A.~P.}\ \bibnamefont {Sutton}},\ }\href
  {\doibase 10.1080/14786430412331320026} {\bibfield  {journal} {\bibinfo
  {journal} {Philos. Mag.}\ }\textbf {\bibinfo {volume} {85}},\ \bibinfo
  {pages} {661} (\bibinfo {year} {2005})}\BibitemShut {NoStop}%
\bibitem [{\citenamefont {Wr{\'{o}}bel}\ \emph {et~al.}(2012)\citenamefont
  {Wr{\'{o}}bel}, \citenamefont {{Hector Jr.}}, \citenamefont {Wolf},
  \citenamefont {Shang}, \citenamefont {Liu},\ and\ \citenamefont
  {Kurzyd{\l}owski}}]{Wrobel2012}%
  \BibitemOpen
  \bibfield  {author} {\bibinfo {author} {\bibfnamefont {J.}~\bibnamefont
  {Wr{\'{o}}bel}}, \bibinfo {author} {\bibfnamefont {L.~G.}\ \bibnamefont
  {{Hector Jr.}}}, \bibinfo {author} {\bibfnamefont {W.}~\bibnamefont {Wolf}},
  \bibinfo {author} {\bibfnamefont {S.~L.}\ \bibnamefont {Shang}}, \bibinfo
  {author} {\bibfnamefont {Z.~K.}\ \bibnamefont {Liu}}, \ and\ \bibinfo
  {author} {\bibfnamefont {K.~J.}\ \bibnamefont {Kurzyd{\l}owski}},\ }\href
  {\doibase 10.1016/j.jallcom.2011.09.085} {\bibfield  {journal} {\bibinfo
  {journal} {J. Alloys Compd.}\ }\textbf {\bibinfo {volume} {512}},\ \bibinfo
  {pages} {296} (\bibinfo {year} {2012})}\BibitemShut {NoStop}%
\bibitem [{\citenamefont {Zhang}\ \emph
  {et~al.}(2007{\natexlab{a}})\citenamefont {Zhang}, \citenamefont {Zhang},
  \citenamefont {Xu},\ and\ \citenamefont {Ji}}]{Zhang2007}%
  \BibitemOpen
  \bibfield  {author} {\bibinfo {author} {\bibfnamefont {J.-M.}\ \bibnamefont
  {Zhang}}, \bibinfo {author} {\bibfnamefont {Y.}~\bibnamefont {Zhang}},
  \bibinfo {author} {\bibfnamefont {K.-W.}\ \bibnamefont {Xu}}, \ and\ \bibinfo
  {author} {\bibfnamefont {V.}~\bibnamefont {Ji}},\ }\href {\doibase
  10.1016/j.jpcs.2007.01.025} {\bibfield  {journal} {\bibinfo  {journal} {J.
  Phys. Chem. Solids}\ }\textbf {\bibinfo {volume} {68}},\ \bibinfo {pages}
  {503} (\bibinfo {year} {2007}{\natexlab{a}})}\BibitemShut {NoStop}%
\bibitem [{\citenamefont {Zhang}\ \emph
  {et~al.}(2007{\natexlab{b}})\citenamefont {Zhang}, \citenamefont {Zhang},
  \citenamefont {Xu},\ and\ \citenamefont {Ji}}]{Zhang2007a}%
  \BibitemOpen
  \bibfield  {author} {\bibinfo {author} {\bibfnamefont {J.-M.}\ \bibnamefont
  {Zhang}}, \bibinfo {author} {\bibfnamefont {Y.}~\bibnamefont {Zhang}},
  \bibinfo {author} {\bibfnamefont {K.-W.}\ \bibnamefont {Xu}}, \ and\ \bibinfo
  {author} {\bibfnamefont {V.}~\bibnamefont {Ji}},\ }\href {\doibase
  10.1016/j.physb.2006.08.008} {\bibfield  {journal} {\bibinfo  {journal}
  {Physica B}\ }\textbf {\bibinfo {volume} {390}},\ \bibinfo {pages} {106}
  (\bibinfo {year} {2007}{\natexlab{b}})}\BibitemShut {NoStop}%
\bibitem [{\citenamefont {Landau}\ and\ \citenamefont
  {Lifshitz}(1969)}]{LandauStatPhys}%
  \BibitemOpen
  \bibfield  {author} {\bibinfo {author} {\bibfnamefont {L.~D.}\ \bibnamefont
  {Landau}}\ and\ \bibinfo {author} {\bibfnamefont {E.~M.}\ \bibnamefont
  {Lifshitz}},\ }\href@noop {} {\emph {\bibinfo {title} {{Statistical
  Physics}}}},\ \bibinfo {edition} {2nd}\ ed.\ (\bibinfo  {publisher} {Pergamon
  Press},\ \bibinfo {address} {Oxford, England},\ \bibinfo {year} {1969})\ pp.\
  \bibinfo {pages} {67--70}\BibitemShut {NoStop}%
\bibitem [{\citenamefont {Piochaud}\ \emph {et~al.}(2014)\citenamefont
  {Piochaud}, \citenamefont {Klaver}, \citenamefont {Adjanor}, \citenamefont
  {Olsson}, \citenamefont {Domain},\ and\ \citenamefont
  {Becquart}}]{Piochaud2014}%
  \BibitemOpen
  \bibfield  {author} {\bibinfo {author} {\bibfnamefont {J.~B.}\ \bibnamefont
  {Piochaud}}, \bibinfo {author} {\bibfnamefont {T.~P.~C.}\ \bibnamefont
  {Klaver}}, \bibinfo {author} {\bibfnamefont {G.}~\bibnamefont {Adjanor}},
  \bibinfo {author} {\bibfnamefont {P.}~\bibnamefont {Olsson}}, \bibinfo
  {author} {\bibfnamefont {C.}~\bibnamefont {Domain}}, \ and\ \bibinfo {author}
  {\bibfnamefont {C.~S.}\ \bibnamefont {Becquart}},\ }\href {\doibase
  10.1103/PhysRevB.89.024101} {\bibfield  {journal} {\bibinfo  {journal} {Phys.
  Rev. B}\ }\textbf {\bibinfo {volume} {89}},\ \bibinfo {pages} {024101}
  (\bibinfo {year} {2014})}\BibitemShut {NoStop}%
\bibitem [{\citenamefont {Kresse}\ and\ \citenamefont {Joubert}(1999)}]{PAW}%
  \BibitemOpen
  \bibfield  {author} {\bibinfo {author} {\bibfnamefont {G.}~\bibnamefont
  {Kresse}}\ and\ \bibinfo {author} {\bibfnamefont {D.}~\bibnamefont
  {Joubert}},\ }\href {\doibase 10.1103/PhysRevB.59.1758} {\bibfield  {journal}
  {\bibinfo  {journal} {Phys. Rev. B}\ }\textbf {\bibinfo {volume} {59}},\
  \bibinfo {pages} {1758} (\bibinfo {year} {1999})}\BibitemShut {NoStop}%
\bibitem [{\citenamefont {Bl\"ochl}(1994)}]{BlochPAW}%
  \BibitemOpen
  \bibfield  {author} {\bibinfo {author} {\bibfnamefont {P.~E.}\ \bibnamefont
  {Bl\"ochl}},\ }\href {\doibase 10.1103/PhysRevB.50.17953} {\bibfield
  {journal} {\bibinfo  {journal} {Phys. Rev. B}\ }\textbf {\bibinfo {volume}
  {50}},\ \bibinfo {pages} {17953} (\bibinfo {year} {1994})}\BibitemShut
  {NoStop}%
\bibitem [{\citenamefont {Kresse}\ and\ \citenamefont
  {Furthm\"uller}(1996)}]{Kresse1}%
  \BibitemOpen
  \bibfield  {author} {\bibinfo {author} {\bibfnamefont {G.}~\bibnamefont
  {Kresse}}\ and\ \bibinfo {author} {\bibfnamefont {J.}~\bibnamefont
  {Furthm\"uller}},\ }\href {\doibase 10.1103/PhysRevB.54.11169} {\bibfield
  {journal} {\bibinfo  {journal} {Phys. Rev. B}\ }\textbf {\bibinfo {volume}
  {54}},\ \bibinfo {pages} {11169} (\bibinfo {year} {1996})}\BibitemShut
  {NoStop}%
\bibitem [{\citenamefont {Kresse}\ and\ \citenamefont
  {Furthm{\"u}ller}(1996)}]{Kresse2}%
  \BibitemOpen
  \bibfield  {author} {\bibinfo {author} {\bibfnamefont {G.}~\bibnamefont
  {Kresse}}\ and\ \bibinfo {author} {\bibfnamefont {J.}~\bibnamefont
  {Furthm{\"u}ller}},\ }\href@noop {} {\bibfield  {journal} {\bibinfo
  {journal} {Comput. Mater. Sci.}\ }\textbf {\bibinfo {volume} {6}},\ \bibinfo
  {pages} {15} (\bibinfo {year} {1996})}\BibitemShut {NoStop}%
\bibitem [{\citenamefont {Perdew}\ \emph {et~al.}(1996)\citenamefont {Perdew},
  \citenamefont {Burke},\ and\ \citenamefont {Ernzerhof}}]{PBE}%
  \BibitemOpen
  \bibfield  {author} {\bibinfo {author} {\bibfnamefont {J.~P.}\ \bibnamefont
  {Perdew}}, \bibinfo {author} {\bibfnamefont {K.}~\bibnamefont {Burke}}, \
  and\ \bibinfo {author} {\bibfnamefont {M.}~\bibnamefont {Ernzerhof}},\ }\href
  {\doibase 10.1103/PhysRevLett.77.3865} {\bibfield  {journal} {\bibinfo
  {journal} {Phys. Rev. Lett.}\ }\textbf {\bibinfo {volume} {77}},\ \bibinfo
  {pages} {3865} (\bibinfo {year} {1996})}\BibitemShut {NoStop}%
\bibitem [{\citenamefont {Nguyen-Manh}\ \emph {et~al.}(2015)\citenamefont
  {Nguyen-Manh}, \citenamefont {Ma}, \citenamefont {Lavrentiev},\ and\
  \citenamefont {Dudarev}}]{Nguyen-Manh2015}%
  \BibitemOpen
  \bibfield  {author} {\bibinfo {author} {\bibfnamefont {D.}~\bibnamefont
  {Nguyen-Manh}}, \bibinfo {author} {\bibfnamefont {P.-W.}\ \bibnamefont {Ma}},
  \bibinfo {author} {\bibfnamefont {M.~Y.}\ \bibnamefont {Lavrentiev}}, \ and\
  \bibinfo {author} {\bibfnamefont {S.~L.}\ \bibnamefont {Dudarev}},\ }\href
  {\doibase 10.1016/j.annucene.2014.10.042} {\bibfield  {journal} {\bibinfo
  {journal} {Annals of Nuclear Energy}\ }\textbf {\bibinfo {volume} {77}},\
  \bibinfo {pages} {246} (\bibinfo {year} {2015})}\BibitemShut {NoStop}%
\bibitem [{\citenamefont {Monkhorst}\ and\ \citenamefont
  {Pack}(1976)}]{Monkhorst}%
  \BibitemOpen
  \bibfield  {author} {\bibinfo {author} {\bibfnamefont {H.~J.}\ \bibnamefont
  {Monkhorst}}\ and\ \bibinfo {author} {\bibfnamefont {J.~D.}\ \bibnamefont
  {Pack}},\ }\href {\doibase 10.1103/PhysRevB.13.5188} {\bibfield  {journal}
  {\bibinfo  {journal} {Phys. Rev. B}\ }\textbf {\bibinfo {volume} {13}},\
  \bibinfo {pages} {5188} (\bibinfo {year} {1976})}\BibitemShut {NoStop}%
\bibitem [{\citenamefont {Castin}\ \emph {et~al.}(2017)\citenamefont {Castin},
  \citenamefont {Messina}, \citenamefont {Domain}, \citenamefont {Pasianot},\
  and\ \citenamefont {Olsson}}]{Castin2017}%
  \BibitemOpen
  \bibfield  {author} {\bibinfo {author} {\bibfnamefont {N.}~\bibnamefont
  {Castin}}, \bibinfo {author} {\bibfnamefont {L.}~\bibnamefont {Messina}},
  \bibinfo {author} {\bibfnamefont {C.}~\bibnamefont {Domain}}, \bibinfo
  {author} {\bibfnamefont {R.~C.}\ \bibnamefont {Pasianot}}, \ and\ \bibinfo
  {author} {\bibfnamefont {P.}~\bibnamefont {Olsson}},\ }\href {\doibase
  10.1103/PhysRevB.95.214117} {\bibfield  {journal} {\bibinfo  {journal} {Phys.
  Rev. B}\ }\textbf {\bibinfo {volume} {95}},\ \bibinfo {pages} {214117}
  (\bibinfo {year} {2017})}\BibitemShut {NoStop}%
\bibitem [{\citenamefont {Dudarev}(2013)}]{Dudarev2013}%
  \BibitemOpen
  \bibfield  {author} {\bibinfo {author} {\bibfnamefont {S.~L.}\ \bibnamefont
  {Dudarev}},\ }\href {\doibase {10.1146/annurev-matsci-071312-121626}}
  {\bibfield  {journal} {\bibinfo  {journal} {Annu. Rev. Mater. Res.}\ }\textbf
  {\bibinfo {volume} {43}},\ \bibinfo {pages} {35} (\bibinfo {year}
  {2013})}\BibitemShut {NoStop}%
\bibitem [{\citenamefont {Willaime}\ \emph {et~al.}(2005)\citenamefont
  {Willaime}, \citenamefont {Fu}, \citenamefont {Marinica},\ and\ \citenamefont
  {{Dalla Torre}}}]{Willaime2005}%
  \BibitemOpen
  \bibfield  {author} {\bibinfo {author} {\bibfnamefont {F.}~\bibnamefont
  {Willaime}}, \bibinfo {author} {\bibfnamefont {C.-C.}\ \bibnamefont {Fu}},
  \bibinfo {author} {\bibfnamefont {M.-C.}\ \bibnamefont {Marinica}}, \ and\
  \bibinfo {author} {\bibfnamefont {J.}~\bibnamefont {{Dalla Torre}}},\ }\href
  {\doibase 10.1016/j.nimb.2004.10.028} {\bibfield  {journal} {\bibinfo
  {journal} {Nucl. Instr. Meth. Phys. Res. B}\ }\textbf {\bibinfo {volume}
  {228}},\ \bibinfo {pages} {92} (\bibinfo {year} {2005})}\BibitemShut
  {NoStop}%
\bibitem [{\citenamefont {Olsson}(2009)}]{Olsson2009}%
  \BibitemOpen
  \bibfield  {author} {\bibinfo {author} {\bibfnamefont {P.}~\bibnamefont
  {Olsson}},\ }\href@noop {} {\bibfield  {journal} {\bibinfo  {journal} {J.
  Nucl. Mater.}\ }\textbf {\bibinfo {volume} {386--388}},\ \bibinfo {pages}
  {86} (\bibinfo {year} {2009})}\BibitemShut {NoStop}%
\bibitem [{\citenamefont {Messina}\ \emph {et~al.}(2020)\citenamefont
  {Messina}, \citenamefont {Schuler}, \citenamefont {Nastar}, \citenamefont
  {Marinica},\ and\ \citenamefont {Olsson}}]{Messina2020}%
  \BibitemOpen
  \bibfield  {author} {\bibinfo {author} {\bibfnamefont {L.}~\bibnamefont
  {Messina}}, \bibinfo {author} {\bibfnamefont {T.}~\bibnamefont {Schuler}},
  \bibinfo {author} {\bibfnamefont {M.}~\bibnamefont {Nastar}}, \bibinfo
  {author} {\bibfnamefont {M.-C.}\ \bibnamefont {Marinica}}, \ and\ \bibinfo
  {author} {\bibfnamefont {P.}~\bibnamefont {Olsson}},\ }\href {\doibase
  10.1016/j.actamat.2020.03.038} {\bibfield  {journal} {\bibinfo  {journal}
  {Acta Mater.}\ }\textbf {\bibinfo {volume} {191}},\ \bibinfo {pages} {166}
  (\bibinfo {year} {2020})}\BibitemShut {NoStop}%
\bibitem [{\citenamefont {Razumovskiy}\ \emph {et~al.}(2011)\citenamefont
  {Razumovskiy}, \citenamefont {Ruban},\ and\ \citenamefont
  {Korzhavyi}}]{Razumovskiy2011}%
  \BibitemOpen
  \bibfield  {author} {\bibinfo {author} {\bibfnamefont {V.~I.}\ \bibnamefont
  {Razumovskiy}}, \bibinfo {author} {\bibfnamefont {A.~V.}\ \bibnamefont
  {Ruban}}, \ and\ \bibinfo {author} {\bibfnamefont {P.~A.}\ \bibnamefont
  {Korzhavyi}},\ }\href {\doibase 10.1103/PhysRevB.84.024106} {\bibfield
  {journal} {\bibinfo  {journal} {Phys. Rev. B}\ }\textbf {\bibinfo {volume}
  {84}},\ \bibinfo {pages} {024106} (\bibinfo {year} {2011})}\BibitemShut
  {NoStop}%
\bibitem [{\citenamefont {Zhang}\ \emph {et~al.}(2009)\citenamefont {Zhang},
  \citenamefont {Johansson},\ and\ \citenamefont {Vitos}}]{Zhang2009}%
  \BibitemOpen
  \bibfield  {author} {\bibinfo {author} {\bibfnamefont {H.}~\bibnamefont
  {Zhang}}, \bibinfo {author} {\bibfnamefont {B.}~\bibnamefont {Johansson}}, \
  and\ \bibinfo {author} {\bibfnamefont {L.}~\bibnamefont {Vitos}},\ }\href
  {\doibase 10.1103/PhysRevB.79.224201} {\bibfield  {journal} {\bibinfo
  {journal} {Phys. Rev. B}\ }\textbf {\bibinfo {volume} {79}},\ \bibinfo
  {pages} {224201} (\bibinfo {year} {2009})}\BibitemShut {NoStop}%
\bibitem [{\citenamefont {{Preston B. A.}}(1932)}]{Preston1932}%
  \BibitemOpen
  \bibfield  {author} {\bibinfo {author} {\bibfnamefont {G.~D.}\ \bibnamefont
  {{Preston B. A.}}},\ }\href {\doibase 10.1080/14786443209461944} {\bibfield
  {journal} {\bibinfo  {journal} {London, Edinburgh Dublin Philos. Mag. J.
  Sci.}\ }\textbf {\bibinfo {volume} {13}},\ \bibinfo {pages} {419} (\bibinfo
  {year} {1932})}\BibitemShut {NoStop}%
\bibitem [{\citenamefont {Sutton}\ and\ \citenamefont
  {Hume-Rothery}(1955)}]{Sutton1955}%
  \BibitemOpen
  \bibfield  {author} {\bibinfo {author} {\bibfnamefont {A.~L.}\ \bibnamefont
  {Sutton}}\ and\ \bibinfo {author} {\bibfnamefont {W.}~\bibnamefont
  {Hume-Rothery}},\ }\href {\doibase 10.1080/14786441208521140} {\bibfield
  {journal} {\bibinfo  {journal} {London, Edinburgh Dublin Philos. Mag. J.
  Sci.}\ }\textbf {\bibinfo {volume} {46}},\ \bibinfo {pages} {1295} (\bibinfo
  {year} {1955})}\BibitemShut {NoStop}%
\bibitem [{\citenamefont {Hill}(1952)}]{Hill1952}%
  \BibitemOpen
  \bibfield  {author} {\bibinfo {author} {\bibfnamefont {R.}~\bibnamefont
  {Hill}},\ }\href@noop {} {\bibfield  {journal} {\bibinfo  {journal} {Proc.
  Phys. Soc. A}\ }\textbf {\bibinfo {volume} {65}},\ \bibinfo {pages} {349}
  (\bibinfo {year} {1952})}\BibitemShut {NoStop}%
\bibitem [{\citenamefont {Speich}\ \emph {et~al.}(1972)\citenamefont {Speich},
  \citenamefont {Schwoeble},\ and\ \citenamefont {Leslie}}]{Speich1972}%
  \BibitemOpen
  \bibfield  {author} {\bibinfo {author} {\bibfnamefont {G.~R.}\ \bibnamefont
  {Speich}}, \bibinfo {author} {\bibfnamefont {A.~J.}\ \bibnamefont
  {Schwoeble}}, \ and\ \bibinfo {author} {\bibfnamefont {W.~C.}\ \bibnamefont
  {Leslie}},\ }\href {\doibase 10.1007/BF02643211} {\bibfield  {journal}
  {\bibinfo  {journal} {Metall. Trans.}\ }\textbf {\bibinfo {volume} {3}},\
  \bibinfo {pages} {2031} (\bibinfo {year} {1972})}\BibitemShut {NoStop}%
\bibitem [{\citenamefont {Rayne}\ and\ \citenamefont
  {Chandrasekhar}(1961)}]{Rayne1961}%
  \BibitemOpen
  \bibfield  {author} {\bibinfo {author} {\bibfnamefont {J.~A.}\ \bibnamefont
  {Rayne}}\ and\ \bibinfo {author} {\bibfnamefont {B.~S.}\ \bibnamefont
  {Chandrasekhar}},\ }\href {\doibase 10.1103/PhysRev.122.1714} {\bibfield
  {journal} {\bibinfo  {journal} {Phys. Rev.}\ }\textbf {\bibinfo {volume}
  {122}},\ \bibinfo {pages} {1714} (\bibinfo {year} {1961})}\BibitemShut
  {NoStop}%
\bibitem [{\citenamefont {Dever}(1972)}]{Dever1972}%
  \BibitemOpen
  \bibfield  {author} {\bibinfo {author} {\bibfnamefont {D.}~\bibnamefont
  {Dever}},\ }\href {\doibase 10.1063/1.1661710} {\bibfield  {journal}
  {\bibinfo  {journal} {J. Appl. Phys.}\ }\textbf {\bibinfo {volume} {43}},\
  \bibinfo {pages} {3293} (\bibinfo {year} {1972})}\BibitemShut {NoStop}%
\bibitem [{\citenamefont {Ghosh}\ and\ \citenamefont
  {Olson}(2002)}]{Ghosh2002}%
  \BibitemOpen
  \bibfield  {author} {\bibinfo {author} {\bibfnamefont {G.}~\bibnamefont
  {Ghosh}}\ and\ \bibinfo {author} {\bibfnamefont {G.}~\bibnamefont {Olson}},\
  }\href {\doibase 10.1016/S1359-6454(02)00096-4} {\bibfield  {journal}
  {\bibinfo  {journal} {Acta Mater.}\ }\textbf {\bibinfo {volume} {50}},\
  \bibinfo {pages} {2655} (\bibinfo {year} {2002})}\BibitemShut {NoStop}%
\bibitem [{\citenamefont {Zhang}\ \emph {et~al.}(2013)\citenamefont {Zhang},
  \citenamefont {Wang}, \citenamefont {Punkkinen}, \citenamefont {Hertzman},
  \citenamefont {Johansson},\ and\ \citenamefont {Vitos}}]{Zhang2013}%
  \BibitemOpen
  \bibfield  {author} {\bibinfo {author} {\bibfnamefont {H.}~\bibnamefont
  {Zhang}}, \bibinfo {author} {\bibfnamefont {G.}~\bibnamefont {Wang}},
  \bibinfo {author} {\bibfnamefont {M.~P.~J.}\ \bibnamefont {Punkkinen}},
  \bibinfo {author} {\bibfnamefont {S.}~\bibnamefont {Hertzman}}, \bibinfo
  {author} {\bibfnamefont {B.}~\bibnamefont {Johansson}}, \ and\ \bibinfo
  {author} {\bibfnamefont {L.}~\bibnamefont {Vitos}},\ }\href {\doibase
  10.1088/0953-8984/25/19/195501} {\bibfield  {journal} {\bibinfo  {journal}
  {J. Phys. Condens. Matter}\ }\textbf {\bibinfo {volume} {25}},\ \bibinfo
  {pages} {195501} (\bibinfo {year} {2013})}\BibitemShut {NoStop}%
\bibitem [{\citenamefont {Derlet}\ and\ \citenamefont
  {Dudarev}(2020)}]{Derlet2020}%
  \BibitemOpen
  \bibfield  {author} {\bibinfo {author} {\bibfnamefont {P.~M.}\ \bibnamefont
  {Derlet}}\ and\ \bibinfo {author} {\bibfnamefont {S.~L.}\ \bibnamefont
  {Dudarev}},\ }\href {\doibase 10.1103/PhysRevMaterials.4.023605} {\bibfield
  {journal} {\bibinfo  {journal} {Phys. Rev. Materials}\ }\textbf {\bibinfo
  {volume} {4}},\ \bibinfo {pages} {023605} (\bibinfo {year}
  {2020})}\BibitemShut {NoStop}%
\bibitem [{\citenamefont {Greenwood}\ and\ \citenamefont
  {Earnshaw}(1997)}]{Greenwood1997}%
  \BibitemOpen
  \bibfield  {author} {\bibinfo {author} {\bibfnamefont {N.~N.}\ \bibnamefont
  {Greenwood}}\ and\ \bibinfo {author} {\bibfnamefont {A.}~\bibnamefont
  {Earnshaw}},\ }\href@noop {} {\emph {\bibinfo {title} {{Chemistry of the
  Elements}}}},\ \bibinfo {edition} {2nd}\ ed.\ (\bibinfo  {publisher}
  {Butterworth-Heinemann},\ \bibinfo {address} {Oxford},\ \bibinfo {year}
  {1997})\ p.\ \bibinfo {pages} {1600}\BibitemShut {NoStop}%
\bibitem [{\citenamefont {Jiao}\ and\ \citenamefont {Was}(2011)}]{Jiao2011}%
  \BibitemOpen
  \bibfield  {author} {\bibinfo {author} {\bibfnamefont {Z.}~\bibnamefont
  {Jiao}}\ and\ \bibinfo {author} {\bibfnamefont {G.~S.}\ \bibnamefont {Was}},\
  }\href {\doibase 10.1016/j.actamat.2011.03.070} {\bibfield  {journal}
  {\bibinfo  {journal} {Acta Mater.}\ }\textbf {\bibinfo {volume} {59}},\
  \bibinfo {pages} {4467} (\bibinfo {year} {2011})}\BibitemShut {NoStop}%
\bibitem [{\citenamefont {Domain}\ and\ \citenamefont
  {Becquart}(2018)}]{Domain2018}%
  \BibitemOpen
  \bibfield  {author} {\bibinfo {author} {\bibfnamefont {C.}~\bibnamefont
  {Domain}}\ and\ \bibinfo {author} {\bibfnamefont {C.~S.}\ \bibnamefont
  {Becquart}},\ }\href {\doibase {10.1016/j.jnucmat.2017.10.070}} {\bibfield
  {journal} {\bibinfo  {journal} {J. Nucl. Mater.}\ }\textbf {\bibinfo {volume}
  {499}},\ \bibinfo {pages} {582} (\bibinfo {year} {2018})}\BibitemShut
  {NoStop}%
\bibitem [{\citenamefont {Fu}\ \emph {et~al.}(2004)\citenamefont {Fu},
  \citenamefont {Willaime},\ and\ \citenamefont {Ordej\'on}}]{Fu2004}%
  \BibitemOpen
  \bibfield  {author} {\bibinfo {author} {\bibfnamefont {C.-C.}\ \bibnamefont
  {Fu}}, \bibinfo {author} {\bibfnamefont {F.}~\bibnamefont {Willaime}}, \ and\
  \bibinfo {author} {\bibfnamefont {P.}~\bibnamefont {Ordej\'on}},\ }\href@noop
  {} {\bibfield  {journal} {\bibinfo  {journal} {Phys. Rev. Lett.}\ }\textbf
  {\bibinfo {volume} {92}},\ \bibinfo {pages} {175503} (\bibinfo {year}
  {2004})}\BibitemShut {NoStop}%
\bibitem [{\citenamefont {Nguyen-Manh}\ \emph {et~al.}(2006)\citenamefont
  {Nguyen-Manh}, \citenamefont {Horsfield},\ and\ \citenamefont
  {Dudarev}}]{NguyenManh2006}%
  \BibitemOpen
  \bibfield  {author} {\bibinfo {author} {\bibfnamefont {D.}~\bibnamefont
  {Nguyen-Manh}}, \bibinfo {author} {\bibfnamefont {A.~P.}\ \bibnamefont
  {Horsfield}}, \ and\ \bibinfo {author} {\bibfnamefont {S.~L.}\ \bibnamefont
  {Dudarev}},\ }\href@noop {} {\bibfield  {journal} {\bibinfo  {journal} {Phys.
  Rev. B}\ }\textbf {\bibinfo {volume} {73}},\ \bibinfo {pages} {020101}
  (\bibinfo {year} {2006})}\BibitemShut {NoStop}%
\end{thebibliography}%

\end{document}